\documentclass[fleqn,usenatbib]{mnras}
\usepackage{appendix}
\usepackage{enumitem}
\usepackage{graphicx,url}
\usepackage{color}
\usepackage{amsmath}
\usepackage{amssymb}	
\usepackage{multirow}
\usepackage[colorinlistoftodos]{todonotes}
\usepackage{xcolor}

\usepackage{newtxtext,newtxmath}
\usepackage[T1]{fontenc}

\def\logxi{\mathrm{log}(\xi/\mathrm{erg}\,\mathrm{cm}\,\mathrm{s}^{-1})}
\def\dcstat{\Delta C\mbox{--}\mathrm{stat}}
\def\omeg{{\tt omeg}}

\def\hot{{\tt hot}}
\def\reds{{\tt reds}}
\def\etau{{\tt etau}}
\def\pow{{\tt pow}}
\def\bb{{\tt bb}}
\def\laor{{\tt laor}}
\def\delt{{\tt delt}}
\def\xabs{{\it xabs}}
\def\pion{{\it pion}}
\def\cie{{\it cie}}

\def\xmm{{\it XMM-Newton}}
\def\chandra{{\it Chandra}}
\def\erosit{{\it eROSITA}}
\def\sr{1H 0707}

\title[]{Wind-luminosity evolution in NLS1 AGN 1H 0707-495}

\author[Y. Xu et al.]
{Yerong Xu$^{1,2},$\thanks{E-mail: yerong.xu@inaf.it}
Ciro Pinto$^{1}$,
Stefano  Bianchi$^{3}$, 
Peter Kosec$^{4}$, 
Michael L. Parker$^{5}$,
Dominic. J. Walton$^{5}$\newauthor
Andrew C. Fabian$^{5}$, 
Matteo Guainazzi$^{6}$,
Didier Barret$^{7}$, 
Giancarlo Cusumano$^{1}$,
\\
$^{1}$INAF - IASF Palermo, Via U. La Malfa 153, I-90146 Palermo, Italy\\
$^{2}$Universit\`a degli Studi di Palermo, Dipartimento di Fisica e Chimica, via Archirafi 36, I-90123 Palermo, Italy\\
$^{3}$Dipartimento di Matematica e Fisica, Università degli Studi Roma Tre, via della Vasca Navale 84, I-00146 Roma, Italy\\
$^{4}$MIT Kavli Institute for Astrophysics and Space Research, Cambridge, MA 02139, USA\\
$^{5}$Institute of Astronomy, Madingley Road, CB3 0HA Cambridge, United Kingdom\\
$^{6}$ESA European Space Research and Technology Centre (ESTEC), Keplerlaan 1, 2201 AZ, Noordwĳk, The Netherlands\\
$^{7}$Universit\'e de Toulouse, CNRS, IRAP, 9 Avenue du colonel Roche, BP 44346, 31028 Toulouse Cedex 4, France\\
}
\date{Accepted XXX. Received YYY; in original form ZZZ}
\pubyear{2021}

\begin{document}
\label{firstpage}
\pagerange{\pageref{firstpage}--\pageref{lastpage}}
\maketitle

\begin{abstract}
Ultra-fast outflows (UFOs) have been detected in the high-quality X-ray spectra of a number of active galactic nuclei (AGN) with fairly high accretion rates and are thought to significantly contribute to the AGN feedback. After a decade of dedicated study, their launching mechanisms and structure are still not well understood, but variability techniques may provide useful constraints. In this work, therefore, we perform a flux-resolved X-ray spectroscopy on a highly accreting and variable NLS1 AGN, 1H 0707-495, using all archival \xmm\ observations to study the structure of the UFO. We find that the wind spectral lines weaken at higher luminosities, most likely due to an increasing ionization parameter as previously found in a few similar sources. Instead, the velocity is anticorrelated with the luminosity, which is opposite to the trend observed in the NLS1 IRAS 13224-3809. Furthermore, the detection of the emission lines, which are not observed in IRAS 13224-3809, indicates a wind with a larger opening angle in 1H 0707-495, presumably due to a higher accretion rate. The emitting gas is found to remain broadly constant with the luminosity. We describe the variability of the wind with a scenario where the strong radiation extends the launch radius outwards and shields the outer emitting gas, similarly to super-Eddington compact objects, although other possible explanations are discussed. Our work provides several hints for a multi-phase outflow in 1H 0707-495.

\end{abstract}

\begin{keywords}
accretion, accretion discs – black hole physics – galaxies: Seyfert - X-rays: individual: 1H 0707-495
\end{keywords}

\section{Introduction} \label{sec:intro}
It is well accepted that active galactic nuclei (AGN) are powered by accretion of matter onto supermassive black holes (SMBHs), producing enormous electromagnetic radiation covering from radio to X-rays, and even up to Gamma rays. The powerful energetic output of AGN can impact the evolution of their host galaxies via the process called AGN feedback \citep[e.g.][and references therein]{2012Fabian}. The energy is released in the form of jets and winds, which are gas outflows launched into the surrounding interstellar medium (ISM). The outflows can expel or heat the surrounding gas, and thus affect the star formation of the host galaxy and further accretion of matter onto the central SMBH \citep{2012Zubovas,2017Maiolino}. Ultra-fast outflows (UFOs) are the most extreme subset of AGN winds with velocities greater than $10\,000\,\mathrm{km}\,\mathrm{s}^{-1}$ and are believed to originate from the inner accretion disk within a few hundred gravitational radii from the black hole \citep[e.g.][]{2015Nardini,2015Tombesi}.

UFOs are thought to carry a large amount of kinetic energy ($\gtrsim0.05\,L_\mathrm{bol}$, where $L_\mathrm{bol}$ is the bolometric luminosity of AGN), which exceeds the minimum required by the simulation of an efficient AGN feedback on the host galaxy \citep{2005DiMatteo,2010Hopkins} if the energy is released into the ISM. UFOs are most commonly identified by strongly blueshifted Fe {\scriptsize XXV/XXVI} absorption features in the $7\mbox{--}10$\,keV energy band \citep{2002Chartas,2006Cappi,2009Cappi,2014Tombesi}, in agreement with highly-ionized gas ($\logxi>3$) at significant fractions of the speed of light, up to $0.4c$ \citep[e.g.][and references therein]{2010Tombesi}. These features are often preferred as they are the strongest absorption lines at high ionization, and will not blend with the lower ionization features from warm absorbers \citep{2000Porquet} or the ISM \citep{2013Pinto} in moderate resolution spectra. However, thanks to the high spectral resolution of the Reflection Grating Spectrometer (RGS) aboard \xmm\ \citep{2001denHerder} and the High Energy Transmission Gratings (HETG) aboard \chandra\ \citep{2005Canizares}, UFO soft X-ray features can be resolved, providing a useful tool to detect and study UFOs \citep[e.g.][]{2015Longinotti,2016Pounds,2018Pinto,2019Boissay}.

Radiation pressure due to a high accretion rate is a popular theory for the origin and acceleration mechanism of UFOs \citep[e.g.][]{2003Reeves,2017Matzeu}. It is therefore of great interest to investigate UFOs in sources accreting near or above their Eddington limit. Narrow Line Seyfert 1 galaxies are an ideal population, characterized by low-mass, high-accretion-rate SMBHs \citep[see review by][]{2007Komossa}. In the past few years, UFOs have been discovered in many NLS1 galaxies, e.g., IRAS 13224-3809 \citep[hereafter IRAS 13224][]{2017Parker}, WKK 4438 \citep{2018Jiang}, 1H 0707-495 \citep[hereafter 1H 0707,][]{2018Kosec}, PG 1448+273 \citep{2020Kosec}, IRAS 00521-7054 \citep[a Seyfert 2 galaxy,][]{2019Walton}, and PG 1211+143 \citep[a narrow emission line quasar,][]{2003Pounds}, showing a blueshifted Fe absorption line above 7\,keV.

Previous studies have reported the possible variable nature of UFOs based on multi-epoch observations \citep{2005Dadina,2012Dauser,2015Zoghbi,2020Kosec}. A series of deep studies on IRAS 13224, on the basis of a 1.5 Ms \xmm\ observation campaign, found an anti-correlation between the equivalent width of the absorption lines and the hard X-ray flux and suggested that the disappearance of absorption lines in the brightest state is due to the almost completely ionized gas \citep{2017Parker,2017Parkerb,2018Pinto}. It is therefore of crucial importance to determine whether such UFO variability trend is also present in other NLS1 AGN and use it to probe the wind nature.

1H 0707 is selected for three reasons: 1) 1H 0707 and IRAS 13224 share many similarities among the broadband spectra \citep{2010Zoghbi}, timing behavior \citep{2013Kara}, and UV emission features \citep{2004Leighly}. 1H 0707 shows strong UFO absorption in the low-flux state, which was suggested to weaken with increasing flux by \citet[][hereafter PK18]{2018Kosec}. 2) 1H 0707 has been observed by \xmm\ for $\sim1.4\,$Ms, providing enough counts for the flux-resolved analysis. 3) The high spectral resolution of the gratings aboard \xmm\ allows us to use also soft X-ray features rather than only iron absorption lines, which is crucial as the Fe K lines are often affected by the instrumental background.

The paper is organized as follows. We report some properties of \sr\ obtained from previous studies in Section~\ref{sec:source}. In Section~\ref{sec:data}, we present the observations and our data reduction process. Details on our analysis and results are shown in Section~\ref{sec:results}. We discuss the results and provide our conclusions in Section~\ref{sec:discussion} and Section~\ref{sec:conclusion}, respectively.

\section{1H 0707-495}\label{sec:source}

1H 0707 is a well-studied NLS1 galaxy at low redshift \citep[$z=0.04057$;][]{2009Jones}, detected by {\it HEAO I} scanning modulation collimator \citep{1986Remillard}. Its optical spectrum shows broad hydrogen lines ($\mathrm{FWHM}_\mathrm{H\beta}=1000\,\mathrm{km/s}$), very strong Fe II lines, and weak forbidden lines \citep{2004Leighly}. It hosts a rapidly spinning SMBH \citep[$a_\star=cJ/GM^2>0.98$;][]{2009Fabian} with a relatively low mass around $2\times10^{6}\,M_\odot$ \citep[e.g.][]{2005Zhou,2013Karab} or below $4\times10^{6}\,M_\odot$ \citep[][]{2016Done} depending on the measurement method adopted, accreting at a super-Eddington rate \citep[e.g.][]{2016Done}. 1H 0707 has been observed by \xmm\ over about 20 years, revealing rapid X-ray variability by a factor of a few in flux \citep[e.g.][]{2021Boller} and relatively steady UV/optical emission \citep{2017Pawar}.

The existence of a UFO has been confirmed in 1H 0707 \citep[e.g.][]{2012Dauser,2016Hagino,2018Kosec} at a velocity of $\sim0.13c$ with an ionization parameter $\logxi\sim4.3$. \citet{2009Blustin} detected Doppler-shifted emission lines and PK18 found that the velocities of the blueshifted emission increase in higher ionization species. This implies that the wind in 1H 0707 is likely stratified and is perhaps slowing down and cooling at larger distances from the SMBH. The large amount of \xmm\ data and high spectral resolution of RGS allows us to search for any correlation between the strength of the UFO absorption lines and the source X-ray luminosity, and to study the wind stratification in different regions, by using both soft and hard X-ray absorption features.

\section{Data Reduction and Products}\label{sec:data}
In this paper, we use the full \xmm\ dataset \citep{2001Jansen} of 1H 0707, including 16 observations ranging from 2000 to 2019, totaling 1.4 Ms gross archival data. We employ the data from the European Photon Imaging Camera (EPIC) with two EPIC-MOS CCDs \citep{2001Turner} and an EPIC-pn CCD \citep{2001Struder}, the Reflection Grating Spectrometers \citep[RGS;][]{2001denHerder} and the Optical Monitor \citep[OM;][]{2001Mason} in order to constrain the spectral energy distribution (SED) of the source and the absorption lines of the wind.

The data are processed with the \xmm\ Science Analysis System (SAS v19.0.0) and calibration files available on February 2021, following the standard SAS threads. Briefly, EPIC-pn and EPIC-MOS data are reduced with {\sc epproc} and {\sc emproc} tools respectively. The background flare contamination are filtered with a standard filtering criterion of 0.5 and 0.35 counts/sec (in the $10\mbox{--}12\,$keV band) for pn and MOS separately. The source spectra are extracted from a circle region with a radius of 20 arcsec centered on the object (X-ray flux peak), and the background spectra are extracted from a nearby circular region with a radius of 60 arcsec away from the source and the copper hole. We adopt a radius of 20 arcsec for the source extraction in order to decrease the background contamination following \citet{2018Chartas}. The RGS data are reduced with the {\sc rgsproc} task, for which high-background contamination is corrected by a threshold of 0.2 counts/sec. We extract the first-order RGS spectra in a cross-dispersion region of 1 arcmin width, and the background spectra are generated by photons beyond the 98 percent of the source point spread function as default. We only use the good time intervals common to both RGS 1 and 2 and stack their spectra. OM has six filters: V (5430\,\AA), B (4500\,\AA), U (3440\,\AA), UVW1 (2910\,\AA), UVM2 (2310\,\AA), UVW2 (2120\,\AA). Only image mode OM data are available, which are reduced using the standard {\sc omichain} pipeline, including all necessary calibration processes.

\subsection{Lightcurve}\label{subsec:lightcurve}
\begin{figure*}
	\includegraphics[width=\textwidth,trim={200 50 80 30}]{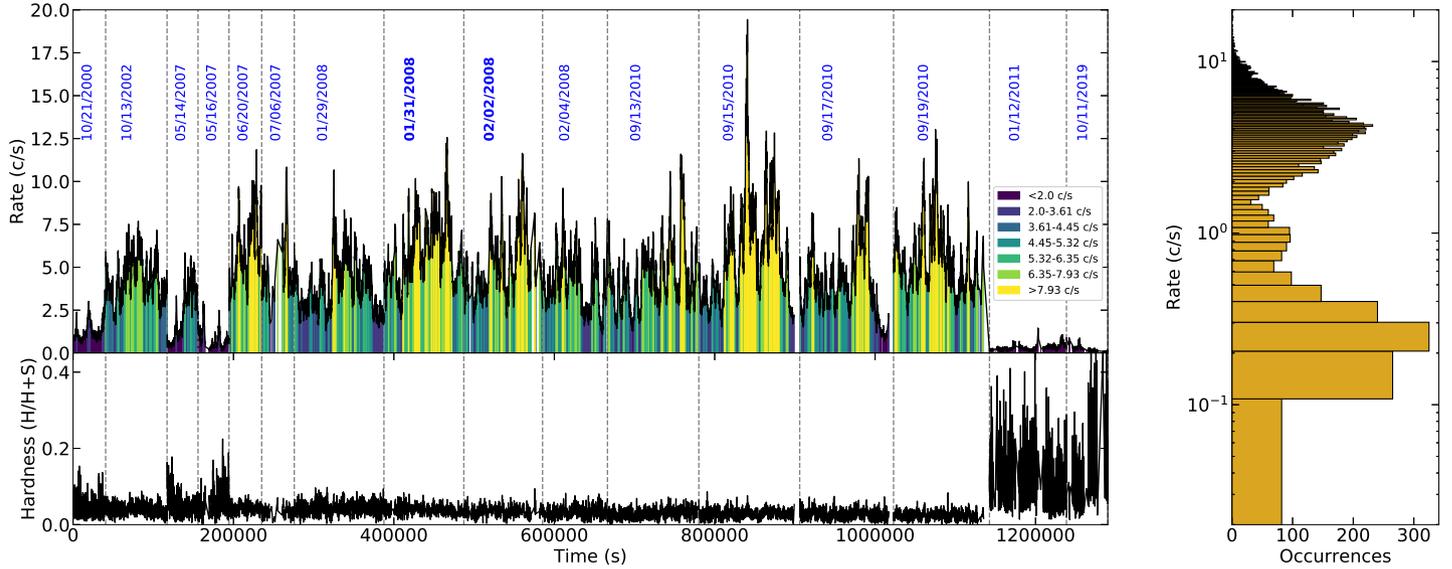}
    \caption{ {\it Upper left:} An overall EPIC-pn light curve ($0.3\mbox{--}10$\,keV) from 2000 to 2019, where the time gaps between observations are ignored and the observation dates are marked. The vertical dashed lines indicate the begin and end time for each observation. The colors represent the different flux intervals, which have comparable counts. {\it Lower left:} The corresponding X-ray hardness-ratio of the EPIC-pn light curve. S and H denote the counts in the soft and hard energy bands defined as $0.3\mbox{--}2$ and $2\mbox{--}10$\,keV, respectively. {\it Right:} The count rate histogram.
    }
    \label{fig:lc_his_hardness}
\end{figure*}

We extract the EPIC-pn ($0.3\mbox{--}10$\,keV) light curve for each observation and create an integrated light curve (see the upper left panel of Fig.~\ref{fig:lc_his_hardness}). For completeness, we have checked that the shape of the light curve for EPIC-MOS and RGS are identical to that for EPIC-pn. The corresponding hardness ratio (HR, H: $2\mbox{--}10$\,keV; S: $0.3\mbox{--}2$\,keV) and the count rate histogram are plotted separately in the lower left and right panel of Fig.~\ref{fig:lc_his_hardness}. 1H 0707 went into a low hard state after 2011. In fact, the change has been reported and explained with the light bending effect \citep{2012Fabianb}. An extreme soft variability was also discovered in the 2019 joint observation with \erosit, interpreted with partial covering absorption \citep[][Parker et al. in prep.]{2021Boller}. To investigate the variability of the UFO with the luminosity, we divide the entire light curve into 7 flux intervals using the good time interval (GTI) files generated by {\sc tabgtigen} script and the thresholds are selected so that the number of counts in each interval is comparable (see Tab.~\ref{tab:log}). The levels from the lowest to highest flux are referred to as F1, F2 ... F7, and their thresholds are marked in different colors in Fig.~\ref{fig:lc_his_hardness}. 

\begin{table}
\caption{EPIC-pn flux levels selection and exposure times}
\centering
\begin{tabular}{lcccccr}
\hline
Level & Minimum & Maximum & Frame$^a$ & Exposure   \\
 & (counts s$^{-1}$) & (counts s$^{-1}$) & (ks) & (ks) \\
\hline
F1 & 0.0 & 2.0 & 1.15 & 215 \\
F2 & 2.0 & 3.61 & 0.54 & 254 \\
F3 & 3.61 & 4.45 & 0.28 & 183\\
F4 & 4.45 & 5.32 & 0.24 & 152\\
F5 & 5.32 & 6.35 & 0.23  & 128\\
F6 & 6.35 & 7.93 & 0.29 & 106\\
F7 & 7.93 & 20.0 & 0.59 & 77\\
\hline
\end{tabular}
\label{tab:log}
\begin{flushleft}
{$^{a}$ Frame means the average duration time for each segment in the EPIC-pn light curve for a given flux level.}
\end{flushleft}
\end{table}

\subsection{Time-average and Flux-resolved Spectra}\label{subsec:spectra}

We stack the EPIC ($0.4\mbox{--}10$\,keV) and RGS ($0.4\mbox{--}1.77$\,keV) spectra from different observations into time-average broad-band spectra using {\sc epicspeccombine} and {\sc rgscombine} scripts. RGS 1 and 2 spectra are combined together to increase the signal-to-noise ratio per energy bin. We cut off the energy coverage below 0.4\,keV because we focus on the wind features, which usually appear well below 30 {\AA} ($\geq0.4$\,keV), and want to avoid the RGS high background regions (see the right panel of Fig.~\ref{fig:spectra}). To obtain flux-resolved spectra, we split each observation according to the defined count rate limits (shown in Fig.~\ref{fig:lc_his_hardness}) by applying the generated GTI files and extract the spectra following the same steps for time-resolved spectra. We then obtain 21 EPIC (one pn and two MOS for each flux level) and 7 RGS spectra by stacking the spectra within same flux level. 

For clarity, only EPIC-pn and RGS source and background spectra are shown in Fig.~\ref{fig:spectra}. The spectra are plotted in the same colors as the thresholds in Fig.~\ref{fig:lc_his_hardness}. The $\sim1.5$\,keV peak in the EPIC background spectra is due to the blend of the strong instrumental Al K and Si K emission lines \citep[e.g.][]{2002Lumb}. The background spectrum is comparable to the source spectra above 8\,keV for EPIC and on both ends ($>30\,\AA$ and $<7\,\AA$) for RGS. The typical UFO feature in EPIC spectra is the broad Fe absorption among $7\mbox{--}8$\,keV, although there are other absorption lines of Si {\scriptsize XIV} and S {\scriptsize XVI} between 1.5 and 3 keV (see PK18). The two dips in the RGS spectra near 13 and 21 {\AA} are not UFO features but RGS instrumental chip gaps. They will not affect our detection of UFOs since these dips are very narrow.

We check the OM light curve for different filters and find that the variability amplitudes of each light curve are less than 20\%, which is orders of magnitude smaller than the X-ray variability (see the light curve in Fig.~\ref{fig:lc_his_hardness}), as commonly observed in NLS1s \citep{2013Ai}. Therefore, we assume that the emission in each OM band is steady and only extract the time-average spectra of different filters. The `canned' response files are used in our spectral analysis\footnote{https://www.cosmos.esa.int/web/xmm-newton/om-response-files}. 

In order to obtain high quality data and avoid over-sampling for spectroscopy, we group EPIC spectra with {\sc specgroup} task, and employ {\tt obin} command for RGS and EPIC spectra in SPEX so that each group is not narrower than 1/3 of the spectral resolution, enabling us to resolve emission and absorption lines in soft energies \citep{2016Kaastra}.

\section{Results}\label{sec:results}
\subsection{EPIC-RGS Continuum Modelling}\label{subsec:EPIC-RGS}
We use the SPEX (v3.06.01) package \citep{1996Kaastra} in this paper to analyze the spectra. All \xmm\ spectra are thus transferred into the format which can be read by SPEX using {\sc ogip2spex} script. We employ the Cash-statistics \citep{1979Cash} and estimate all parameter uncertainties at the 68\% confidence level. The calibration differences between the detectors are taken into account by allowing a relative instrument normalization free to vary (with that of EPIC-pn fixed to unity). The Galactic ISM absorption is described by the \hot\ model, using the default proto-solar abundances of \citet{2009Lodders}. The temperature is fixed at $0.02$\,eV to have a quasi-neutral ISM. The Galactic column density is set at $N_\mathrm{H}^\mathrm{Gal}=4\times10^{20}\,\mathrm{cm}^{-2}$ \citep{2005Kalberla} during the spectral fitting. The redshift of the source is also taken into account by a \reds\ model with a fixed redshift $z=0.04057$.

We begin the broadband X-ray spectroscopy by fitting the high-quality time-average spectra (EPIC+RGS). After checking the consistency of the residuals between EPIC and RGS spectra to the continuum model in soft energies, we adopt EPIC and RGS data between $1.77\mbox{--}10$ and $0.4\mbox{--}1.77$\,keV respectively, because EPIC has a relatively low spectral resolution in the soft band but complements the RGS in the hard band. The physical properties of the spectral components in 1H 0707 continuum have been discussed in many papers with great detail \citep[e.g.][]{2012Dauser,2012Fabianb,2021Boller} and are beyond the purpose of this paper. Therefore, we just describe the continuum with phenomenological models as we mainly focus on the outflow properties. The spectral shape of 1H 0707 has been reported to be characterized by a power-law component, a soft excess, two strong skewed and broad emission lines \citep{2009Fabian}. We initially applied a typical continuum model ({\tt hot*reds*(pow+bb)}) to the time-average spectra, where the blackbody reproduces the soft excess, and clearly found two broad emission lines peaking at around 0.9 and 6.5\,keV.

Therefore, the data can be described by a continuum model plus two broad lines, which are related to the inner-disk reflections: \hot*\reds*(\pow+\bb+\laor*(\delt+\delt)). This model was also adopted by \citet{2009Fabian}, which physically means {\tt ISM*redshift*(continuum+rel-broad*(Fe L+Fe K))}. The result reveals a power-law component with a photon index of $\Gamma=2.75^{+0.01}_{-0.01}$, a soft excess described by a black body with a temperature of $T_\mathrm{in}=0.123^{+0.001}_{-0.001}$\,keV, and two relativistically-broadened emission lines characterized by energies of $\sim0.9$ and $\sim6.5$\,keV, an innermost radius of $\sim1.2\,r_\mathrm{g}$ ($r_\mathrm{g}=GM_\mathrm{BH}/c^{2}$), a fixed outermost radius of $400\,r_\mathrm{g}$, an emissivity index of $q=5.3^{+0.1}_{-0.1}$ and an inclination angle of $50.7^\circ$. The residuals of this model are depicted on the top panel of Fig.~\ref{fig:ratio}. The residuals in the RGS band unveil several possible P-cygni-like profiles, where the emission is combined with a blueshifted absorption, interpreted by an expanding gaseous envelope. The main emission features are located at around 0.5, 0.65, 0.85, 1\,keV and are very close to the rest-frame positions of O {\scriptsize VIII}, N {\scriptsize VII}, Fe {\scriptsize XVII} and Ne {\scriptsize X} transitions. 

This phenomenological model is then applied to the seven flux-resolved spectra. Since the inclination angle, the innermost and outermost radius are not expected to significantly change within the observation period, we fix them at the best-fit values derived from the fit to time-average spectra. The residuals and the parameters are summarized in Fig.~\ref{fig:ratio} and Tab.~\ref{tab:ratios}, respectively. The spectral slope gradually increases with the flux level, consistent with the `softer-when-brighter' trend shown in Fig.~\ref{fig:lc_his_hardness}. The statistics suggest a general tendency of a better fit appearing at a higher flux level (i.e. the C-statistics are smaller for comparable degrees of freedom). This indicates that the residuals are weaker when the flux is higher (see Fig.~\ref{fig:ratio}), compatible what was previously seen in IRAS 13224 \citep[e.g.][]{2017Parker,2018Pinto}.

\subsection{Gaussian Line Scan}\label{subsec:Line-Scan}

To confirm the weakening trend for the absorption lines with the luminosity, following the approach used in \citet{2018Pinto}, we search for narrow spectral features by fitting a Gaussian line spanning the $0.4\mbox{--}10$\,keV energies. We adopt a logarithmic grid of 1000 energy steps between 0.4 and 10\,keV, which maintains the balance between the computational cost and the resolving power. The resolution of the line search is comparable to the RGS spectral resolution ($R_\mathrm{RGS}\sim100\mbox{--}500$) and a few times over the EPIC resolution ($R_\mathrm{pn}\sim20\mbox{--}50$). In each step, we fix the line energy and width at given values, and allow the normalization of the Gaussian to be positive or negative, to reproduce emission and absorption lines, respectively. And the parameters of the broadband model are left free. The improved $\dcstat$ compared to the best-fit continuum model is recorded at each step. For the high S/N ratio and easy recognition of the lines, we start the scan for the time-average spectra and test several line widths between 500\,km/s and 10000\,km/s.

\begin{figure*}
\centering
\includegraphics[width=\textwidth,trim={0 0.0cm 0 0}]{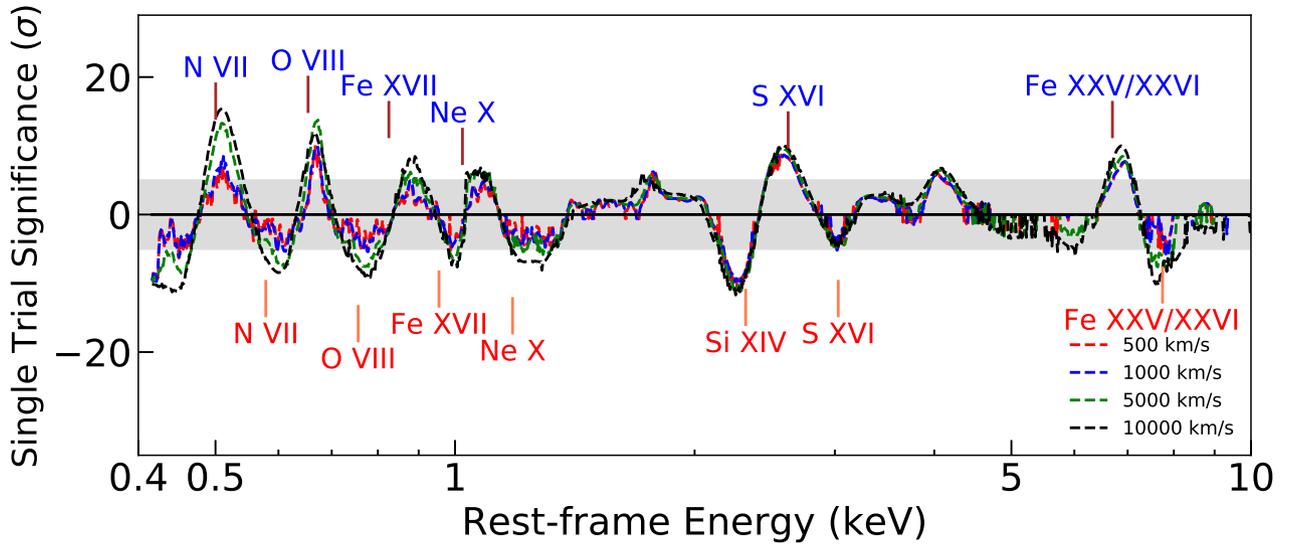}
\caption{
The results of a line search over the \xmm\ broadband time-average spectra ($0.4\mbox{--}1.77$\,keV from RGS and $1.77\mbox{--}10$\,keV from EPIC) in the AGN rest frame. Four different line widths $\sigma_v$ (FWHM$=2.355\sigma_v$) of 500, 1000, 5000, and 10000\,km/s are adopted. Thesingle trial significance is calculated as the square root of $\dcstat$ times the sign of the Gaussian normalization to distinguish the emission and absorption features. The emission/absorption lines are identified as blue/red. The emission lines are labeled at the rest-frame of the transition, while the absorption lines matches the blueshift around $z\sim-0.15$. The shade region corresponds to $5\sigma$ of the single trial significance.
}
\label{fig:gaussian}
\end{figure*}

The results of the line scan for the time-average spectra are expressed as the square root of $\dcstat$ times the sign of the Gaussian normalization in Fig.~\ref{fig:gaussian}. This quantity approximately corresponds to the single trial detection significance of each Gaussian feature (the look-elsewhere effect ignored) \citep[][]{1979Cash,2016Kaastra}. Based on SPEX (\citealt{2008Kaastra}) and {\tt ATOMDB}\footnote{http://www.atomdb.org/Webguide/webguide.php}, the emission lines are identified and the corresponding rest-frame energies are labelled. Since the UFO has been confirmed by PK18, the drops next to peaks are assumed to be the blueshifted absorption lines corresponding to the identified emissions. After several trials of shifting the rest-frame absorption lines, we find that many of absorption lines match a typical UFO blueshift of $z\sim-0.15$: N {\scriptsize VII}, O {\scriptsize VIII}, Fe {\scriptsize XVII}, Ne {\scriptsize X}, Si {\scriptsize XIV}, S {\scriptsize XVI}, and Fe {\scriptsize XXV/XXVI}. The Si {\scriptsize XIV} emission line is not detected possibly due to the superposition of the absorption line from another ion. Our results are consistent with most of the absorption lines already reported in PK18, which focused on the \xmm\ observations before 2010. As we cannot rule out the possibility that the features in the $2\mbox{--}3$\,keV range are affected by instrumental effects, we prefer to pay more attention onto other strong UFO signatures in this paper. Details of the identified lines are listed in Tab.\ref{tab:lines}. Ideally, a Monte Carlo simulation should be done to rule out the possibility of the look-elsewhere effect. However, we do not do this because of the extremely high significance of the features (each feature has a $\Delta C\mbox{--}\mathrm{stat}>25$ detection) with the Monte Carlo simulations basically being redundant to corroborate the detection, and the strong wind detection from the following photonionization models on time-average and flux-resolved spectra (see Fig.\ref{fig:XABSflux}, \ref{fig:PIONflux}, and Tab.\ref{tab:evolution}).

\begin{figure}
\centering
\includegraphics[width=0.48\textwidth,trim={0 90 0 0}]{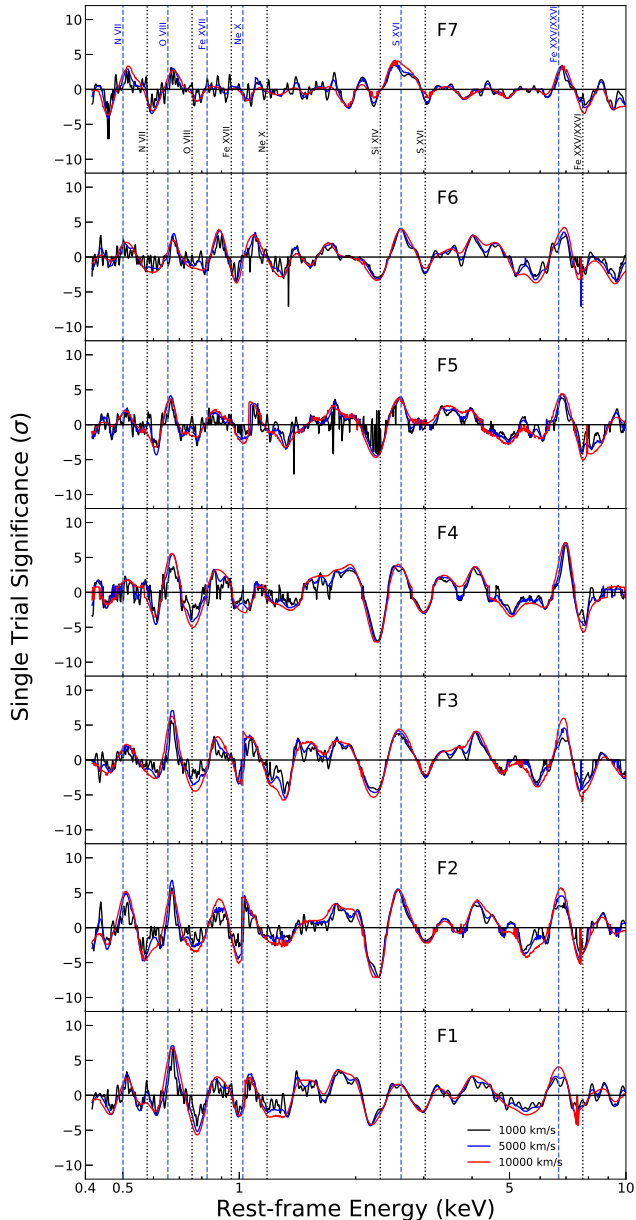}
\caption{The results of a line search over the flux-resolved RGS-EPIC spectra. Three different line widths (1000, 5000, 10000\,km/s) are adopted. The emission and absorption features detected in Fig.~\ref{fig:gaussian} are labelled by vertical blue and black dash lines separately.
}
\label{fig:gaussian_flux_resolved}
\end{figure}

The Gaussian line scan with the same primary settings is then applied to the flux-resolved spectra and the results are shown in Fig.~\ref{fig:gaussian_flux_resolved}. As previously reported UFOs in other sources \citep[e.g.][]{2018Pinto}, there is a general trend that the significance of the flux-resolved spectra residuals weaken at high fluxes. For individual lines, only Fe {\scriptsize XVII} and Ne {\scriptsize X} features nearly disappear in the highest flux spectra, while the others, no matter absorption or emission lines, are weak but are still recognizable in the spectrum. This implies that the UFO does not completely disappear in 1H 0707, as PK18 suggested. The variable position of the absorption residuals with respect to the vertical dashed lines indicates that the velocity of the UFO changes with the luminosity, which is investigated with the aid of physical models in Section~\ref{subsubsec:fullwind}.

\begin{table}
\centering
\caption{The strongest transition residuals fitted with a Gaussian.}\label{tab:lines}
\begin{tabular}{lccccccr}
\hline
Transition & $\Delta$C-stat & Rest-frame & Best-fitting$^{a}$ & Norm & FWHM   \\
           &                &  energy  (keV)     &   energy (keV)                 & ($10^{50}$ ph/s)    & (keV)   \\
\hline
\multicolumn{6}{c}{Absorption}\\
\hline
N VII  & 69 & 0.500 &  $0.601^{+0.004}_{-0.004}$ & $-1.7^{+0.2}_{-0.3}$ & $0.047^{+0.006}_{-0.007}$ \\
O VIII & 110 & 0.653 & $0.767^{+0.003}_{-0.003}$ & $-1.7^{+0.2}_{-0.2}$ & $0.093^{+0.008}_{-0.007}$\\
Fe XVII & 65 & 0.826 &  $0.996^{+0.003}_{-0.003}$ & $-1.0^{+0.2}_{-0.2}$ & $0.062^{+0.009}_{-0.008}$ \\
Ne X & 122 & 1.022 & $1.216^{+0.009}_{-0.010}$ & $-1.7^{+0.4}_{-0.2}$ & $0.28^{+0.02}_{-0.02}$  \\
Fe XXV/XXVI & 106 & 6.699/6.973 & $7.70^{+0.02}_{-0.02}$ & $-0.12^{+0.02}_{-0.01}$ & $0.81^{+0.04}_{-0.08}$\\
\hline
\multicolumn{6}{c}{Emission}\\
\hline
N VII  & 251 & 0.500 & $0.508^{+0.001}_{-0.002}$ & $3.8^{+0.5}_{-0.3}$ & $0.046^{+0.005}_{-0.003}$  \\
O VIII & 202 & 0.653 & $0.669^{+0.001}_{-0.001}$ & $1.1^{+0.1}_{-0.1}$ & $0.021^{+0.003}_{-0.002}$ \\
Fe XVII & 79 & 0.826 &  $0.884^{+0.003}_{-0.004}$ & $1.1^{+0.2}_{-0.2}$ & $0.086^{+0.008}_{-0.011}$ \\
Ne X & 49 & 1.022 &  $1.075^{+0.003}_{-0.007}$ & $0.58^{+0.10}_{-0.05}$ & $0.066^{+0.008}_{-0.006}$ \\
Fe XXV/XXVI & 121 & 6.699/6.973 & $6.86^{+0.02}_{-0.02}$ & $0.08^{+0.02}_{-0.01}$ & $0.66^{+0.07}_{-0.05}$\\
\hline
\end{tabular}
\begin{flushleft}
{$^{a}$ The best-fitting energy of the Gaussian is in the AGN rest-frame.}
\end{flushleft}
\end{table}
\subsection{Wind Modelling}\label{subsec:windmodel}
In this section, we employ the photoionization code {\it pion} in the SPEX package to model the wind in 1H 0707 and to study the emission/absorption lines produced by the ionized gas. This model self-consistently calculates the transmission and emission of a slab of matter, where the ionic column densities are linked through photoionization equilibrium. The relevant parameter is the ionization parameter: $\xi=L_\mathrm{ion}/n_\mathrm{H}R^2$, with $L_\mathrm{ion}$ the ionizing luminosity, $n_\mathrm{H}$ the hydrogen volume density and $R$ the distance from the ionizing source to the matter. Once a radiation field is provided, this model reproduces the X-ray spectrum of a photoionized plasma. 

\subsubsection{SED and thermal stability}\label{subsubsec:SEDandion}

\begin{figure}
\centering
\includegraphics[width=0.49\textwidth,trim={0 0.0cm 0 0}]{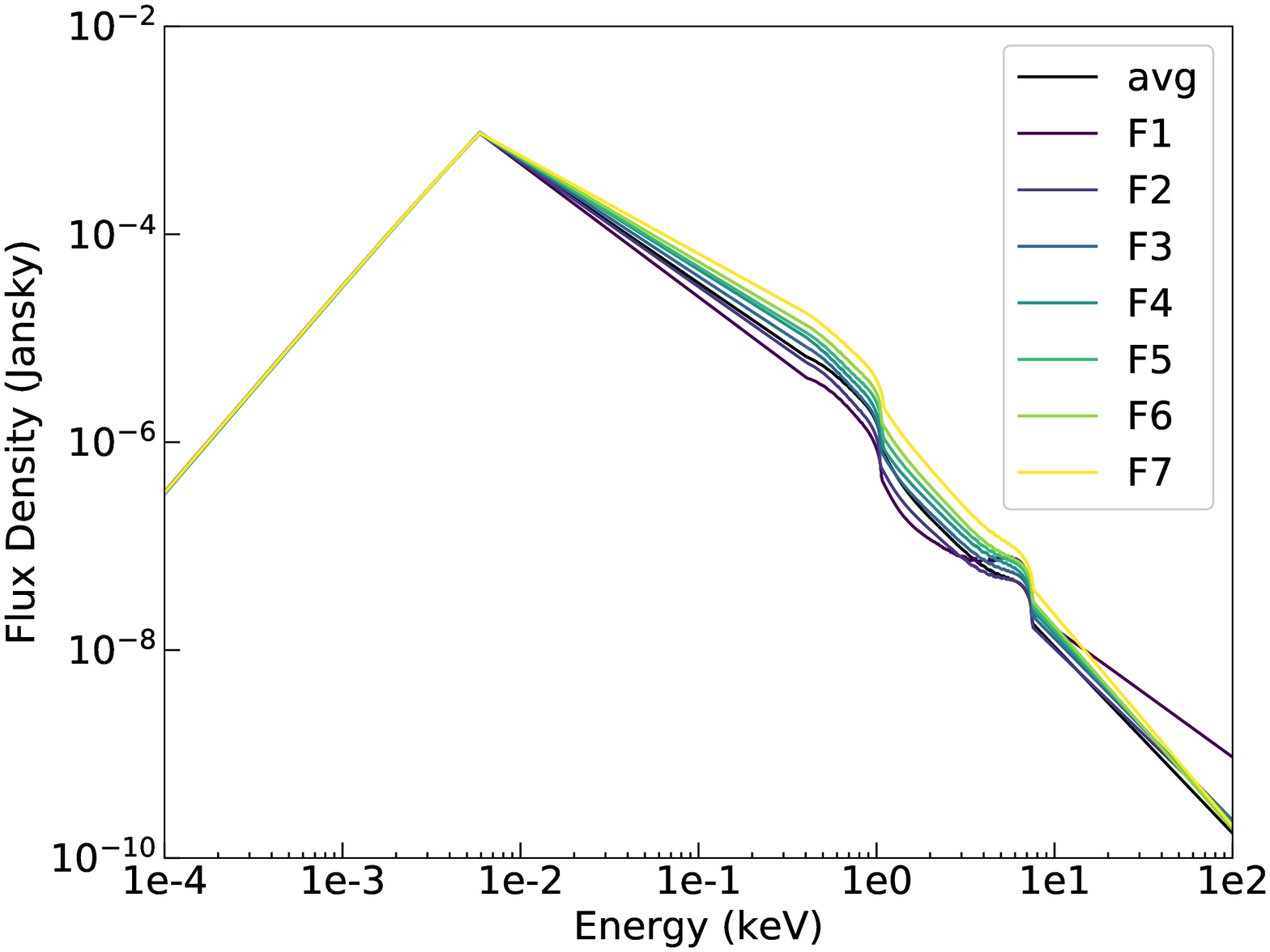}
\includegraphics[width=0.49\textwidth,trim={0 0.0cm 0 0}]{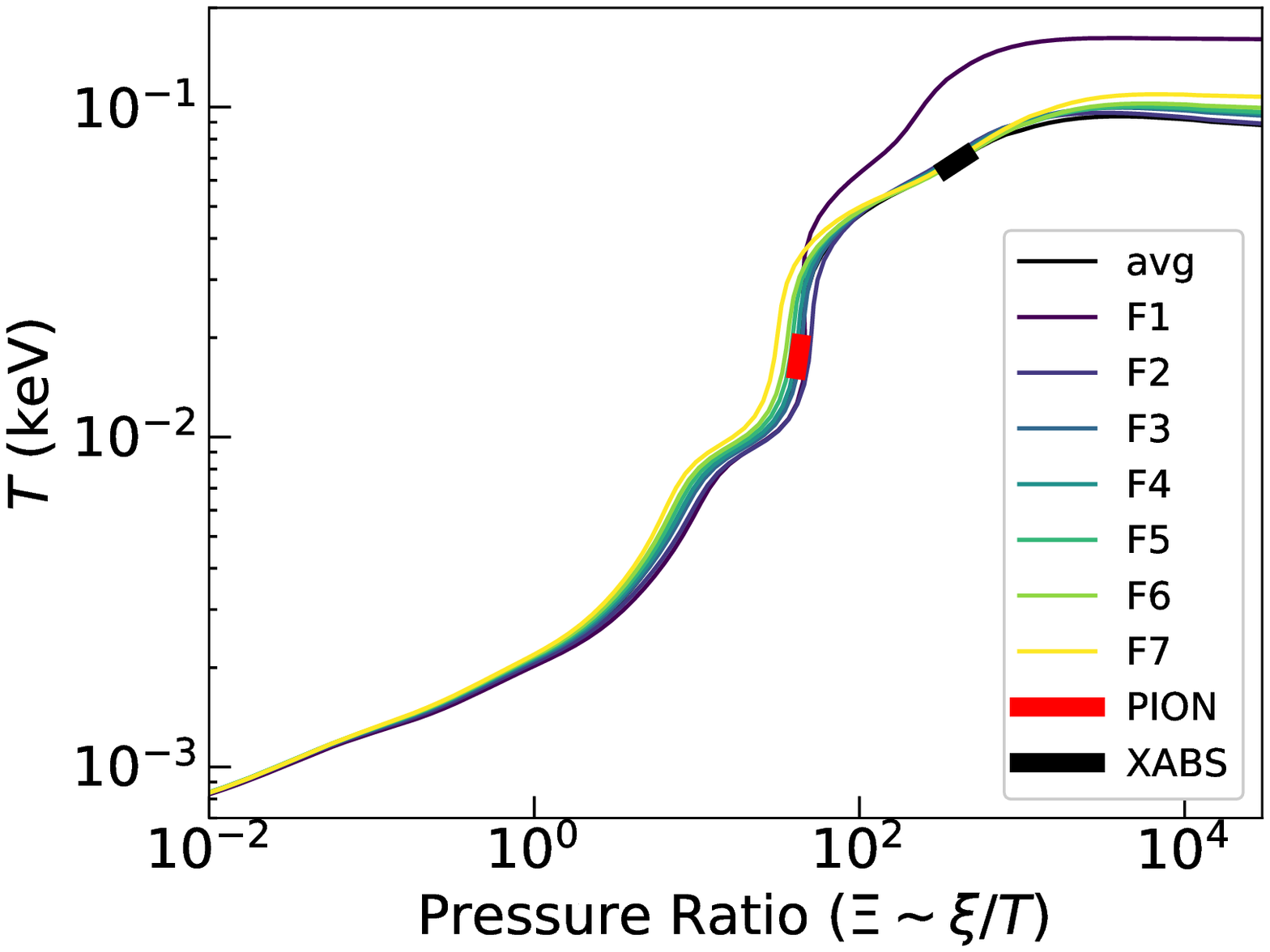}
\caption{
The SEDs ({\it Upper}) and the corresponding stability curves ({\it Lower}) of a plasma in the photoionization equilibrium for different 1H 0707 flux levels. The {\it red} and {\it black} bold lines in lower panel indicate the best solutions ($\pm1\sigma$ errors) of the photoionized emitter and absorber respectively for time-average spectrum, where the fits are described in section \ref{subsec:evolution}. The positive gradient of the curves indicates the winds of 1H 0707 are thermally stable.
}
\label{fig:SEDstability}
\end{figure}

The intrinsic spectral energy distribution (SED) of 1H 0707 is computed from UV/optical to hard X-ray energies by taking advantage of the time-average OM fluxes. The UV/optical spectra are described by an additional blackbody component to the best-fit continuum, \hot*\reds*(\etau*\pow+\bb+\bb+\laor*(\delt+\delt)), where the additional thermal emission is characterized by a temperature of $\sim14$\,eV, consistent with \citet{2017Pawar}. Here, we apply a convolution model \etau\ (cut-off at 0.136\,keV) to the power-law component, so as to avoid the unphysical divergence at low energies of the \pow\ component. The X-ray continuum parameters remain unaltered after including OM data. The redshift effect and ISM absorption are removed to obtain the intrinsic and rest-frame source spectrum as seen by the winds. Owing to the missing far-ultraviolet (FUV) data due to the interstellar absorption, the SEDs covering this domain are interpolated (i.e. assuming a straight line to link the UVW2 and the 0.4\,keV X-ray data points) to avoid over-estimating the UV flux, but later on we test how the UV bump affects the ionization balance. The time-average and flux-resolved SEDs of 1H 0707 are shown in the upper panel of Fig.~\ref{fig:SEDstability}. From the soft to hard energies, the SEDs deviate above 0.01\,keV and get close at high energy ($\geq7$\,keV), which has already been indicated by the steady UV/optical fluxes and smaller X-ray variability above 8\,keV (see in Fig.~\ref{fig:spectra}). We caution that the SEDs above 10\,keV are predicted by a power-law model without any constraint from \xmm\ data. An SED, cut at 10\,keV, has thus been input in the following analysis to estimate the influence of the high-energy tail and we find it does not make any significant difference for our results due to soft spectrum.

Following the approach used in \citet{2020Pintostability}, we then compute the photoionization balance of ionized gas irradiated by the radiation field from 1H 0707 with the {\it pion} code. We also test the {\it xabsinput} tool in SPEX, which is a pre-calculated model and performs the same calculation. The results obtained from {\it pion} and {\it xabsinput} are consistent. In the lower panel of Fig.~\ref{fig:SEDstability}, we illustrate the stability curve, which is the plasma temperature $kT$ ($k$, Boltzmann constant) as a function of the pressure ratio $\Xi$. The pressure ratio is defined as the ratio between the radiation pressure ($F/c$) and thermal pressure ($n_\mathrm{H}kT$), $\Xi=\frac{F}{n_\mathrm{H}ckT}=19222\frac{\xi}{T}$, where $F$ is the ionizing flux \citep{1981Krolik}. It shows the impact of each flux-resolved SED on the ionization balance of the plasma. All curves are compatible below $kT\sim0.03\,$keV and F1 displays a slightly different behavior at high temperature states. The bold lines (PION and XABS) indicate the best solutions ($\pm1\sigma$ errors) of the photoionized absorber ({\it black}) and emitter ({\it red}) for the time-average spectrum (see Tab.~\ref{tab:evolution}), where the fits are described in detail in following section~\ref{subsubsec:fullwind}. Along the curve, the plasma is in thermal balance, where heating equals cooling. To the left of the curve, cooling dominates over heating, while to the right of the curve, heating dominates over cooling. On the branches of the curve with a positive gradient, the plasma is thermally stable, because small perturbations upwards/downwards in temperature will be balanced by the increase in cooling/heating. Instead, on the branches with negative gradient, the plasma is thermally unstable. In Fig.~\ref{fig:SEDstability}, the stability curves of the investigated spectra are very similar with a positive gradient, which means that winds of 1H 0707 are likely thermally stable.

\subsubsection{Full wind model}\label{subsubsec:fullwind}
To determine the physical properties of UFOs in 1H 0707, we employ the \xabs\ and \pion\ model to, respectively, describe the absorption and emission components of the photoionized gas by inputting the calculated SED and ionization balance. \xabs\ is a fast version of \pion\ only modelling the absorption features. Ideally, the parameters of these two photoionization models, such as the column density $N_\mathrm{H}$, the ionization parameter $\xi$, the line-of-sight velocity $v_\mathrm{LOS}$ and the turbulent velocity $\sigma_v$, could be obtained by direct fits to the spectra. However, in order to locate the global best-fit solution, following the method implemented in \citet{2020Pinto}, we build a routine to automatically scan a large grid in the parameter space. This method can efficiently locate the potential solutions and prevent getting stuck in local minima at the cost of computational time. The systematic scan on \xabs\ and \pion\ are launched individually and the technical details are described in appendix~\ref{app:modelscan}.

From the photoionzation model scan, we find an increasing ionization parameter of the absorber and emitter as the source luminosity rises. However, we have to caution that the wind properties of the model scan might not be ultimate, as they are obtained by scanning an individual photoionization model, either \xabs\ or \pion, without simultaneously taking both of them into account. Therefore, we apply a full wind model, including both \pion\ and \xabs, on the top of the continuum model and link their line width, $\sigma_v$, for saving computational time and avoiding further degeneracy, as it has insignificant effects on the fit. Then we directly fit the spectra starting from the solutions obtained with the emission and absorption model scans. The line width is assumed at 5000\,km/s during F7 spectrum fitting due to its loose constraint at the highest flux level.

\begin{figure*}
\centering
\includegraphics[width=0.99\textwidth,trim={100 10 50 0}]{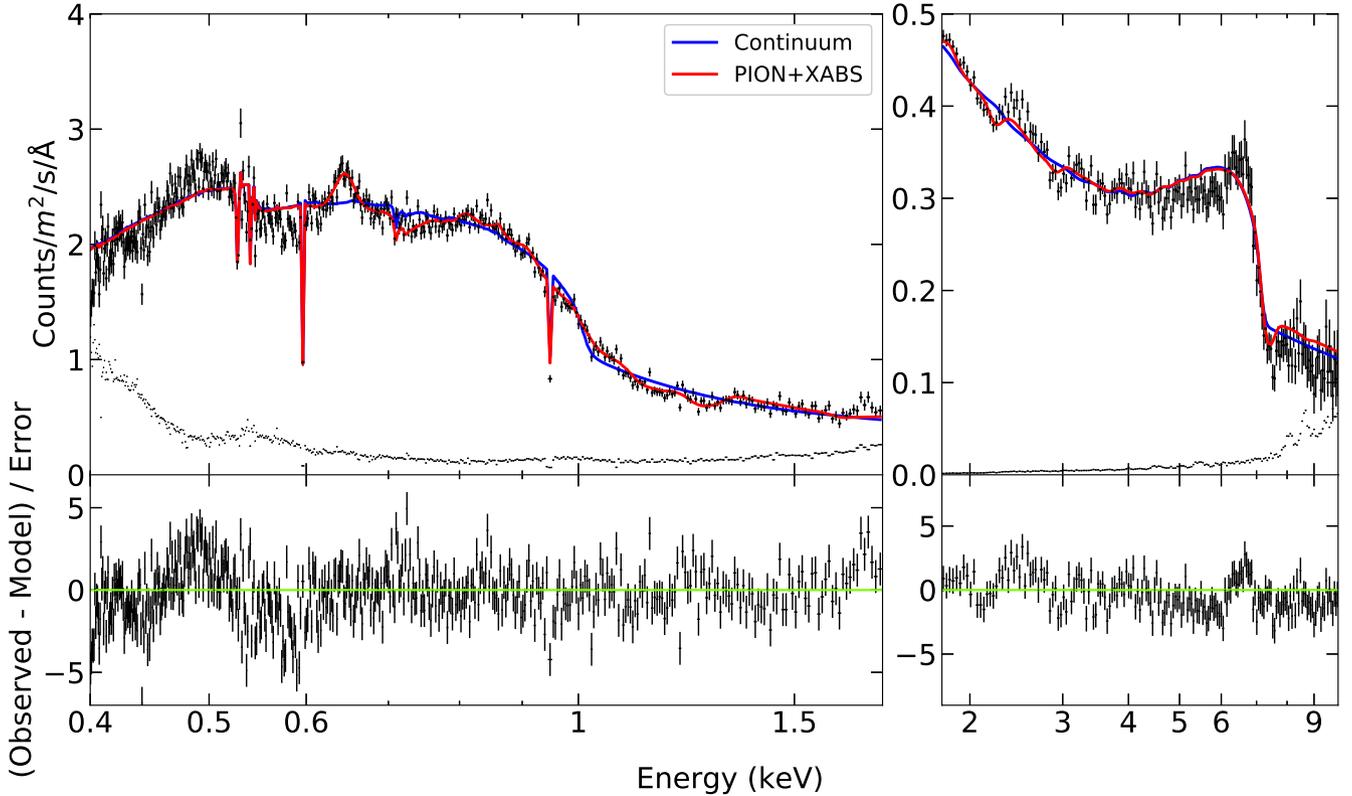}
\caption{
The time-average X-ray spectrum ({\it top}) of 1H0707 using RGS ({\it left}) and EPIC-pn ({\it right}) data. The top two panels contain fits with the baseline continuum model ({\it blue}) and the continuum plus a \pion\ and \xabs\ model ({\it red}). The background is shown in the dots without errorbars. The bottom two panels are the corresponding residuals to the broadband continuum fit with the blueshifted emission and absorption included. The spectrum is well explained except the residuals among $0.5\mbox{--}0.6$ and $6\mbox{--}7$\,keV, which are discussed in section~\ref{subsec:straitified}.
}
\label{fig:spectrum+fit}
\end{figure*}

The parameters of the absorbing/emitting gas at different flux levels are listed in Tab.~\ref{tab:evolution}. The wind properties of the emitting and absorbing gas as derived for the time-average spectrum are similar to those of PK18. Some differences may come from the data selection and the SED inputted into \pion\ as they used the data before 2010 and the default SED in SPEX (i.e. NGC 5548 SED). The most significant change in the full wind model with respect to the individual \xabs\ or \pion\ scan is that all of the spectra prefer the low-ionization solution of the emission component, while no change happens on the absorber. The fitting result to the time-average spectrum, as an example, and the corresponding residuals are shown in Fig.~\ref{fig:spectrum+fit}. Most of the lines detected in section~\ref{subsec:Line-Scan} are well explained while the N {\scriptsize VII} feature and Fe {\scriptsize XXV/XXVI} emission are still unmodelled.

As for the flux-resolved results, the absorber displays an increasing ionization state, a slightly higher column absorption and a slower velocity with the brighter source. The properties of the emitter are instead not that variable at different luminosities, indicating a stable emitting gas, although there is a weakly decreasing trend of the ionization parameter and column density. We summarize the results in Fig.~\ref{fig:evolution}. The line width is relatively stable and supports our assumption of $\sigma_v=5000\,\mathrm{km/s}$ for F7. In terms of statistics, the full wind model provides of course better fits than the individual photoionization models (see for a comparison the $\dcstat$ between Tab.~\ref{tab:evolution} and Tab.~\ref{tab:ratios} and the individual model scans in Fig.~\ref{fig:XABSflux} or \ref{fig:PIONflux}). The extent of the improvement decreases with the increasing fluxes as expected due to the weakening wind features. Furthermore, the statistical improvement of a full wind model $\Delta$C-stat$_\mathrm{pion+xabs}$, is not simply the sum of that of applying an individual \xabs\ and \pion\ on the continuum, $\Delta$C-stat$_\mathrm{xabs}+\Delta$C-stat$_\mathrm{pion}$, but usually smaller than the sum, which indicates that some residuals are explained by the overlapping of the emission and absorption components.

\section{Discussion}\label{sec:discussion}

\subsection{Systematic Effects}\label{subsec:systematics}

\subsubsection{Eddington Accretion Rate}\label{subsubsec:accretion}
UFOs are expected to be mainly driven by the radiation pressure when the black hole is accreting at a high, super-Eddington, accretion rate \citep[e.g.][]{2013Takeuchi}. 1H 0707 has been reported to accrete either just below \citep{2009Fabian} or above the Eddington limit \citep{2016Done}. Accordingly, we estimate the accretion rate by measuring the bolometric luminosity of 1H 0707 from our interpolated time-average SED ($0.01\,\mathrm{eV}\mbox{--} 1000\,\mathrm{keV}$), which is $L_\mathrm{bol}\sim2\times10^{44}\,$erg/s. We adopt a BH mass of $2\times10^{6}\,M_\odot$ \citep[e.g.][]{2005Zhou,2013Kara,2016Done} and the corresponding Eddington limit is $L_\mathrm{Edd}\sim2.8\times10^{44}\,$erg/s, where $L_\mathrm{Edd}=4\pi G M_\mathrm{BH}m_\mathrm{p}c/\sigma_{T}$. This means 1H 0707 is accreting close to the Eddington limit ($\dot{m}\sim0.7$).

However, we caution that the interpolation between the FUV and the soft X-ray bands only provides a lower limit onto the bolometric luminosity, as we do not know the actual SEDs of 1H 0707. To investigate the effect of the interpolation, we calculate the non-interpolated SED predicted by continuum models (i.e. using the additional blackbody spectrum instead of a straight line between the UV and soft X-rays) on the left panel of Fig.~\ref{fig:nonchop}. The corresponding bolometric luminosity is estimated at $L_\mathrm{bol}\sim5.5\times10^{45}\,$erg/s, which is a super-Eddington accretion rate for 1H 0707, ($\dot{m}\sim20$), as suggested by \citet{2016Done}. As a result, the high accretion rate of 1H 0707 has been confirmed whether we adopt the interpolated or non-interpolated SED. We also repeat the ionization balance calculation (see the right panel of Fig.~\ref{fig:nonchop}) and the full wind modelling. The solutions of the photoionization models are also illustrated in the stability curve. The inclusion of the FUV bump cools the plasma down, while the temperature of the emitter remains unchanged. Although the stability curves are different, the photoionized plasma exposed to a non-interpolated SED is still thermally stable. Furthermore, we have tried to fit the photoionization models on the flux-resolved spectra with non-interpolated SEDs, and still found the trends we discovered in Tab.~\ref{tab:evolution}, apart from a systematic shift in the absolute values. Therefore we are able to confirm that the interpolation of SED has negligible influence on our conclusion onto the UFO variability, but it will likely underestimate the actual accretion rate.

\subsubsection{Screening Effect}\label{subsubsec:screen}
It is worth noting that there might be differences between the SED that we observe and the SED portions that ionize all the different regions of the wind. In the case of super-Eddington accretion, the thick inner disk \citep[e.g.][]{1973Shakura,1980Abramowicz} and the wind itself may self-screen the inner regions of the wind from some UV/optical emission from the outer part of the disk. Therefore, we also simulate the self-screening effect for 1H 0707 by manually constructing pseudo-SEDs with just 10\% and 1\% of the UV/optical emission received by winds (see the upper panel of Fig.~\ref{fig:ionUV}) by following \citet{2020Pintostability}. The stability curves are also calculated on the lower panel of Fig.~\ref{fig:ionUV}. This shows that the variation in the optical-UV flux by a factor larger than $10\mbox{--}100$ does not lead to a thermally unstable wind. As a result, the effect of UV/optical screening will not affect the current wind modelling.

\subsubsection{Relativistic Effect}\label{Relativistics}
It has been recently reported that the special relativistic effects are of importance for UFO modelling \citep{2020Luminari,2021Luminari}, because the radiation received by the ultra-fast outflows would decrease for increasing outflow velocity. They argued that most of UFO velocities from the literature cannot be reproduced in a radiative-driven scenario within the relativistic treatment, unless the winds are launched at a radii $>50\,r_\mathrm{g}$, with a velocity up to $0.15\,c$, in a Eddington or Super-Eddington luminosity case. 

Coincidentally, 1H 0707 potentially satisfies all of these conditions. Its relativistically corrected outflow velocity is $\sim0.145\,c$ derived from the time-average spectrum, which is not extreme like the $0.3-0.4\,c$ seen in some quasars \citep[e.g. PDS 456 and IRAS 00521-7054][]{2018Reeves,2019Walton}. Furthermore, if we assume the launching radius to be equal to the escape radius, $r_\mathrm{esc}=2GM_\mathrm{BH}/v_\mathrm{out}^2$, the lower limit on the location of the wind is $>95\,r_\mathrm{g}$. Moreover, the accretion rate of 1H 0707 ($\dot{m}>0.7$, as estimated in section~\ref{subsubsec:accretion}) likely surpasses Eddington, meaning that the radiation force must be very high and that magnetic fields might not be necessarily required although still potentially helping to accelerate the wind at the observed speeds.

\subsubsection{F1" spectrum (low state)}\label{subsubsec:F1"}
We note that F1 spectrum also includes the full observations affected by the AGN intrinsic variability event after 2010 and, therefore, we need to investigate how much the variability affects our result. We produce F1" spectra by excluding the latest two observations, and find that the observations before 2010 contribute about 85\% of the total counts in the F1 spectra. This implies that F1" is not very different from F1 spectrum. However, we perform the same analysis for F1" as done for F1 spectrum, i.e., the spectral fitting, ionization balance calculation, and photoionization modelling. Fig.~\ref{fig:comparison} illustrates the comparison of the SED and stability curves between F1 and F1" spectra. The X-ray spectroscopy demonstrates that F1" has a softer spectrum ($\Delta\Gamma\sim0.2$) with other parameters similar to the F1. The SED and stability curve of F1" also do not dramatically change and only show a slightly steeper spectrum and a hotter plasma, which should not affect our results. The solutions of the photoionization models are slightly hotter due the need for more heating in the softer SED. We conclude that the inclusion of the final observations in the low-flux spectrum does not significantly affect the wind modelling.

\subsection{Evolution of the wind components}\label{subsec:evolution}

\begin{figure*}
\centering
\includegraphics[width=0.49\textwidth,trim={40 40 0 30}]{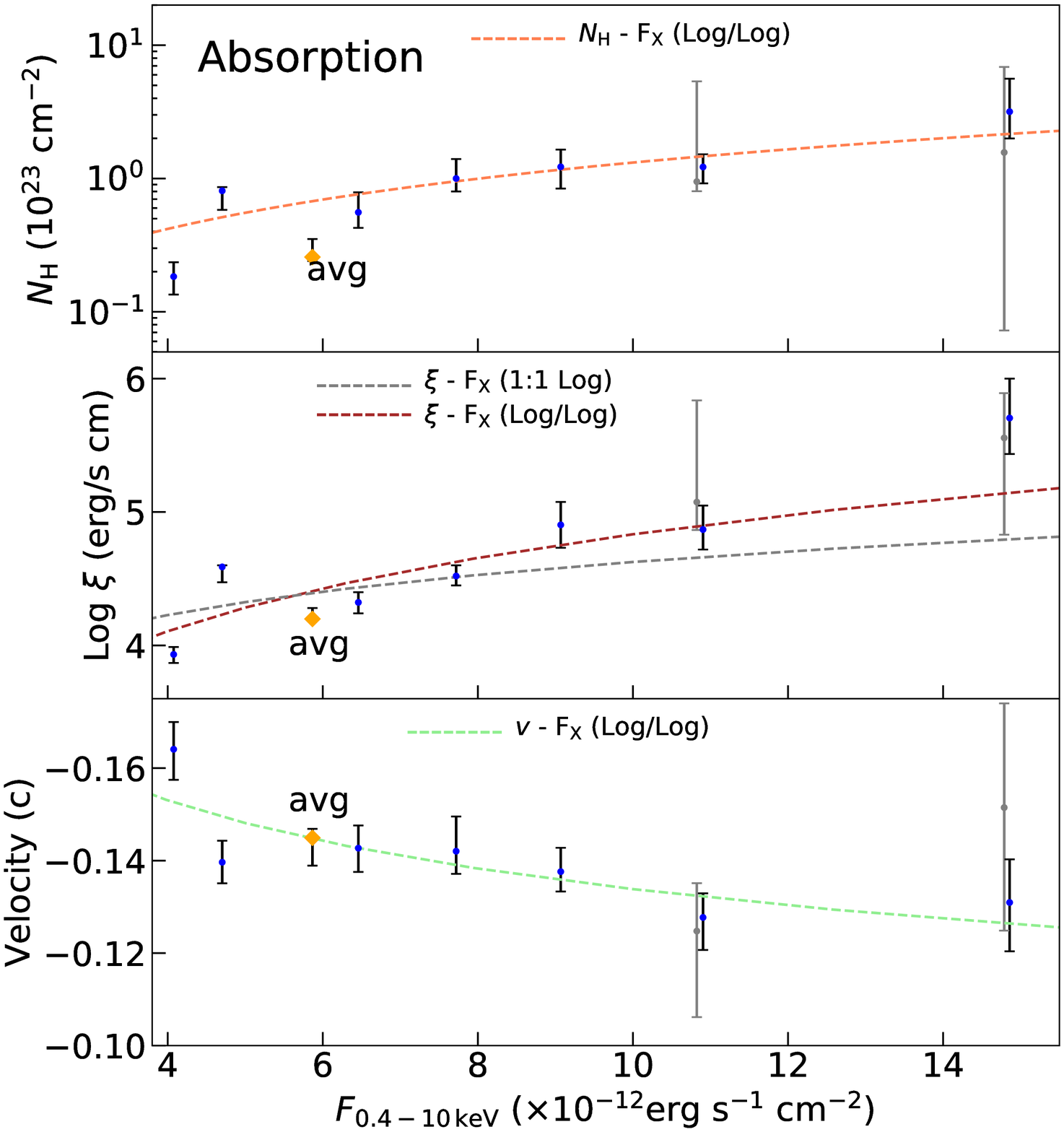}
\includegraphics[width=0.49\textwidth,trim={40 40 0 30}]{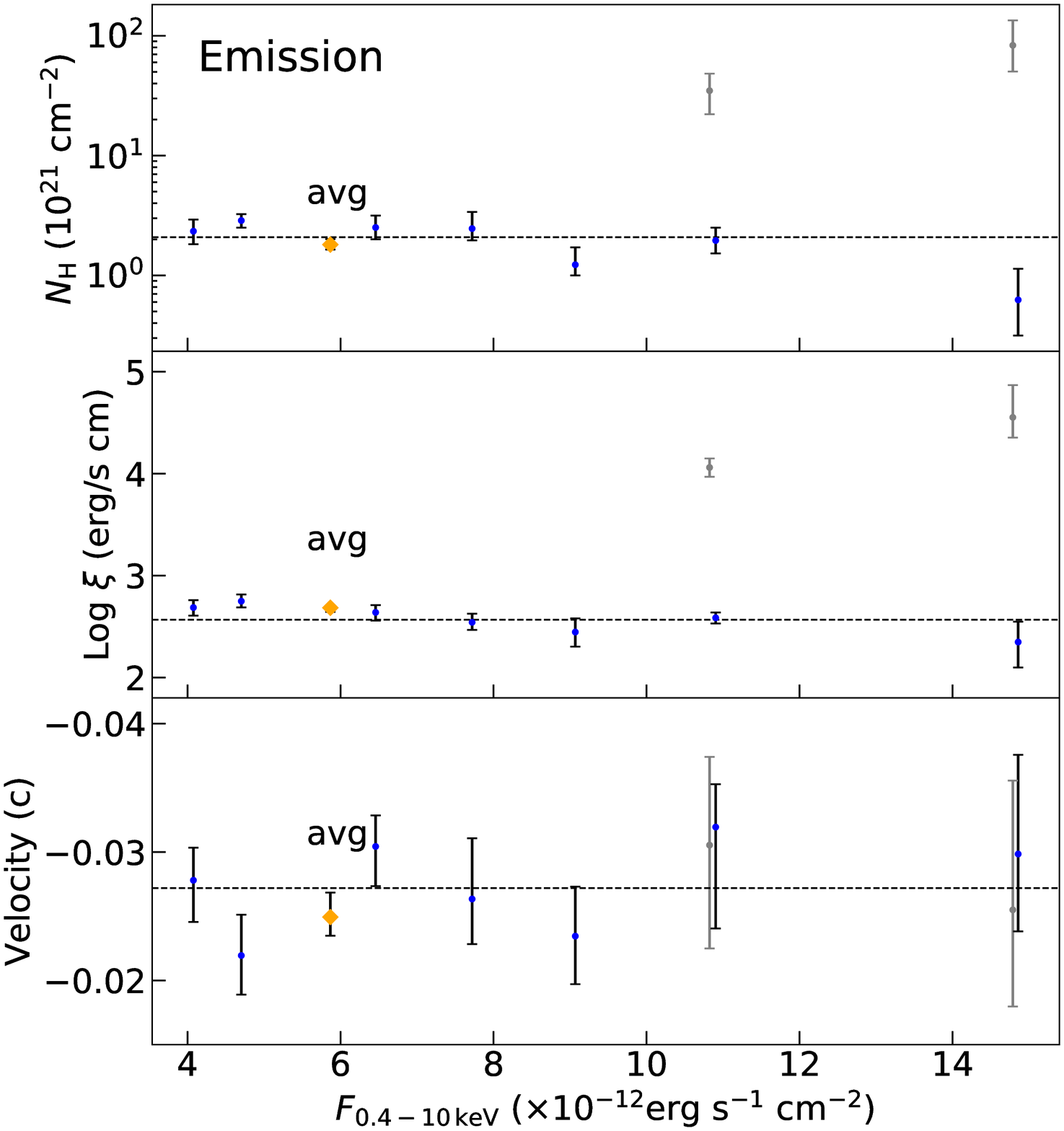}
\caption{
The column density ({\it Top}), ionization parameters ({\it Middle}) and velocities ({\it Bottom}) of the photoionization absorbing ({\it Left}) and emitting ({\it Right}) plasmas versus the absorbed X-ray fluxes of the flux-resolved spectra. The uncertainties within the 68\% confidence level are shown in {\it black}. The time-average results are also marked and shown with {\it orange} points. The linear function fits with (1:1 Log) and without (Log/Log) a slope fixed at unity are performed in a logarithmic space for the ionization parameters of the absorbing gas, and also for the logarithmic of the column density and the absolute value of the velocity of the absorber (Log/Log). The black horizontal dash lines on right panels display the average values of parameters calculated from those of flux-resolved points, indicating a more or less stable emitting gas. Another solution ({\it grey}), which is comparable in statistics, is also shown in the plots, and its flux is slightly shifted for clarity.
}
\label{fig:evolution}
\end{figure*}
In section~\ref{subsubsec:fullwind}, we have measured the wind properties at different flux levels and discovered several potential trends of the parameters.  We thus plot the column density, ionization parameter, and the relativistically corrected velocity of the absorption and emission components in Fig.~\ref{fig:evolution}. We take the observed fluxes measured by the continuum model as the X-axis because they are not affected by the detailed modelling of the photoionized gas and supply the lower limits of the source luminosity. The results obtained for the time-average spectrum are also marked for a comparison. The grey points in the figure present another solution, which has a high ionization parameter ($\log\xi>4$) and column density ($N_\mathrm{H}>10^{22}\,\mathrm{cm}^{-2}$) of the emitting gas, indicated by the \pion\ model scan (see Fig.~\ref{fig:evolution}). The other parameters have loose constraints and are compatible with the blue points within uncertainties. Their fitting statistics are comparable with that of the low-ionization solution in F6 spectrum ($\dcstat<1$) and marginally better fits in the F7 spectrum ($\dcstat\sim4$), while in the other spectra, the high-ionization solution is preferred to the low-ionization one. This suggests that there might be another emission component, which becomes more important at high-flux levels, although we cannot exclude the possibility of a degeneracy between the two different solutions \citep[see e.g.][]{2018Pinto}.

\subsubsection{Absorbing gas}\label{subsubsec:absorbing}
Fig.~\ref{fig:evolution} suggests a possible correlation between the ionization parameter of the absorbing gas and the X-ray flux, while the velocity is anticorrelated with the X-ray flux. The Pearson correlation coefficient of the ionization parameter, column density and velocity versus the flux are 0.93, 0.92, 0.75 respectively, implying the strong correlation. Here, we perform a linear fit with a slope fixed at unity in a logarithmic (1:1 Log) space among the ionization parameter and the flux, in order to compare with the definition of the ionization parameter ($\xi=L_\mathrm{ion}/n_\mathrm{H}R^2=4\pi r^2F_\mathrm{ion}/n_\mathrm{H}R^2$, where $r$ is the distance between the source and the earth). We also fit another linear function with a free slope (Log/Log), which supplies a slightly better fitting result. The Log/Log fit provides 
\begin{equation}
    \log\xi= (3.00\pm0.58)+(1.83\pm0.77)\log(\frac{F_\mathrm{0.4-10}}{10^{-12}}), 
\end{equation}
corresponding to $\chi^2/\nu=13/5$. The positive slope means the absorbing gas is increasingly ionized with the enhanced radiation field, which would lead to a decline of the equivalent widths of the lines at the same column density of the wind. Furthermore, the slope is consistent with unity within less than $2\,\sigma$, indicating that $n_\mathrm{H}R^2$ is relatively constant with the luminosity. It also suggests that the wind in 1H 0707 is able to respond to the ionizing luminosity instantaneously, which is not achieved by IRAS 13224 \citep{2018Pinto}, perhaps due to a more severe degeneracy between the two solutions (Fe {\scriptsize XXV} and Fe {\scriptsize XXVI} dominated UFO). Regarding the possibility of a steeper slope ($>1$), this might be due to small, local, changes in the density and the position of the absorbing gas.

In addition, we apply the linear fit in a logarithmic space (Log/Log) on the column density and velocity of the absorbing matter, $\chi^2/\nu=4/5$ and $\chi^2/\nu=7/5$, respectively. The Log/Log fit gives
\begin{equation}
    \log\frac{N_\mathrm{H}}{10^{23}\,\mathrm{cm}^{-2}}= (-1.13\pm0.42)+(1.25\pm0.51)\log(\frac{F_\mathrm{0.4-10}}{10^{-12}}),
\end{equation}
\begin{equation}
    \log\frac{|v|}{\mathrm{km/s}}= (4.75\pm0.05)+(-0.15\pm0.05)\log(\frac{F_\mathrm{0.4-10}}{10^{-12}}),
\end{equation}
where $c$ is the speed of light. 

This anticorrelation between the wind velocity and X-ray flux is opposite to the correlation observed in PDS 456 \citep{2017Matzeu} and IRAS 13224 \citep{2018Pinto}, where they interpret that the winds are mainly accelerated by the strong radiation field in high-accretion sources. Alternatively, based on the simulations for a purely magnetohydrodynamic (MHD) driven wind with some parameter assumptions, \citet{2018Fukumura} indeed predicted a weak anticorrelation in their Fig. 2a when the flux/accretion rate scaling parameter $s=3$, which is defined by $L_\mathrm{ion}\propto\dot{m}^{s}$. In their model, they construct several relations between the ionizing luminosity and wind parameters, such as the ionization parameter, $\xi\propto L_\mathrm{ion}^{1-2/s}$ and the density $n_w\propto L_\mathrm{ion}^{2/s}$. For $s=3$, the wind density increases slower than the X-ray flux and therefore, an increase in $L_\mathrm{ion}$ will bring the wind ionization front to larger radii, yielding slower plasma velocities. However, the $s=3$ case is not quite physical because this scaling parameter tends to be below unity for a high accretion rate ($\dot{m}>1$) region due to the photon trapping effects \citep[e.g.][]{2009Takeuchi}.

In fact, given the high accretion rate estimated in section~\ref{subsubsec:accretion}, it is difficult to neglect the role of the strong radiation field. For this reason, we come up with two possible explanations to account for the decreasing trend of velocity. The first is that the gas is driven by the line radiation pressure \citep[e.g.][]{2000Proga} and is being over-ionized in the higher flux states with a consequent decrease in the driving force and thus the wind velocity. On the other hand, the second explanation is that a stronger radiation field beyond the Eddington limit, presumably due to a region in the disk with higher local $\dot{m}$, increases the wind driving force, thus expanding the wind launching region into disk radii with lower Keplerian velocities, in turn leading to lower outflow velocities. This explanation was also invoked by \citet{2020Pinto} for super-Eddington stellar-mass compact object (ultraluminous X-ray source), in which at higher fluxes the wind showed lower velocities.

As another option, \citet{2020Fabian} suggested that when the disk reflection component is strong relative to the primary continuum, the highly blueshifted absorption lines are likely a consequence of the reflection component passing through an optically thin, highly ionized absorbing layer at the surface of the inner disk rather than a fast wind. The relativistic velocity is interpreted as the orbital motion of the inner disk. In this case, a higher hot corona would reduce the effect of light bending, resulting in a brighter luminosity, a smaller reflection fraction and therefore weaker absorption lines. Moreover, the greater coronal height means that the reflection peak would come from the larger radius of the disk, leading to a slower velocity, compatible with the anticorrelation observed in 1H 0707. Nevertheless, the ionization state is not determined because it should decrease due to the more distant location in a constant disk, while the density of disk may drop outwards due to the reduced radiation pressure, perhaps leading to an opposite trend. The validation of this scenario requires a precise calculation of the density structure of the outer Thomson depth of the disk and will be done elsewhere. Therefore, for the rest of the paper, we will keep the UFO explanation for the absorption features.

\subsubsection{Emitting gas}\label{subsubsec:emitting}
For the parameters of the emitting gas shown in the right panel of Fig.~\ref{fig:evolution}, we find that they are more or less constant at different fluxes, which is illustrated by the consistency with a constant function fitted to the flux-resolved points, despite the potential decrease in ionization parameter and column density. The stable velocity implies that the emitting gas is located at constant radii from the center. The relatively constant emitting gas is compatible with the argument by PK18 that the emission might come from a relatively large scale wind. The large-scale emission origin is also suggested by \citet{2021Parker} who find the Fe and Ne complex emission lines around 1 keV are less variable than the continuum. The emission lines are probably located on large enough scales that they are relatively unaffected by the variability of the absorbed continuum. 

Furthermore, if the wind is launched isotropically, in principle, both of the blueshifted and redshifted emission lines are detectable and we should observe a centrally peaked or flat-topped emission line profile. However, only the blueshifted emission lines are detected. One explanation is that if the wind is indeed launched at a large opening angle, either there might be a particular MHD configuration breaking the symmetry of the wind or the redshifted part is partially obscured by the blueshifted absorbing gas for some geometrical configurations. The assumption of a large open angle is plausible for a high-Eddington AGN as the matter pushed outwards is likely to form a shield to outer gas and thus to absorb the redshifted emission. This scenario also gives an explanation for the potentially cooler emitting gas in the high flux states, because the stronger radiation field will push more matter away, leading to a stronger screening of the hot inner gas. 

Interestingly, in the high flux states (i.e. F6 and F7), there is another solution for the emitting gas, comparable in statistics, with a distinct ionization state and column density, while the other parameters are consistent within the uncertainties. This distinction is also suggested by what we obtained in \pion\ model scans, where a $\log\xi$ gap appears for the emission component. This occurs because an ionization parameter $\log\xi\sim3$ would lead to Fe complex emission lines around 0.95\,keV that are not observed in the spectrum. The Fe {\scriptsize XXV/XXVI} emission lines are not taken into account in the time-average spectrum (see Fig.~\ref{fig:spectrum+fit}), but are accounted for by a $\log\xi\sim4.5$ emitting gas in F7 spectrum. As a result, the alternative high-ionization solution implies that either another weak emission component is hidden in the spectrum or the emitting gas is not constant with luminosity.

\subsection{A stratified wind}\label{subsec:straitified}
\subsubsection{Multi-wavelength Information}\label{subsubsec:multiwavelength}

\begin{figure}
\centering
\includegraphics[width=0.49\textwidth,trim={0 30 0 0}]{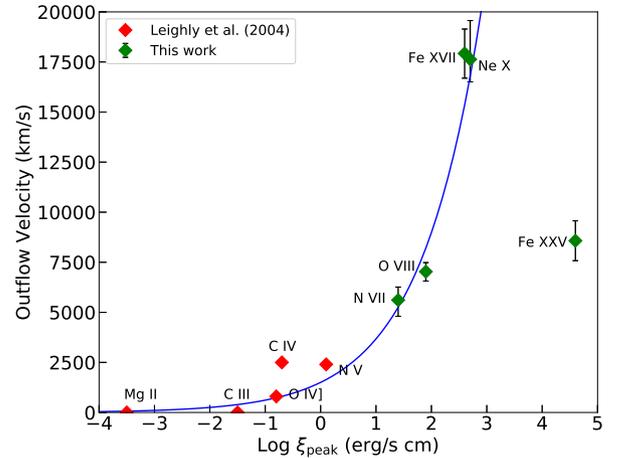}
\caption{
The outflow velocity of different ions in 1H 0707 versus the ionization parameter at the peak abundance of each ion in a photoionized plasma. The velocities for the UV ion are taken from \citet{2004Leighly} and \citet{2004Leighlyb}, where the uncertainty is assumed at 500\,km/s. The ions up to Ne {\scriptsize X} are characterized by a power-law function ({\it blue}), while Fe {\scriptsize XXV} is not along with the curve.
}
\label{fig:stratified}
\end{figure}
\begin{figure}
\centering
\includegraphics[width=0.49\textwidth,trim={0 30 0 0}]{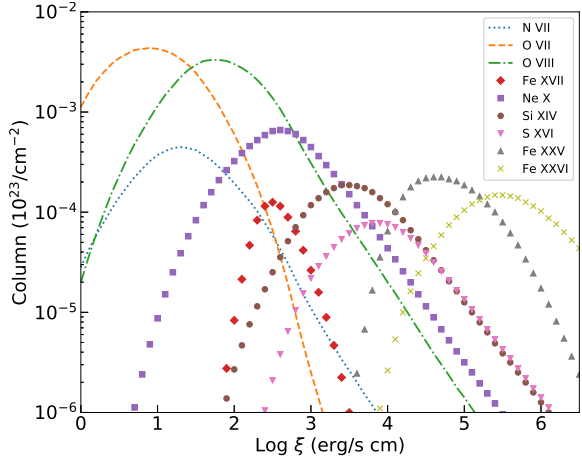}
\caption{
The predicted column densities of different partially ionized ions in a photoionized plasma versus the ionization parameter assuming $N_\mathrm{H}=10^{24}\,\mathrm{cm}^{-2}$ and the time average SED (see Fig.~\ref{fig:SEDstability}).
}
\label{fig:column}
\end{figure}
PK18 discovered a trend for which the higher ionization emission lines are more blueshifted in 1H 0707 by combining the N {\scriptsize VII} and O {\scriptsize VIII} lines with the low ionization lines observed in the UV spectrum from \citet{2004Leighly}. Here we complement other identified ionization lines in X-ray band and reproduce the result in Fig.~\ref{fig:stratified}. The outflow velocity of each ion is obtained by the shift obtained in Tab.\ref{tab:lines}. The ionization parameter is the value at which the abundance of each ion peaks in a plasma photonionized by 1H 0707 SED.  Our result confirms the previous finding and demonstrates that the velocities of ions up to Ne {\scriptsize X} are monotonically increasing. A power-law function is thus applied to fit and provides, \begin{equation}
v_\mathrm{outflow}=(1502\pm437)\xi^{(0.39\pm0.06)}\,\mathrm{km/s},
\end{equation}
which is compatible with the result from PK18 within the uncertainties. As they suggested, the $\frac{v^3}{\xi}$ is consistent with being constant and hence the outflow energy is conserved if the ion emission results from the same wind producing the UFO absorption lines.

However, the Fe {\scriptsize XXV/XXVI} ions deviate from the power-law curve with a relatively slow velocity. According to the full wind model fit, we find that the majority of Fe {\scriptsize XXV/XXVI} features originate from Fe {\scriptsize XXV} and thus only show Fe {\scriptsize XXV} ions in the figure. The deviation suggests that the high ionization lines are mainly related to another component. This finding also supports a secondary emission component suggested in section~\ref{subsubsec:emitting}. On the other hand, we caution that the Fe K band is modelled with a simple \laor\ component rather than a physical reflection model, which might explain the Fe K emission residuals.

To investigate the abundance distribution of each ion at different ionization states, building on the calculated ionization balance for 1H 0707, we plot the column density of each identified ion line versus the ionization parameter in Fig.~\ref{fig:column}. In terms of the emitting gas, it is unlikely to produce both the O {\scriptsize VIII} and Fe {\scriptsize XXV} emission using one component with a ionization state of $\log\xi\sim2.7$ without over-predicting lower-ionization iron species. This confirms the need for another emission component to produce the Fe {\scriptsize XXV}.

\subsubsection{Zooming on 2008 Epoch}\label{subsubsec:2008}
In principle, the secondary emission component might be an artefact of the flux-resolved integration. When we merge the spectra of the same fluxes but different epochs, the time variability could wash out or broaden the emission and absorption features. \citet{2012Dauser} indeed discovered evidence for a faster UFO from the 2010 observation than that of the 2008 spectrum. To understand the general needs of the spectrum and how many phases could be there, we extract a combined spectrum from the 01/31/2008 and 02/02/2008 observations (see the bold dates in Fig.~\ref{fig:lc_his_hardness}), for the sake of the plenty of counts and closeness in time, and fit it with double photoionization models.

Here we employ double emission or absorption components over the baseline continuum model rather than applying an additional \xabs\ or \pion\ upon the full wind model, because of the statistical limit. The emitting gas is explained by the precalculated model of $\sigma_v=5000\,\mathrm{km/s}$ for saving the computational cost. Including one emission component leads to a statistical improvement of $\dcstat\sim55$ against the baseline continuum and provides an emitting gas characterized by an ionization parameter of $\log\xi=2.5$ and a LOS velocity of $-0.012^{+0.003}_{-0.004}\,c$. This component is mainly required by the strongest feature, O {\scriptsize VIII}. The inclusion of an additional highly ionized ($\log\xi=4.0$) emission component improves the statistics of $\dcstat/\Delta\mathrm{dof}=43/2$ with a high velocity of $-0.025^{+0.004}_{-0.003}\,c$. This fast component chiefly models the Fe {\scriptsize XXV} line. The improvement of including a secondary emission model is comparable to that of the first emission component, confirming the existence and increasing significance of the high ionization emission component with the source luminosity, since the observations we have chosen are dominated by the highest flux level (see Fig.~\ref{fig:lc_his_hardness}). This thus might be a hint for a more structured wind.

Regarding the possible secondary absorption, the remaining absorption residuals between 0.5 and 0.6\,keV seen in Fig.~\ref{fig:spectrum+fit} provide a hint for a low ionization component. We thus apply double \xabs\ models to the baseline continuum of the 2008 spectrum. The primary absorption, similar to the results in Tab.~\ref{tab:evolution}, requires a LOS velocity of $-0.122^{+0.006}_{-0.006}\,c$, an ionization parameter of $\log\xi=4.86^{+0.12}_{-0.08}$, a line width of $\sigma_v=7990^{+1295}_{-1263}\,\mathrm{km/s}$ and the column density of $N_\mathrm{H}=1.5^{+0.5}_{-0.3}\times10^{23}\mathrm{cm}^{-2}$, with a $\dcstat=62$ improvement. As shown in Fig.~\ref{fig:spectrum+fit}, this component mostly explains the absorption above 0.65\,keV. Interestingly, the secondary absorption, which is characterized by a broad line profile ($\sigma_v=10532^{+1018}_{-1014}\,\mathrm{km/s}$), a modest velocity ($v_\mathrm{LOS}=-0.075^{+0.005}_{-0.004}\,c$), a low ionization state ($\log\xi=1.46^{+0.06}_{-0.09}$) and a low column density ($N_\mathrm{H}=4.6^{+0.7}_{-0.8}\times10^{20}\mathrm{cm}^{-2}$), has a similar statistical improvement of $\dcstat=50$ to the primary one. The reason why this component is not detected before, although \xabs\ scan at some fluxes show local peaks in this area (see Fig.~\ref{fig:XABSflux}), is due to the limit of the line width. We indeed find a secondary peak through the \xabs\ model scan with $\sigma_v=10000\,$km/s on F7 spectrum. This component makes a great contribution to the residuals between 0.5 and 0.6\,keV, which are mainly N {\scriptsize VII} and O {\scriptsize VII} lines.

Resolving in time and flux by only looking at the 2008 spectra removes some degeneracies and makes us find a slower absorber which could fit in between the emitter and the fast absorber, therefore describing the complex structure of winds. From this slow absorption component, we might see the outer layers of the wind. Basically, this component with the intermediate velocity could be the absorption part of the P-cygni profile where the emission comes from the primary emitter.

After modelling the $0.5\mbox{--}0.6$\,keV absorption residuals, a natural question is where the N {\scriptsize VII} emission comes from. We find that there is no way to fit the N {\scriptsize VII} by adding additional \pion\ component without over-predicting the O {\scriptsize VII} or Ne {\scriptsize IX} and Fe complex emission between $0.9\mbox{--}1.0$\,keV. Therefore, either the N {\scriptsize VII} originates from another mechanism such as the shock, or the abundances are not solar with over-abundant nitrogen, presumably due to enrichment by Asymptotic giant branch (AGB) stars \citep[e.g.][]{1983Wood}. The collisional ionization might facilitate a high amount of O {\scriptsize VIII} and N {\scriptsize VII} without O {\scriptsize VII} if the temperature is high (e.g. 1\,keV). Hence, as a test, we replace the \pion\ with a \cie\ model in SPEX assuming the solar abundance and free line width, but find it cannot explain N {\scriptsize VII} and O {\scriptsize VIII} with a high temperature of $1.4$\,keV, because a lot of Fe L complex ions are also expected. On the other hand, we allow the nitrogen abundance of \pion\ to vary and find it requires an extremely high value of $\sim17$ solar abundance, suggesting that the AGB stars (main generators of ISM nitrogen) might be more abundant in the galactic central regions.

\subsection{Implications for Feedback and Wind properties}\label{subsec:feedback}
It has been well established that UFOs at small scales might carry enough power to affect the evolution of their host galaxies at larger scales such as quenching or triggering star formation \citep[e.g.][]{2005DiMatteo,2010Hopkins,2013Tombesi,2017Maiolino}. According to simulations, the deposition of a few percent of the source luminosity into the ISM is sufficient to prompt a considerable effect on the host galaxy. The kinetic power of the UFO in 1H 0707 is estimated by:
\begin{equation}\label{eq:feedback}
    L_w=0.5\dot{M}_wv^2_w=0.5\Omega R^2\rho v_w^3 C,
\end{equation}
where $\dot{M}_w=\Omega R^2\rho v_wC$ is the mass outflow rate, $\Omega$ is the solid angle, $R$ is the distance from the ionizing source, $\rho$ is the density, $v_w$ is the wind velocity, and $C$ is the volume filling factor (or clumpiness). The density is defined as $\rho=n_\mathrm{H}m_\mathrm{p}\mu$, where $n_\mathrm{H}$ is the number density, $m_\mathrm{p}$ is the proton mass and $\mu=0.6$ is the average particle weight of a highly ionized plasma. Here $n_\mathrm{H}R^2$ could be replaced by the ionization parameter definition ($\xi=L_\mathrm{ion}/n_\mathrm{H}R^2$). 

To estimate the solid angle, we utilize the \omeg\ parameter of \pion\ code in the full wind spectral analysis, which includes direct information on the fraction of solid angle covered by the reprocessing gas shell ({\tt omeg}$=\Omega/4\pi$). We adopt the time-average spectrum and allow \omeg\ to vary. If we assume the emission and absorption comes from the same gas, i.e. a P-cygni profile, we could force \omeg\ to explain the difference between \pion\ and \xabs\ flux by coupling the ionization parameter and column density between the two components, where \omeg\ provides a lower limit of $>0.98$. Such a high solid angle indicates a full covered gas shell, which is likely due to strong radiation field. If we remove that assumption and decouple the parameters, the obtained solid angle is $>0.725$, still suggesting a large opening angle. We would utilize the conservative value of \omeg$>0.725$ during the estimation.

We estimate the volume filling factor by adopting the Eq. 23 in \citet{2018Kobayashi}, on the basis of assuming that most of the mass of the outflow is contained in the clump. Although they focus on stellar mass systems, the volume filling factor is independent of the black hole mass and hence the equation is applicable to AGN. By assuming that the mass outflow rate is comparable to the accretion rate and the relativistically corrected LOS velocity of absorbing gas is the wind velocity, we obtain $C\sim8\times10^{-3}$, which is similar to a typical value of the outflow from the supercritical accretion flow.

The ionizing luminosity ($1\mbox{--}1000\,$Rydberg) is estimated from the interpolated SED at $L_\mathrm{ion}\sim1.42\times10^{44}$\,erg/s, which is a lower limit for $L_\mathrm{ion}$. Therefore, the lower limit on the mass outflow rate is $\dot{M}_w>0.046\,M_\odot/yr$ and the kinetic energy of the UFO wind is then $L_w>2.74\times10^{43}\,\mathrm{erg/s}\approx13.7\%L_\mathrm{Bol}$. This lower limit meets the theoretical condition and demonstrates that 1H 0707 has enough influence on the surrounding medium.

Furthermore, we attempt to constrain the region from which the UFOs originates with different methods. As mentioned in section~\ref{Relativistics}, the wind is at least launched at $95\,r_\mathrm{g}$ if we assume that the outflow velocity is larger than or equal to the escape velocity at the launch radius. Alternatively, if the absorption wind responds on time to continuum variations, the light needs to travel a distance within 230--1150\,s (see frame times in Tab.~\ref{tab:log}), which corresponds to $R_\mathrm{t}=c\Delta t\sim23-117\,r_\mathrm{g}$. This should be an approximate distance between the wind and the X-ray emitting source, which is consistent with the former estimate. For the upper limit, if we assume the ionized gas is able to recombine within each segment at a given flux level, the lower limit of the density could be estimated by using {\tt rec\_time} code in SPEX, which is $n_\mathrm{H}\geq4.16\times10^{5}\,\mathrm{cm}^{-3}$ by adopting $t_\mathrm{rec}=1150\,$s (frame time for F1) and the heating/cooling rate of O {\scriptsize VIII} line, $n_\mathrm{H}t=4.78\times10^{8}\,\mathrm{cm^{-3}s^{-1}}$. This could also be used to estimate the maximal launch radius through $R=\sqrt{L_\mathrm{ion}/\xi n_\mathrm{H}}\leq1.5\times10^{17}\,\mathrm{cm}\approx5\times10^{5}\,r_\mathrm{g}$. On the other hand, if we presume the thickness of the absorbing gas is lower than or equal to its maximal distance from the source ($N_\mathrm{H}=Cn_\mathrm{H}\Delta R\leq Cn_\mathrm{H}R$), the maximal launch radius is $R\leq CL_\mathrm{ion}/\xi N_\mathrm{H}=5.7\times10^{14}\,\mathrm{cm}\approx1947\,r_\mathrm{g}$, when we take the average column density calculated from the flux-resolved results ($N_\mathrm{H}\sim1.27\times10^{23}\,\mathrm{cm}^{-2}$). The same procedures are applied to the emission component, obtaining a location estimation of $3\times10^{3}\mbox{--}8\times10^{5}\,r_\mathrm{g}$. As a result, the launch radius of UFO is estimated between $95-1947\,r_\mathrm{g}$ and the density towards the base of the wind is thus constrained within the range $n_\mathrm{H}=L_\mathrm{ion}/\xi R^{2}=5\times10^{10}\mbox{--}1\times10^{13}\,\mathrm{cm}^{-3}$. The emission component arises from the outer region of $3\times10^{3}\mbox{--}8\times10^{5}\,r_\mathrm{g}$ and the corresponding density is $n_\mathrm{H}\sim9.5\times10^{8}\mbox{--}5.9\times10^{13}\,\mathrm{cm}^{-3}$.

\subsection{Comparison with other AGN}\label{subsec:comparison}
Thanks to the high resolution grating instruments, the UFO has been discovered among many other AGN that accrete near the Eddington limit. The velocity of the UFO in 1H 0707 ($\sim0.145\,c$) does not stand out if compared to other AGN, such as PG 1211+143 \citep[0.2\,c,][]{2003Pounds}, Mrk 1044 \citep[0.08\,c,][]{2021Krongold}, IRAS 13224-3809 \citep[0.24\,c,][]{2017Parker}, PG 1448+273 \citep[0.09\,c,][]{2020Kosec}, and IRAS 00521-7054 \citep[0.4\,c,][]{2019Walton}. The other parameters also fall in the typical UFO region, where the ionization parameter spans from $\logxi\sim3\mbox{--}6$, and column density $\log(N_\mathrm{H/\mathrm{cm}^{-2}}\sim22\mbox{--}24)$ \citep[][]{2010Tombesi}.

IRAS 13224 is a source sharing many similarities with 1H 0707. In particular, the ionization parameter and column density of the UFO observed in IRAS 13224 are similar to our results, despite a faster velocity. However, the trend of the UFO velocity with AGN flux is opposite between these `twin' sources as discussed in section~\ref{subsubsec:absorbing}. Another main difference is that the strong emission lines are only detected in 1H 0707 \citep{2018Pinto}. A plausible origin is that the emitting gas has a large opening angle in 1H 0707 ($\Omega/4\pi>0.725$). If such a large solid angle results from the magnetic field, it would require a peculiar MHD configuration, since MHD-driven winds tend to be along with the polar direction \citep[e.g.][]{2010Fukumura}. Alternatively, the high accretion rate is likely to launch more matter at a wide-ranging angle. According to \citet{2019Alston}, the accretion rate of IRAS 13224 is around $\dot{m}\sim1\mbox{--}3$, while in our estimation, 1H 0707 probably has a much higher accretion rate ($\dot{m}=0.7\mbox{--}20$). This explanation is also supported by the discovery of photoionized emission lines in many ultraluminous X-ray sources (ULXs), accreting at super-Eddington rates \citep[e.g.][]{2016Pinto,2018Kosecb,2018Pinto,2020Pinto}.

The number of AGN with multiphase outflows is increasing in the last decade. For example, the detection of two fast UFOs at $0.25\,c$ and $0.46\,c$ was reported in the spectrum of PDS 456 \citep{2018Reeves}. Similarly, up to 4 UFO phases are reported in IRAS 17020+4544 \citep{2018Sanfrutos} and Mrk 1044 \citep{2021Krongold}. Our secondary UFO in absorption is similar to the UFO2 in Mrk 1044, of which ionization parameter is $\log\xi\sim1.6$, velocity $v_\mathrm{out}\sim0.086\,c$ and column density $\log(N_\mathrm{H})\sim21.2$, although our turbulent velocity is larger than their fixed value of 10\,km/s. However, as \citet{2018Kosec} stated, the difference between these objects and 1H 0707 is that, to our knowledge, 1H 0707 is the currently unique source, which shows significantly blueshifted, probably multiphase X-ray absorption and emission at the same time, likely driven by a higher accretion rate.

\subsection{Athena Simulation}\label{subsec:athena}
\begin{figure}
\centering
\includegraphics[width=0.49\textwidth,trim={0 0 0 0}]{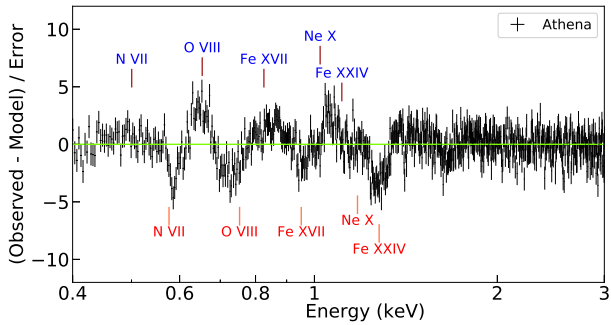}
\caption{
The residuals for the Athena/X-IFU data simulated by F4 full wind model assuming 10\,ks exposure with respect to the baseline continuum model. The detected lines are labeled.
}
\label{fig:athena}
\end{figure}
Future missions with unprecedented spectral resolution and effective area are expected to solve the systematic limitations of the flux-resolved spectroscopy, as they will collect enough photons for spectroscopy within a short timescale. This means that we will be able to trace the variability of the wind properties if we split a single observation into several time-resolved epochs with different count rates. Here we simulate data from the {\it Advanced Telescope for High-Energy Astrophysics} \citep[Athena,][]{2013Nandra} by using the {\tt simulate} command in SPEX with the full wind model. The X-ray Integral Field Unit \citep[X-IFU,][]{2018Barret} instrument on Athena has an
effective energy range of $0.2\mbox{--}12$\,keV with 2.5\,eV spectral resolution up to 7\,keV. We choose the intermediate flux level F4 and assume an exposure time of 10\,ks. The residuals with respect to the baseline continuum model are shown in Fig.~\ref{fig:athena}.The properties of the lines present in the Athena X-IFU spectra will be investigated on much shorter integration times, and may appear narrower than currently detected. There is a remarkable statistical difference between the full wind and the baseline models, $\dcstat\sim828$ ($\dcstat=120$ for the full \xmm\ F4 spectrum), implying that Athena will reach seven times larger significance for wind detection within one order of magnitude shorter time (152\,ks for F4 spectrum). To obtain the similar significance of detection obtained from \xmm\ data, Athena only needs 1\,ks exposure time, which is two orders of magnitude shorter than F4 exposure time. This implies that tracing the wind variability on the necessary timescales in AGN will be possible with Athena.

\section{Conclusions}\label{sec:conclusion}
In this work, we perform a flux-resolved X-ray spectroscopy on all available archival \xmm\ data of the highly accreting NLS1 1H 0707 to investigate the variability of UFO at different flux levels. We have found that the ionization parameter of the UFO is correlated and the velocity is anticorrelated with the luminosity. In the brightest state, the wind features are weak, implying the gas is highly ionized, similar to the phenomenon seen in NLS1 IRAS 13224. The emitting gas is nearly constant with a potential cooling tendency with luminosity, indicating that it originates from a large-scale wind. We propose that all of these results might be consistent with a scenario where the stronger radiation field pushes the wind launching region outwards, leading to a slower and broader wind. The inner wind portions would shield the outer gas, resulting in a cooler plasma. We also discuss alternative solutions. Our study confirms the existence of a stratified wind discovered by \citet{2018Kosec} and provides hints for multiple phases in both absorption and emission components. The simulation of an observation by the future mission Athena suggests that the study of disk winds in AGN will be revolutionized thanks to its high effective area and high spectral resolution.

\section*{Acknowledgements}
S.B. acknowledges financial support from the Italian Space Agency under grant ASI-INAF I/037/12/0 and from the PRIN MIUR project "Black Hole winds and the Baryon Life Cycle of Galaxies: the stone-guest at the galaxy evolution supper", contract \#2017PH3WAT. D.J.W. acknowledges support from STFC in the form of an Ernest Rutherford fellowship (ST/N004027/1).

\section*{Data Availability}
The data underlying this article are available in ESA’s XMM-Newton Science Archive (https://www.cosmos.esa.int/web/xmm-newton/xsa).

\bibliographystyle{mnras}
\bibliography{ref}

\appendix
\newpage
\section{Photoionization Model Scan}\label{app:modelscan}

\subsection{Absorption}\label{subsec:XABS}

We create a multidimensional grid of $v_\mathrm{LOS}$, $\xi$ and $\sigma_v$, while allowing $N_\mathrm{H}$ to be free. We also adopt a grid of turbulent velocities with the same values as used in section~\ref{subsec:Line-Scan} ($\sigma_v=500$, 1000, 5000 km/s). $v_\mathrm{LOS}$ is varied between 0 and 105000\,km/s ($-0.35c$, which corresponds to an outflow velocity of $-0.3c$ once special relativistic effects are accounted for; note that the minus sign here implies blueshift), with steps that depend on the turbulent velocity ($\Delta v_\mathrm{LOS}=500$, 700, 2500\,km/s for $\sigma_v=500$, 1000, 5000\,km/s respectively). This choice has been made in order not to over-sample broadened features, and save computational time. The grid of ionization parameters $\log\xi$ spans between 0 and 6 with a step of $\Delta\log\xi=0.12$. The search is based on the broadband continuum obtained in section~\ref{subsec:EPIC-RGS} and the continuum parameters are left free during the scan. The \xabs\ scanned each flux-resolved spectrum with the corresponding flux-resolved SED and ionization balance. Afterwards, the $\dcstat$ statistical improvements for each fit are recorded, as done in section~\ref{subsec:Line-Scan}. The improvement of $\dcstat$ indicates the significance of the wind absorption at this point over the baseline continuum model. 

The results for the time-average and flux-resolved spectra are respectively shown in Fig.~\ref{fig:XABSflux}. The results with different turbulent velocities are compatible with each other. Therefore, we present the significance distributions with $\sigma_v=5000\,$km/s in this paper, which has the largest statistical improvement. We then correct the velocity in our line-of-sight for the Doppler relativistic effect through the standard equation: $v/c=\sqrt{(1+v_\mathrm{LOS}/c)/(1-v_\mathrm{LOS}/c)}-1$. As expected, our search for the time-average spectrum achieves a strong detection of the highly ionized ($\logxi\sim4.2$) plasma with a blueshift velocity peaking around $-0.14c$ and a statistical improvement of up to $\dcstat=320$. Our result confirms the best-fit solution ($v\sim-0.13c$, $\logxi=4.3$) found in PK18. In addition, our results indicate a secondary less significant ($\dcstat\sim160$) peak of a lower ionization state ($\logxi\sim3$), appearing also in flux-resolved results, which itself could be considered significant if it is detected alone. The presence of the secondary peak could suggest a multi-phase absorbing wind in 1H 0707, consistent with the discovery in PK18. Given the huge $\dcstat$, we do not need to run Monte Carlo simulations to probe the significance of the wind components (both absorption $\dcstat=320$ and emission $\dcstat=200$ in following section).

\subsection{Emission}\label{subsec:PION}

To accelerate the speed of our model scan, we use the code\footnote{https://github.com/ciropinto1982/Spectral-fitting-with-SPEX/tree/master/SPEX-physical-grid-scan} employed in \citet{2020Pinto} to pre-calculate the \pion\ model for all 1H 0707 SEDs. We construct a simple grid of \pion\ models with the ionization parameter, $\log\xi$, between 0 and 6 with 0.1 steps and several line widths between 100\,km/s and 5000\,km/s ($\sigma_v=100$, 500, 1000, 5000\,km/s). Due to the lack of well-resolved He-like triplets (e.g. O {\scriptsize VII}), we are unable to estimate the volume density and it will degenerate with the column density of photoionized gas \citep{2000Porquet}. Consequently, we assume a low constant plasma density $n_\mathrm{H}=1\,\mathrm{cm}^{-3}$ as \pion\ calculates fast at a low volume density and we have checked that this parameter has little effect on emission. For the emission line scan, we also set a full solid angle $\Omega=4\pi$ for saving computation time.

The results of the emission model scan are shown in Fig.~\ref{fig:PIONflux} using a line width of 5000\,km/s. The peak in the time-average result corresponds to a solution at $\logxi\sim3.6$ with a blueshift velocity of $v\sim-0.02c$. The ionization parameter is over one order of magnitude larger than that ($\sim2.4$) of PK18, while the velocity is consistent with theirs. Their results were derived by explaining both emission and absorption components, while here we just consider the emission component. Moreover, the plot shows two potential solutions of $\log\xi<3$ and $>3$. The peak of the \pion\ scan shows that the emitting gas is increasingly ionized at a higher luminosity ($\log\xi$ ranging from $\sim2$ to 4).

\begin{figure}
\centering
\includegraphics[width=0.49\textwidth,trim={0 0 0 0}]{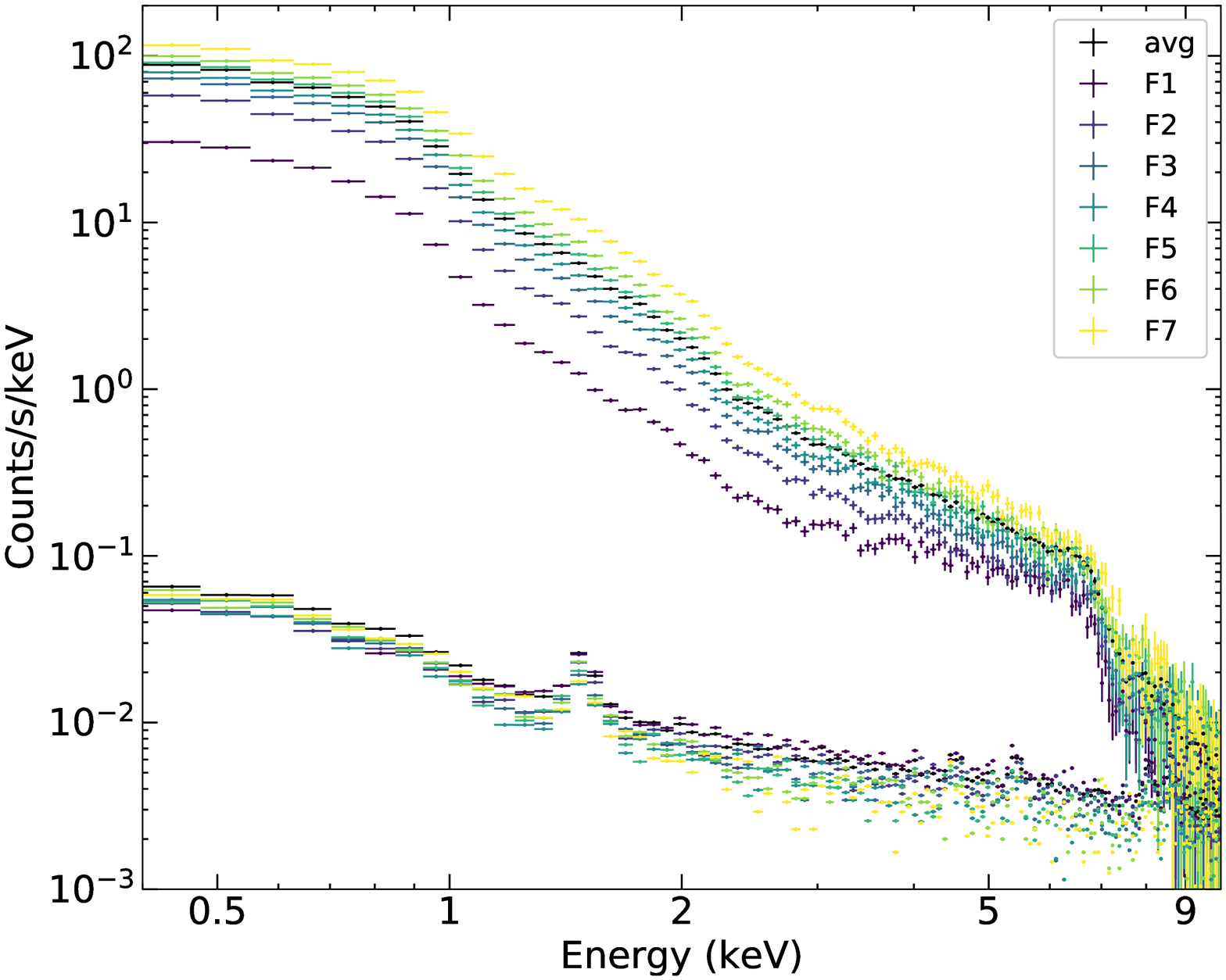}
\includegraphics[width=0.48\textwidth,trim={0 30 0 0}]{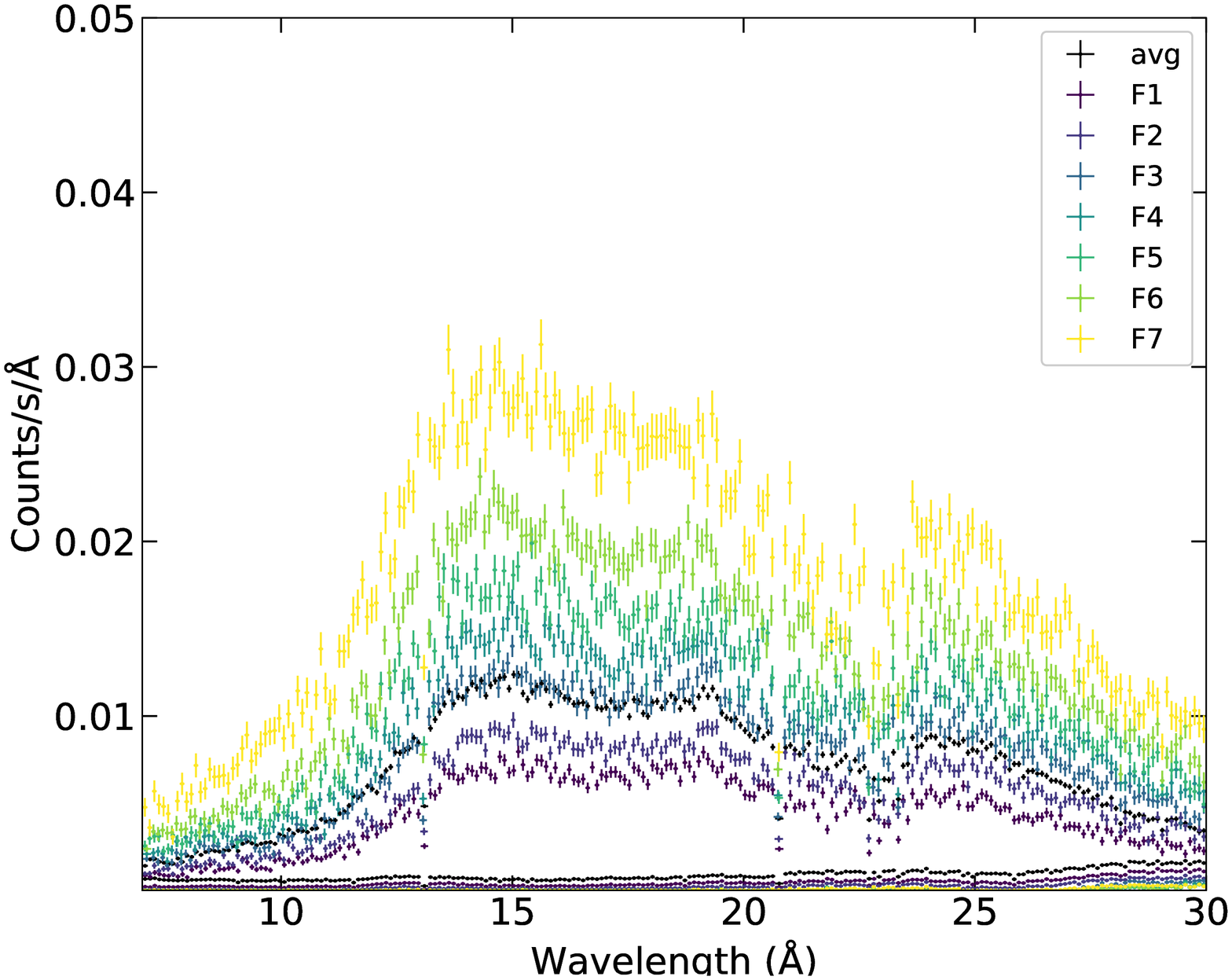}
\caption{
The time-average and flux-resolved source spectra of 1H 0707 using EPIC-pn ({\it Upper}) and RGS ({\it Lower}) instruments. The background spectra are also plotted, comparable to the source spectra above 8\,keV for EPIC-pn and on both ends of the spectral range of RGS. The spectra have been rebinned for plotting purpose.
}
\label{fig:spectra}
\end{figure}

\begin{figure}
\centering
\includegraphics[width=0.49\textwidth,trim={0 0.0cm 0 0}]{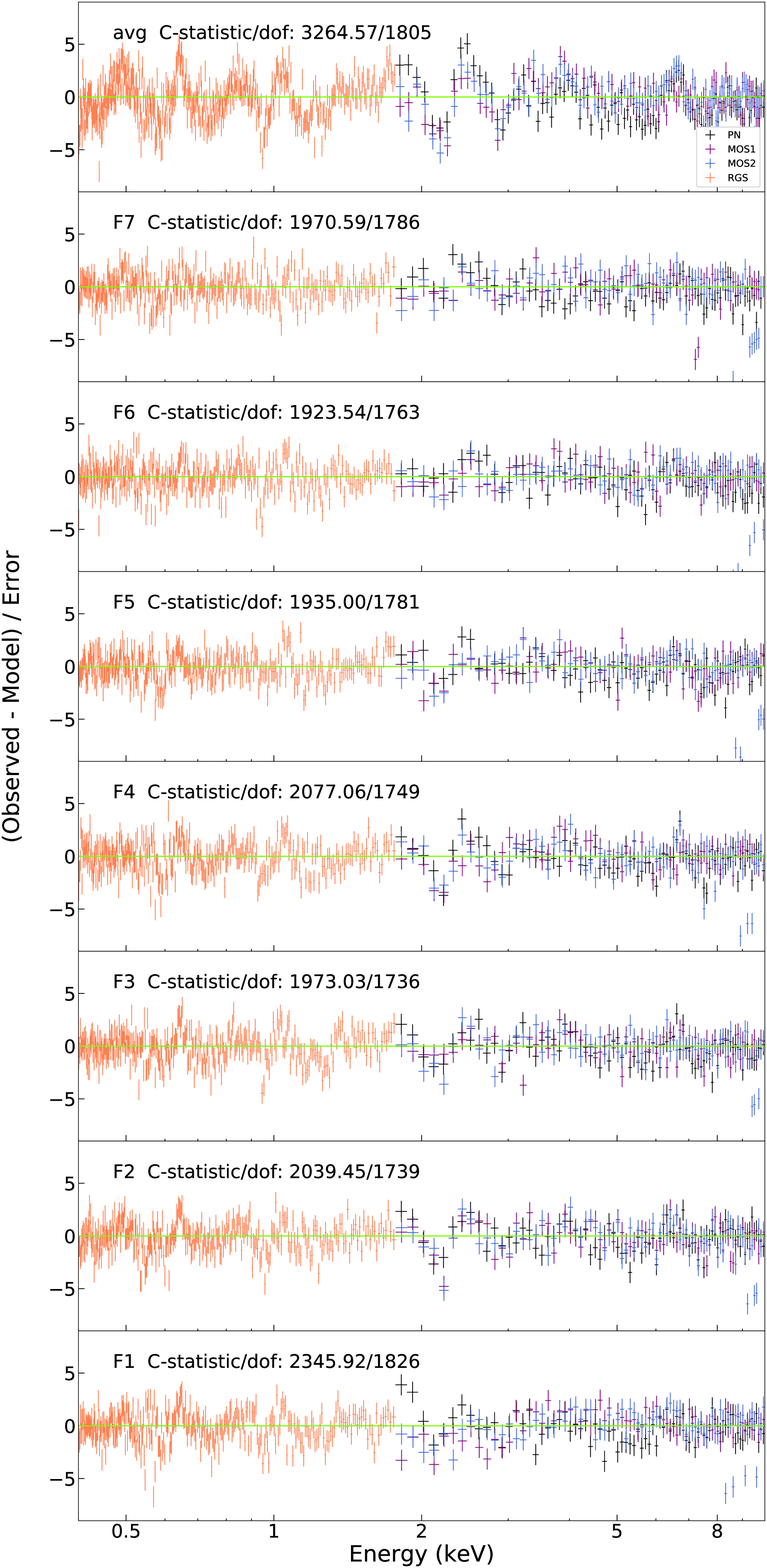}
\caption{
The residuals for the time-average ({\it top}) and flux-resolved (ordered according to the count rates from bottom to the second panel) spectra with respect to a phenomenological model {\tt hot*reds*(pow+bb+laor*(delt+delt))}. The Cash statistics and the degree of freedom (dof) are marked as well. 
}
\label{fig:ratio}
\end{figure}

\begin{table*}
\caption{Best-fit parameters of the phenomenological model {\tt hot*reds*(pow+bb+laor*(delt+delt))} for the time-average and flux-resolved spectra.}
\begin{tabular}{lcccccccccccr}
\hline\hline
Description & Parameter & avg & F1 & F2 & F3 & F4 & F5 & F6 & F7    \\
\hline
{\tt hot} &    $N_\mathrm{H}^\mathrm{Gal}$ ($10^{20}$ cm$^{-2}$)  & \multicolumn{8}{|c|}{$4^\star$}\\
            &    kT ($10^{-5}$\,keV)    & \multicolumn{8}{|c|}{$2^\star$}\\
\hline
{\tt pow}        &  $\Gamma$   &      $2.75^{+0.01}_{-0.01}$    &      $2.20^{+0.04}_{-0.04}$         & $2.61^{+0.03}_{-0.03}$  & $2.71^{+0.03}_{-0.02}$ & $2.76^{+0.03}_{-0.02}$ & $2.84^{+0.03}_{-0.02}$ & $2.91^{+0.02}_{-0.02}$ & $2.99^{+0.02}_{-0.02}$\\
                  & $N_\mathrm{PL}$  ($10^{51}$\,ph/s/keV)  &  $3.52^{+0.05}_{-0.04}$ &  $1.49^{+0.05}_{-0.05}$  & $2.49^{+0.07}_{-0.07}$ & $3.87^{+0.10}_{-0.10}$ & $4.97^{+0.12}_{-0.12}$ &
                  $6.34^{+0.15}_{-0.15}$ & $8.0^{+0.2}_{-0.2}$ & $12.6^{+0.03}_{-0.03}$ \\
\hline
{\tt bb}      & $T_\mathrm{in}$ (keV)  &  $0.123^{+0.001}_{-0.001}$  &  $0.117^{+0.001}_{-0.001}$    & $0.111^{+0.002}_{-0.002}$ & $0.109^{+0.002}_{-0.002}$ & $0.107^{+0.002}_{-0.002}$ & $0.110^{+0.002}_{-0.002}$  & $0.117^{+0.003}_{-0.003}$ & $0.121^{+0.004}_{-0.004}$ \\
                &  $N_\mathrm{BB}$ ($10^{20}\,\mathrm{cm}^{2}$)   &     $841^{+22}_{-18}$    &     $922^{+51}_{-47}$     & $1322^{+91}_{-86}$ & $1774^{+140}_{-124}$  & $2173^{+220}_{-190}$ & $1906^{+223}_{-171}$ & $1566^{+155}_{-134}$ &$1121^{+175}_{-147}$  \\
\hline
{\tt de1t1} & $\mathrm{LineE1}$ (keV) & $0.914^{+0.001}_{-0.001}$ & $0.922^{+0.004}_{-0.009}$ & $0.905^{+0.002}_{-0.002}$ & $0.909^{+0.005}_{-0.003}$ & $0.922^{+0.003}_{-0.003}$ & $0.935^{+0.002}_{-0.002}$ & $0.920^{+0.003}_{-0.003}$ & $0.976^{+0.003}_{-0.002}$\\
          & $N_\mathrm{delt1}$ ($10^{51}$\,ph/s)  &  $2.5^{+0.1}_{-0.1}$ &  $1.7^{+0.1}_{-0.1}$ & $2.2^{+0.2}_{-0.2}$ & $3.0^{+0.2}_{-0.2}$ & $4.0^{+0.3}_{-0.3}$ & $4.5^{+0.4}_{-0.3}$ & $4.7^{+0.4}_{-0.4}$ & $7.3^{+0.7}_{-0.7}$   \\
\hline
{\tt delt2} & $\mathrm{LineE2}$ (keV)  & $6.47^{+0.02}_{-0.02}$ & $6.40^{+0.04}_{-0.04}$ & $6.48^{+0.04}_{-0.03}$ & $6.49^{+0.05}_{-0.05}$ & $6.53^{+0.06}_{-0.06}$ & $6.54^{+0.05}_{-0.04}$ & $6.47^{+0.05}_{-0.05}$ & $6.72^{+0.09}_{-0.07}$\\
          & $N_\mathrm{delt2}$ ($10^{50}$\,ph/s)   &  $0.67^{+0.02}_{-0.02}$ &  $1.65^{+0.09}_{-0.09}$ & $0.77^{+0.06}_{-0.05}$ & $0.80^{+0.05}_{-0.05}$ & $0.88^{+0.06}_{-0.06}$ & $1.00^{+0.07}_{-0.06}$ & $0.95^{+0.07}_{-0.07}$ & $1.32^{+0.11}_{-0.10}$   \\
\hline
{\tt laor} 	& $q$ &  $5.3^{+0.1}_{-0.1}$ &  $5.6^{+0.1}_{-0.1}$&  $5.3^{+0.1}_{-0.1}$&  $5.3^{+0.2}_{-0.1}$&  $5.5^{+0.2}_{-0.2}$ &  $5.5^{+0.1}_{-0.1}$ &  $5.2^{+0.1}_{-0.1}$ &  $6.4^{+0.2}_{-0.2}$ \\
			& $i$ (deg)    & $50.7^{+0.3}_{-0.1}$ & $50.7^\star$ & $50.7^\star$ & $50.7^\star$ & $50.7^\star$ & $50.7^\star$ & $50.7^\star$  & $50.7^\star$\\
\hline
                    & C-statistic/dof & 3264.57/1805 & 2345.92/1826 & 2039.45/1739 & 1973.03/1736 & 2077.06/1749 & 1935.00/1781 & 1923.54/1763 & 1970.59/1786  \\
\hline
\end{tabular}

\begin{flushleft}
{$^{\star}$ The parameter is pegged at this value.}
\end{flushleft}
\label{tab:ratios}
\end{table*}

\begin{table*}
\caption{Best-fit parameters of the photoionization absorption (\xabs) and emission (\pion) models applied on the continuum model for the time-average and flux-resolved spectra. The line width, $\sigma_v$, of them are linked together. Here the velocity, $v_\mathrm{LOS}$, are the absolute values of the line-of-sight velocity shown in SPEX, which has not been relativistically corrected yet. The uncertainties are estimated at 68\% confidence level.}
\begin{tabular}{lcccccccccccr}
\hline\hline
Description & Parameter & avg & F1 & F2 & F3 & F4 & F5 & F6 & F7    \\
\hline
{\tt xabs} &   $N_\mathrm{H}$ ($10^{23}$\,cm$^{-2}$)    &  $0.26^{+0.09}_{-0.02}$  & $0.18^{+0.05}_{-0.05}$ &   $0.80^{+0.05}_{-0.23}$  &  $0.6^{+0.2}_{-0.1}$ & $1.0^{+0.4}_{-0.2}$  &   $1.2^{+0.4}_{-0.4}$ & $1.2^{+0.3}_{-0.3}$ &  $3.2^{+1.1}_{-2.4}$  \\
        &    $\log\xi$ (erg\,cm\,s$^{-1}$)   &  $4.20^{+0.08}_{-0.01}$  & $3.93^{+0.05}_{-0.06}$ & $4.59^{+0.01}_{-0.11}$    &  $4.32^{+0.08}_{-0.08}$ & $4.52^{+0.07}_{-0.08}$  & $4.90^{+0.17}_{-0.17}$& $4.87^{+0.18}_{-0.15}$ &  $5.70^{+0.30}_{-0.27}$ \\
        &    $|v_\mathrm{LOS}|$ (km/s)   &  $46594^{+577}_{-1806}$  & $53197^{+1765}_{-1985}$ &   $44788^{+1395}_{-1364}$  &  $45827^{+1475}_{-1543}$ & $45602^{+2250.64}_{-1466.45}$  &   $44094^{+1548}_{-1290}$ & $40738^{+1563}_{-2110}$ &  $41832^{+3169}_{-2796}$  \\
        &    $\sigma_v$ (km/s)   &  $4986^{+709}_{-276}$  & $4630^{+973}_{-774}$ &   $4630^{+787}_{-981}$  &  $4535^{+1088}_{-622}$ & $8373^{+1279}_{-1194}$  &   $4800^{+2767}_{-1529}$ & $7077^{+1400}_{-1258}$ &  $5000^{\star}$  \\

\hline
{\tt pion}       &   $N_\mathrm{H}$ ($10^{21}$\,cm$^{-2}$)    &  $1.8^{+0.2}_{-0.2}$  & $2.3^{+0.6}_{-0.5}$ &   $2.9^{+0.4}_{-0.4}$  &  $2.5^{+0.6}_{-0.5}$ & $2.5^{+1.0}_{-0.5}$  &   $1.2^{+0.5}_{-0.2}$ & $1.9^{+0.5}_{-0.4}$ &  $0.6^{+0.5}_{-0.3}$  \\
        &    $\log\xi$ (erg\,cm\,s$^{-1}$)   &  $2.68^{+0.02}_{-0.04}$  & $2.69^{+0.07}_{-0.08}$ & $2.75^{+0.06}_{-0.06}$    &  $2.64^{+0.07}_{-0.08}$ & $2.54^{+0.08}_{-0.07}$  & $2.44^{+0.13}_{-0.14}$& $2.59^{+0.05}_{-0.05}$ &  $2.35^{+0.20}_{-0.25}$ \\
        &    $|v_\mathrm{LOS}|$ (km/s)   &  $7574^{+572}_{-437}$  & $8457^{+760}_{-973}$ &   $6654^{+954}_{-913}$  &  $9269^{+726}_{-928}$ & $8007^{+1418}_{-1055}$  &   $7117^{+1157}_{-1122}$ & $9736^{+1001}_{-2369}$ &  $9088^{+1808}_{-2316}$  \\
\hline
                    & C-statistic/dof & 2639/1800 & 2233/1820 & 1906/1733 & 1801/1730 & 1957/1743 & 1866/1775 & 1873/1757 & 1936/1781 \\
\hline
\end{tabular}
\begin{flushleft}
{$^{\star}$ The parameter is pegged at this value.}
\end{flushleft}
\label{tab:evolution}
\end{table*}

\begin{figure*}
\centering
\includegraphics[width=0.30\textwidth,trim={0 0.0cm 0 0}]{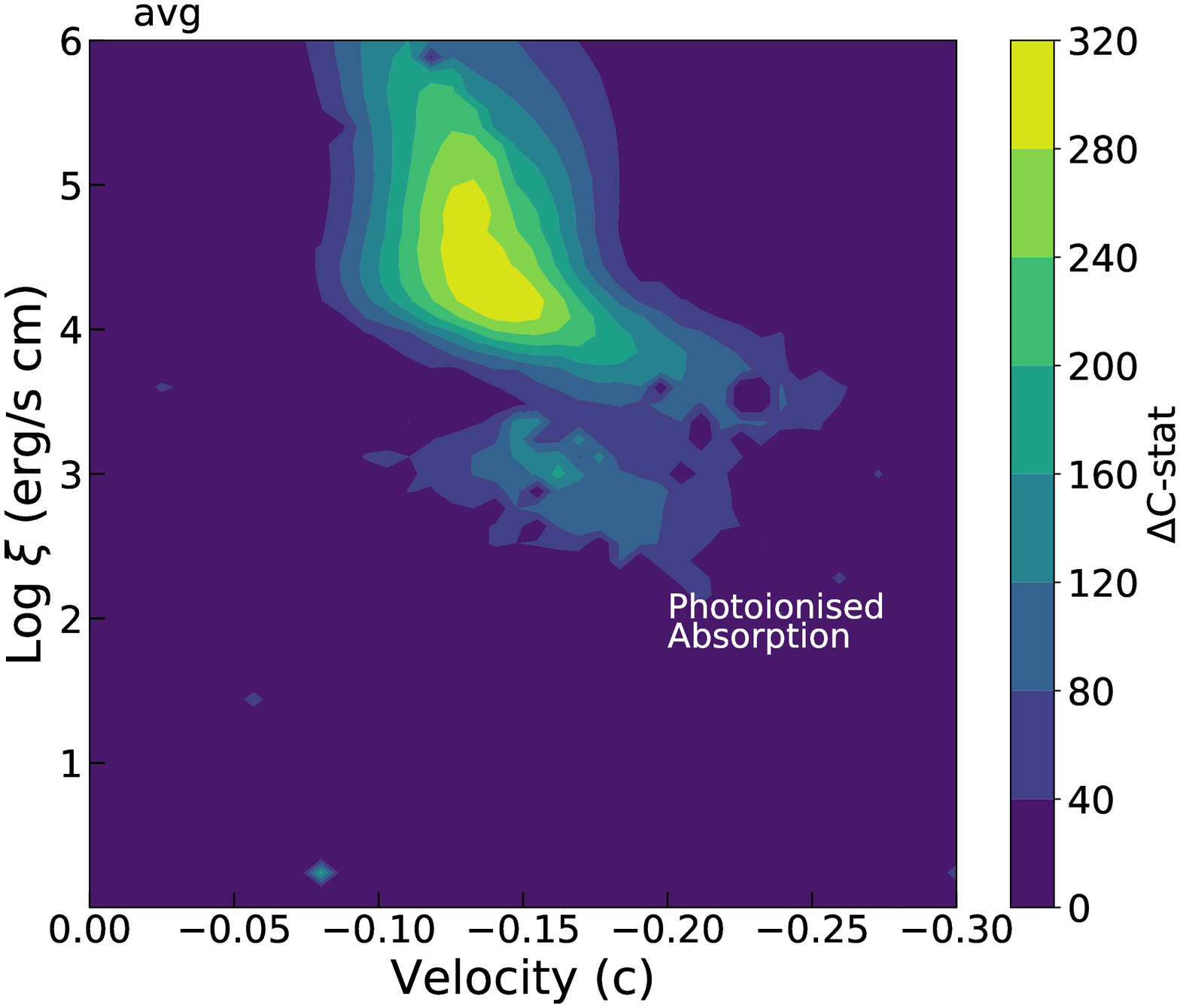}
\includegraphics[width=0.30\textwidth,trim={0 0.0cm 0 0}]{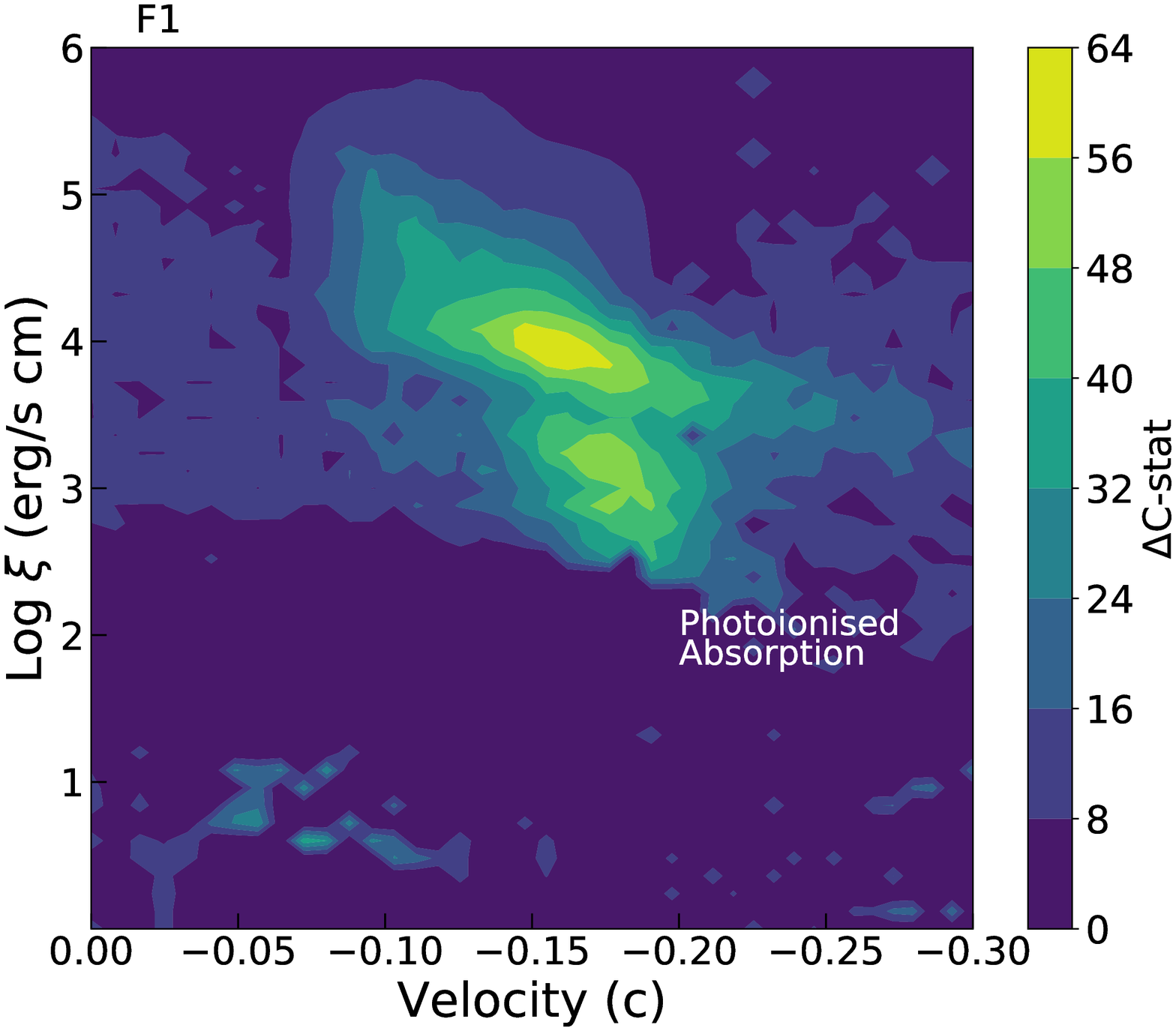}
\includegraphics[width=0.30\textwidth,trim={0 0.0cm 0 0}]{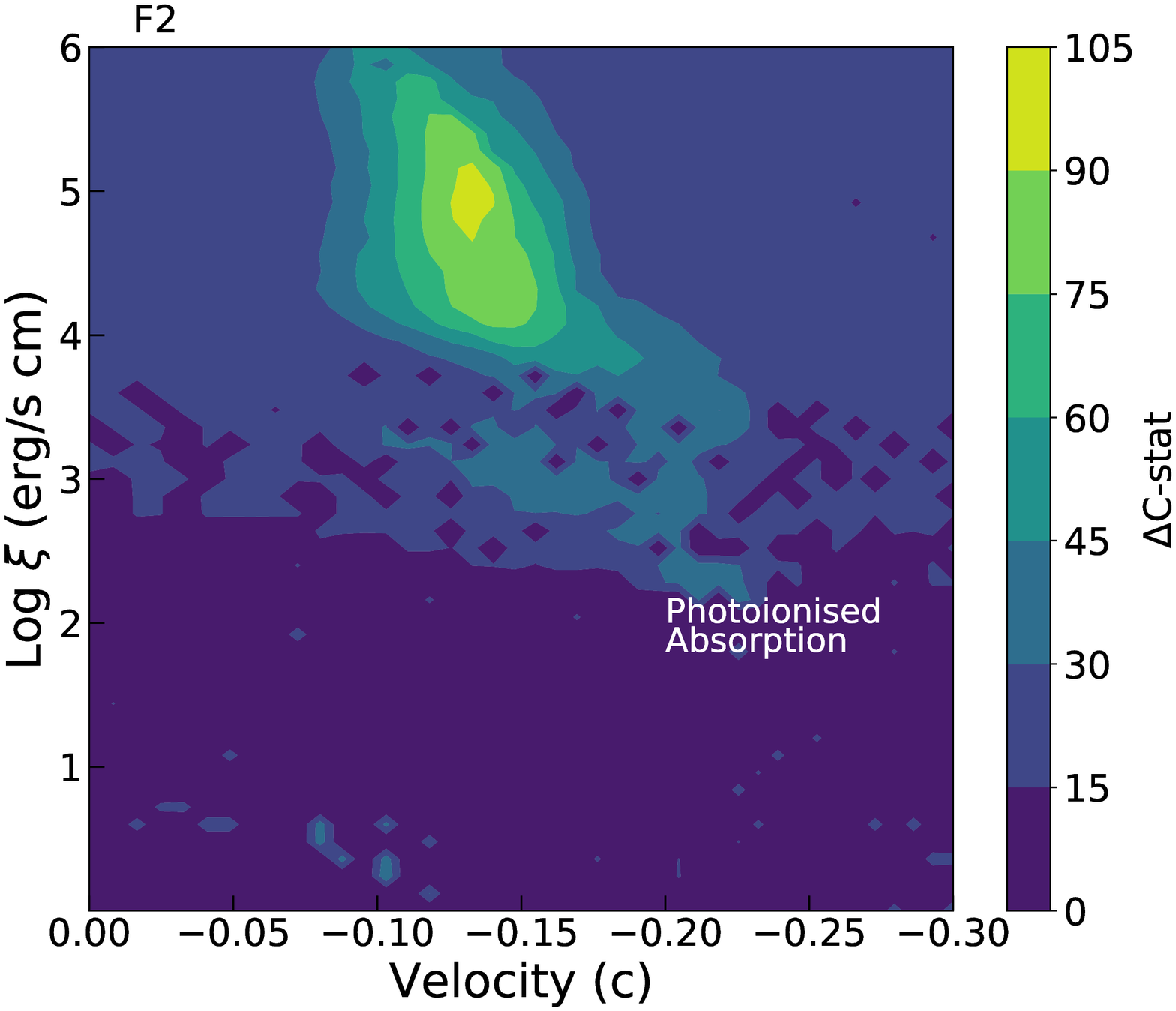}
\includegraphics[width=0.30\textwidth,trim={0 0.0cm 0 0}]{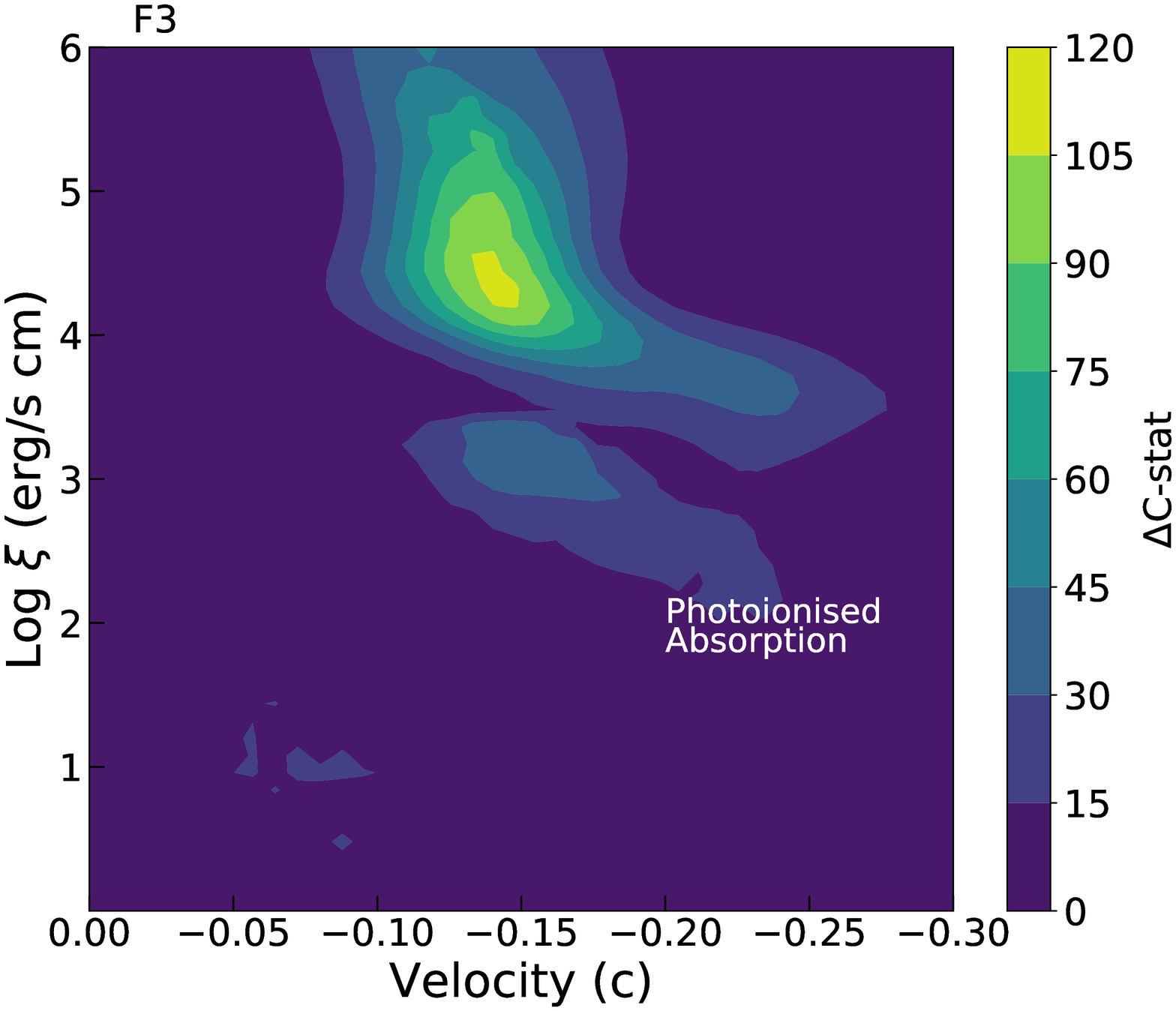}
\includegraphics[width=0.30\textwidth,trim={0 0.0cm 0 0}]{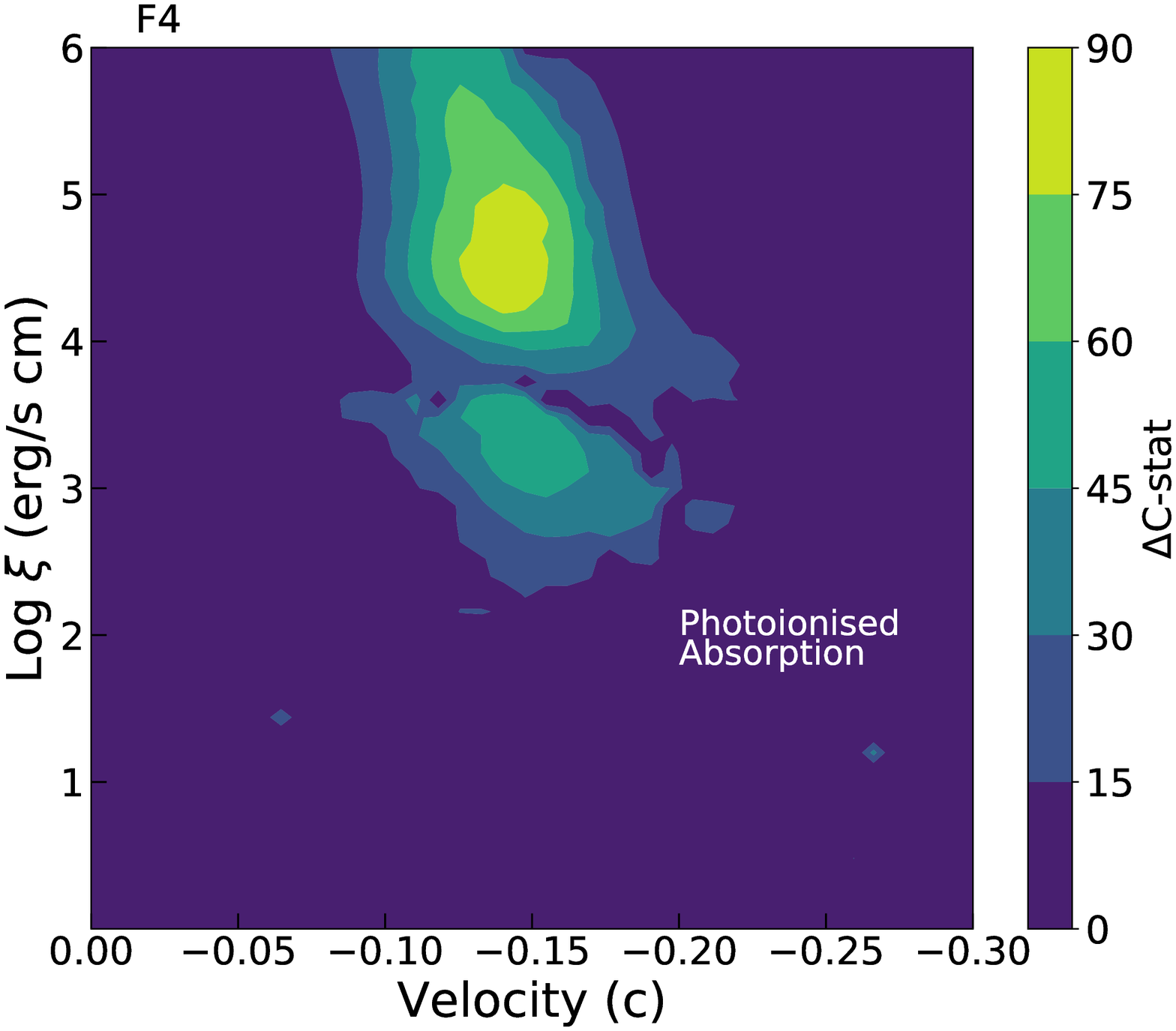}
\includegraphics[width=0.30\textwidth,trim={0 0.0cm 0 0}]{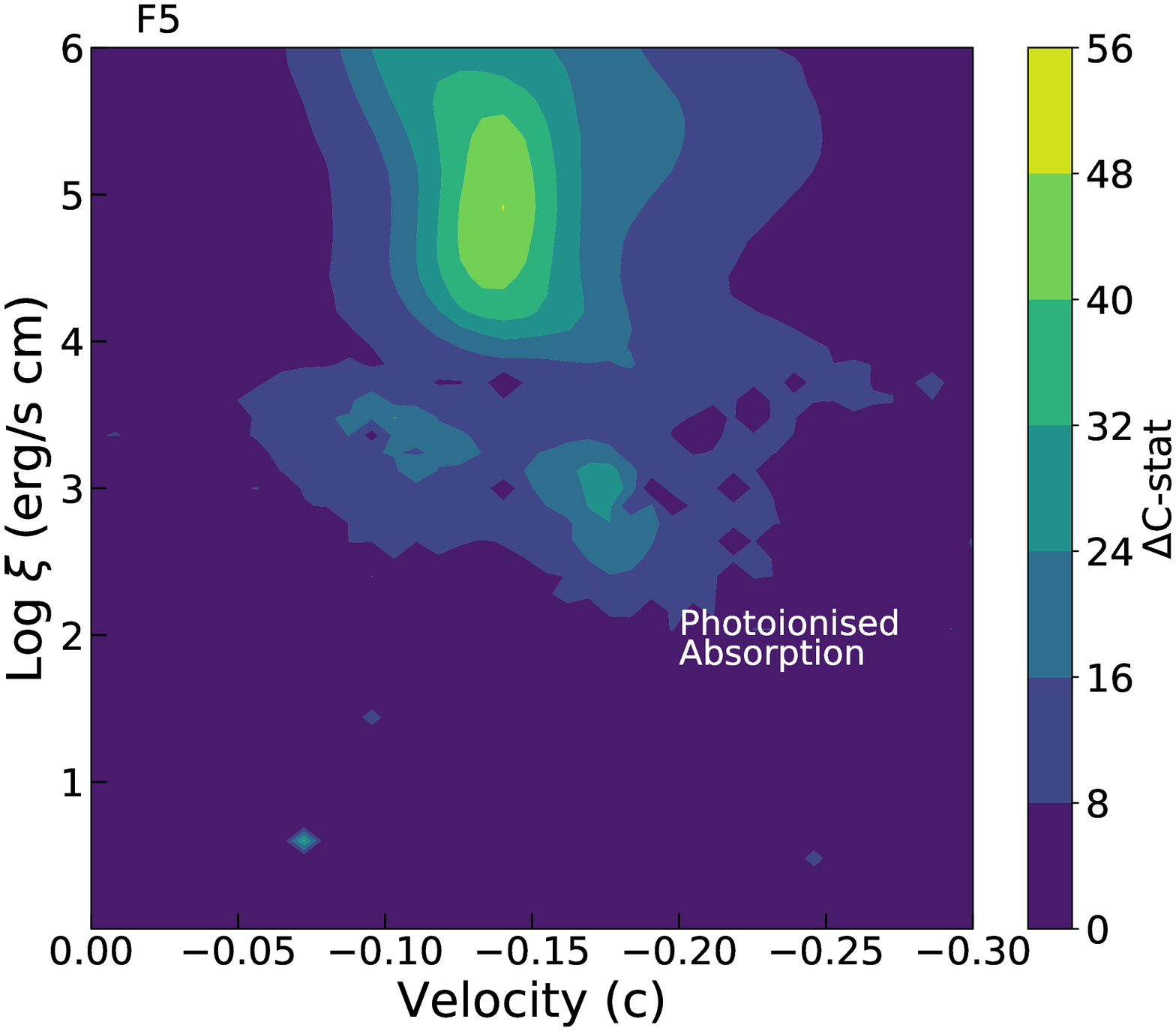}
\includegraphics[width=0.30\textwidth,trim={0 0.0cm 0 0}]{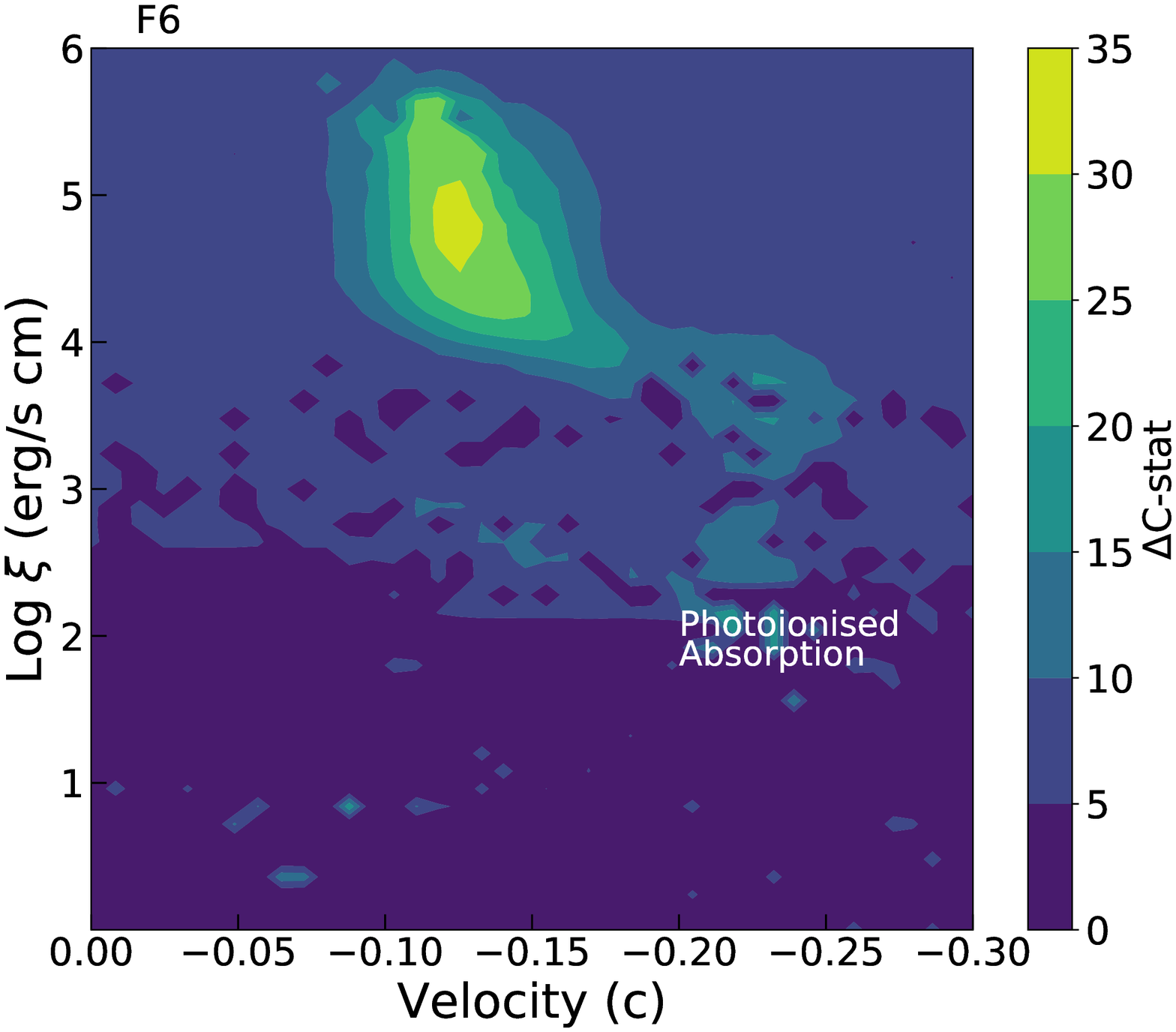}
\includegraphics[width=0.30\textwidth,trim={0 0.0cm 0 0}]{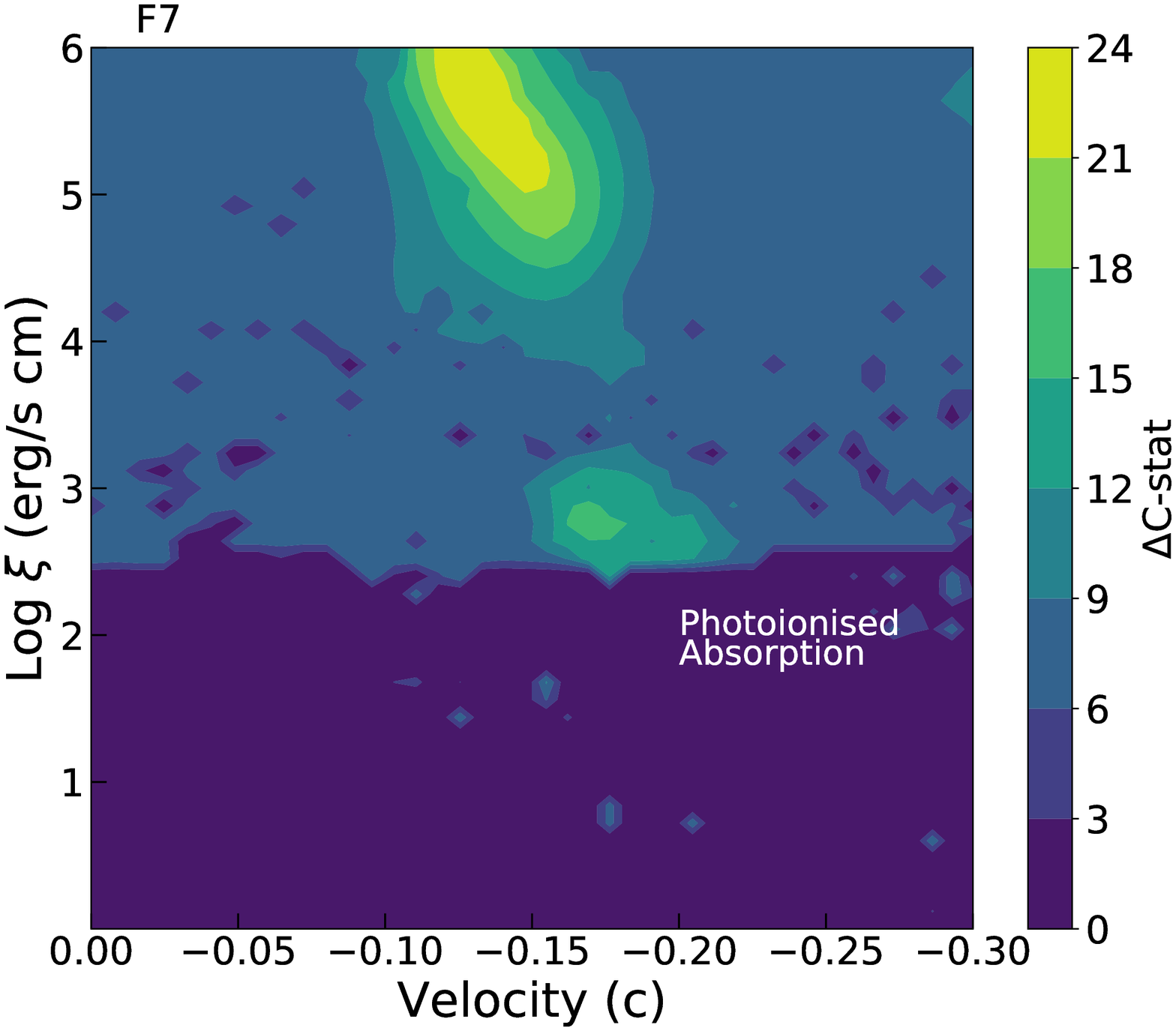}
\caption{
Photoionization absorption model search for the time-average and flux-resolved spectrum of 1H 0707. Each spectrum is scanned in a grid of ionization parameters $\log\xi$ (on Y axis) and the line-of-sight velocities $v_\mathrm{LOS}$, with turbulent velocities of 5000\,km/s. Here the velocities on the X axis are the relativistically corrected velocities. The color shows the statistical improvement $\dcstat$ after adding the absorption model to the continuum model. The best solution for each flux-resolved spectra illustrates an increasing trend of the ionization state with the source luminosity.
}
\label{fig:XABSflux}
\end{figure*}

\begin{figure*}
\centering
\includegraphics[width=0.30\textwidth,trim={0 0.0cm 0 0}]{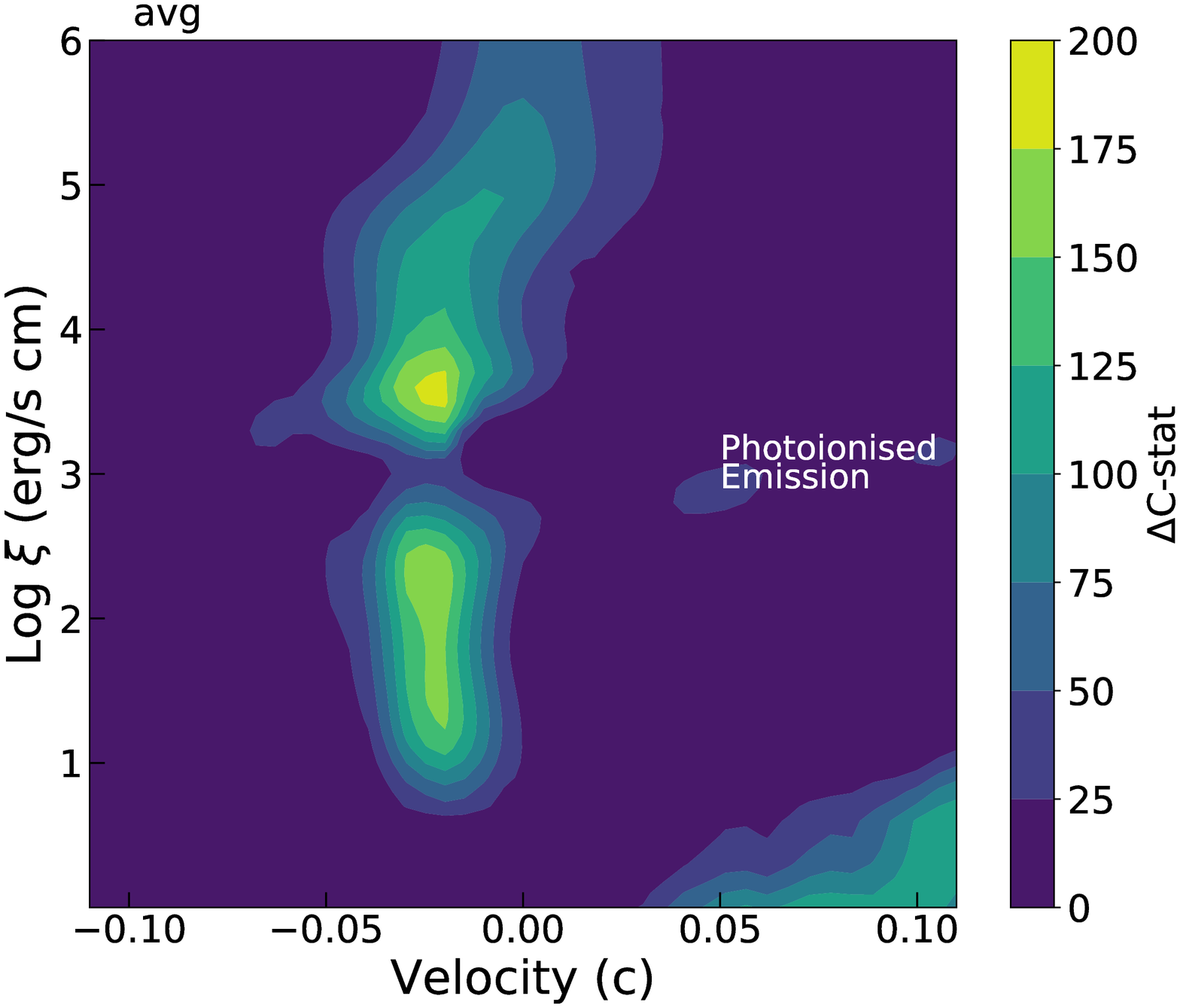}
\includegraphics[width=0.30\textwidth,trim={0 0.0cm 0 0}]{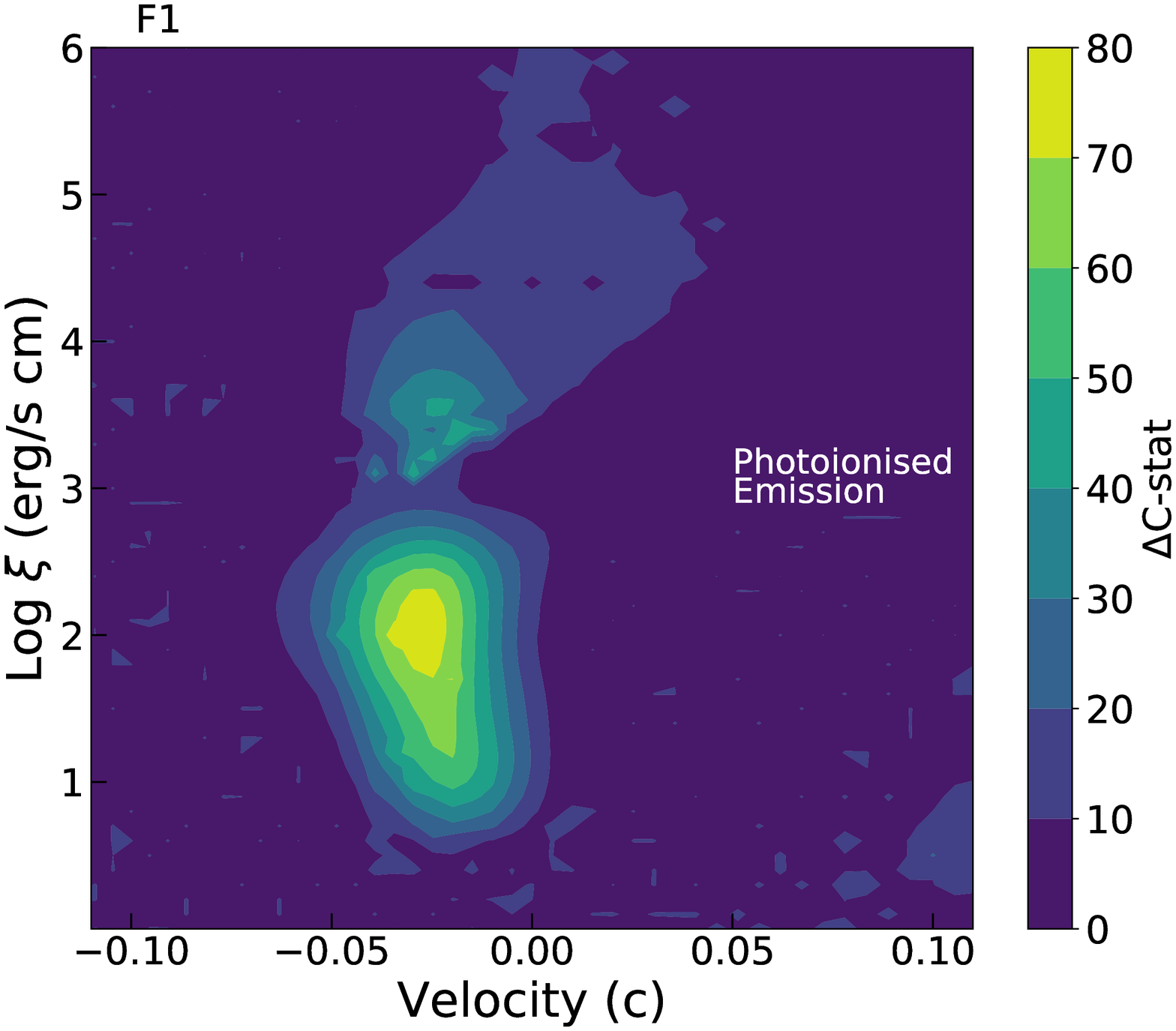}
\includegraphics[width=0.30\textwidth,trim={0 0.0cm 0 0}]{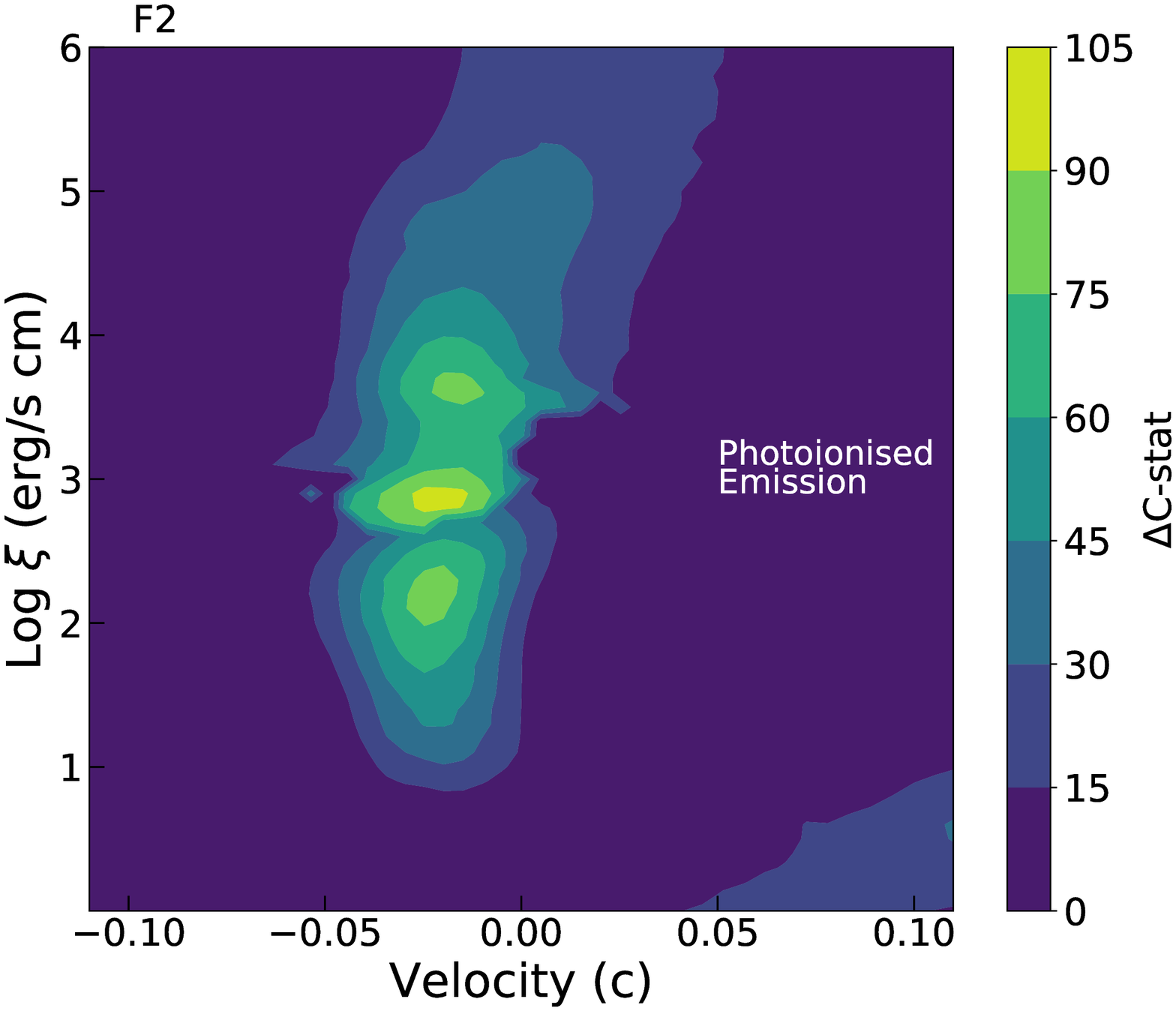}
\includegraphics[width=0.30\textwidth,trim={0 0.0cm 0 0}]{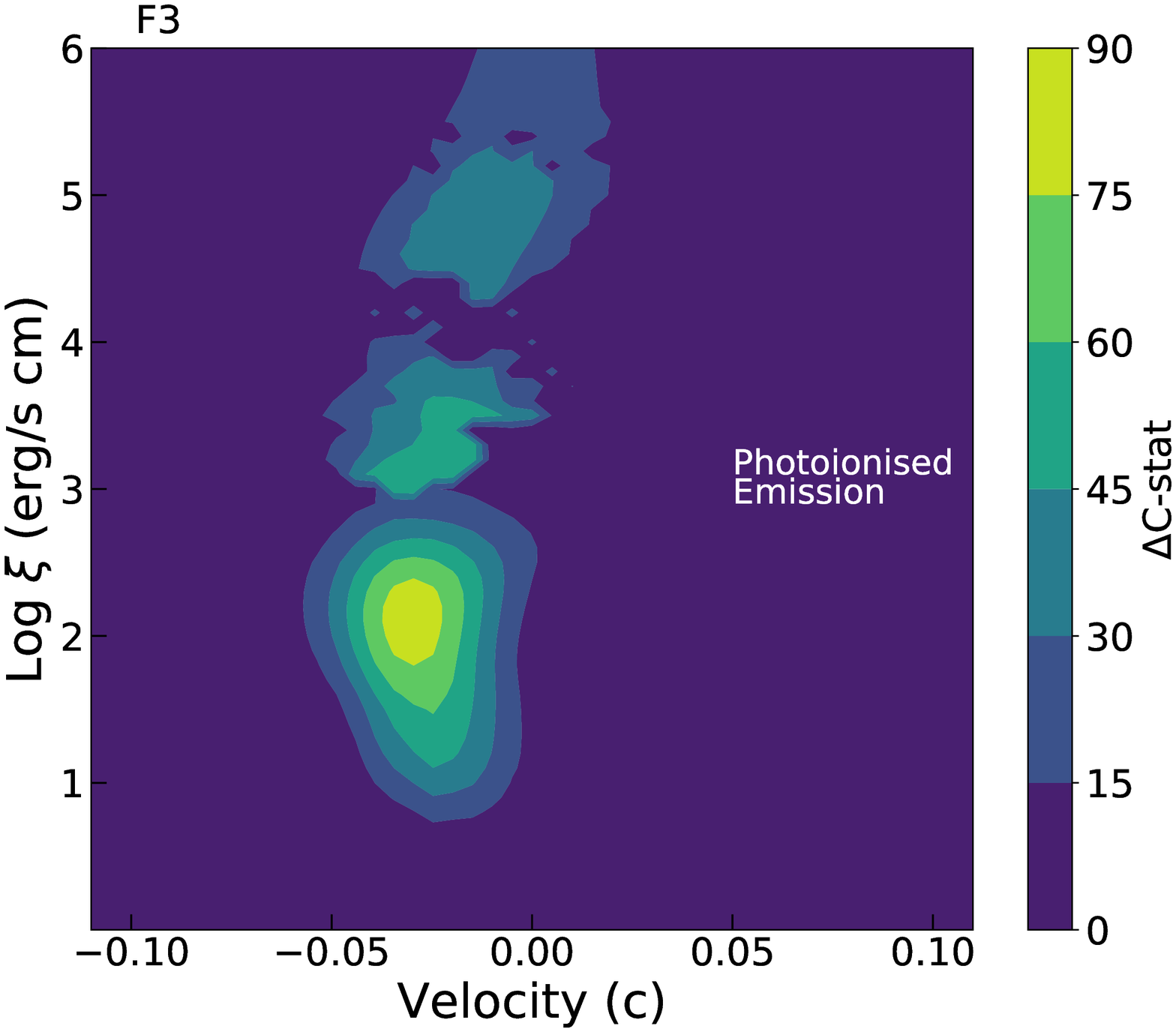}
\includegraphics[width=0.30\textwidth,trim={0 0.0cm 0 0}]{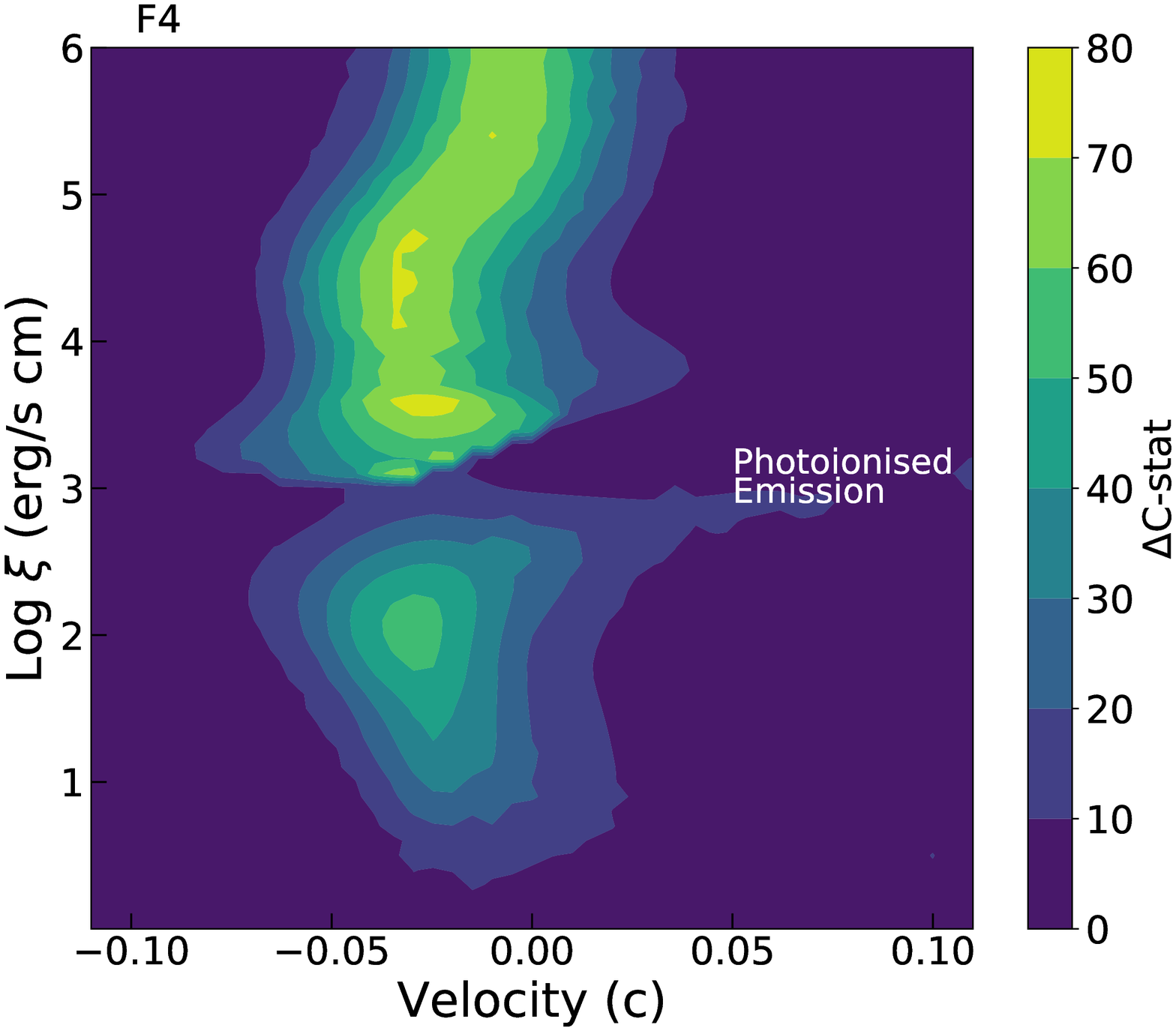}
\includegraphics[width=0.30\textwidth,trim={0 0.0cm 0 0}]{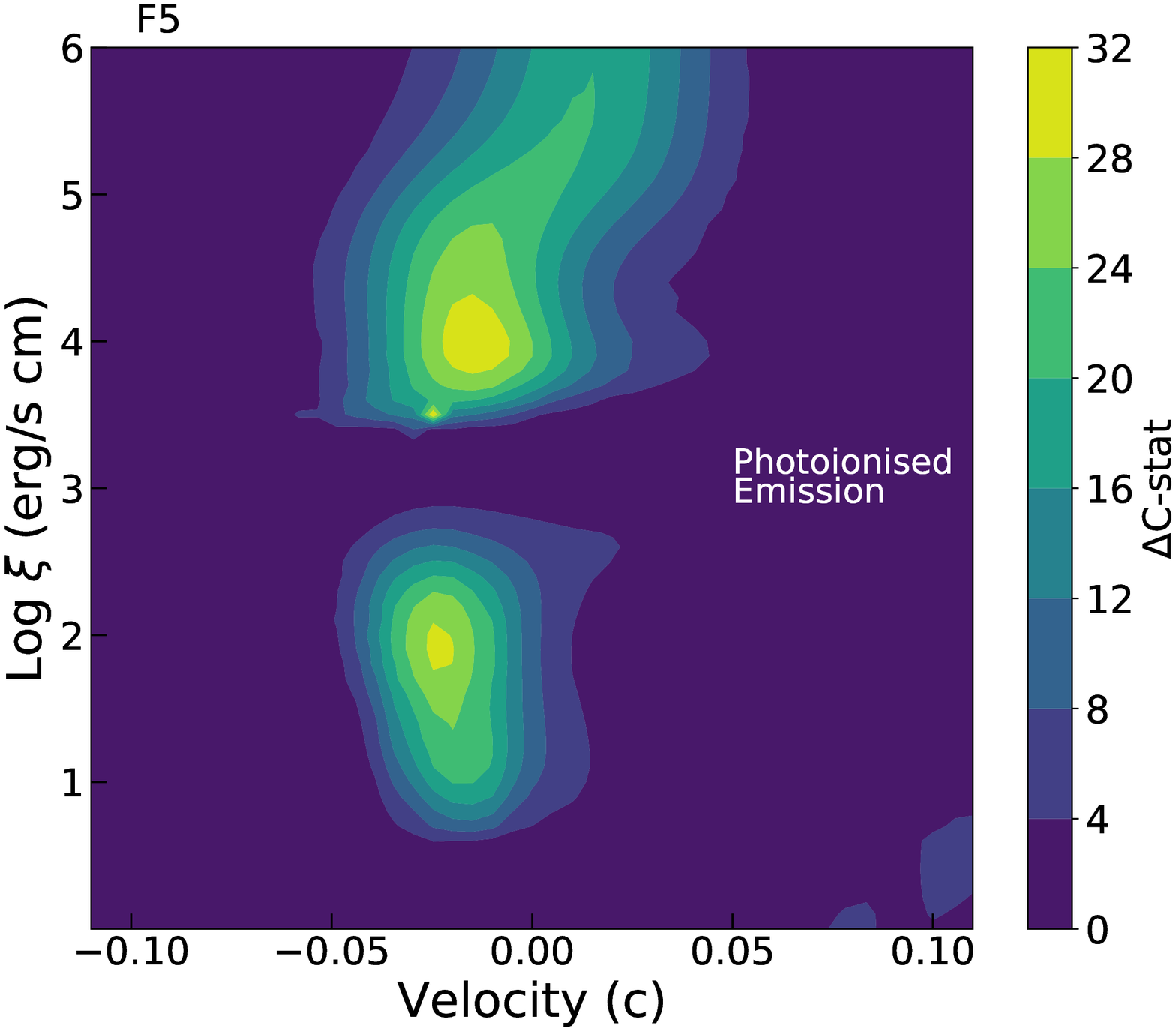}
\includegraphics[width=0.30\textwidth,trim={0 0.0cm 0 0}]{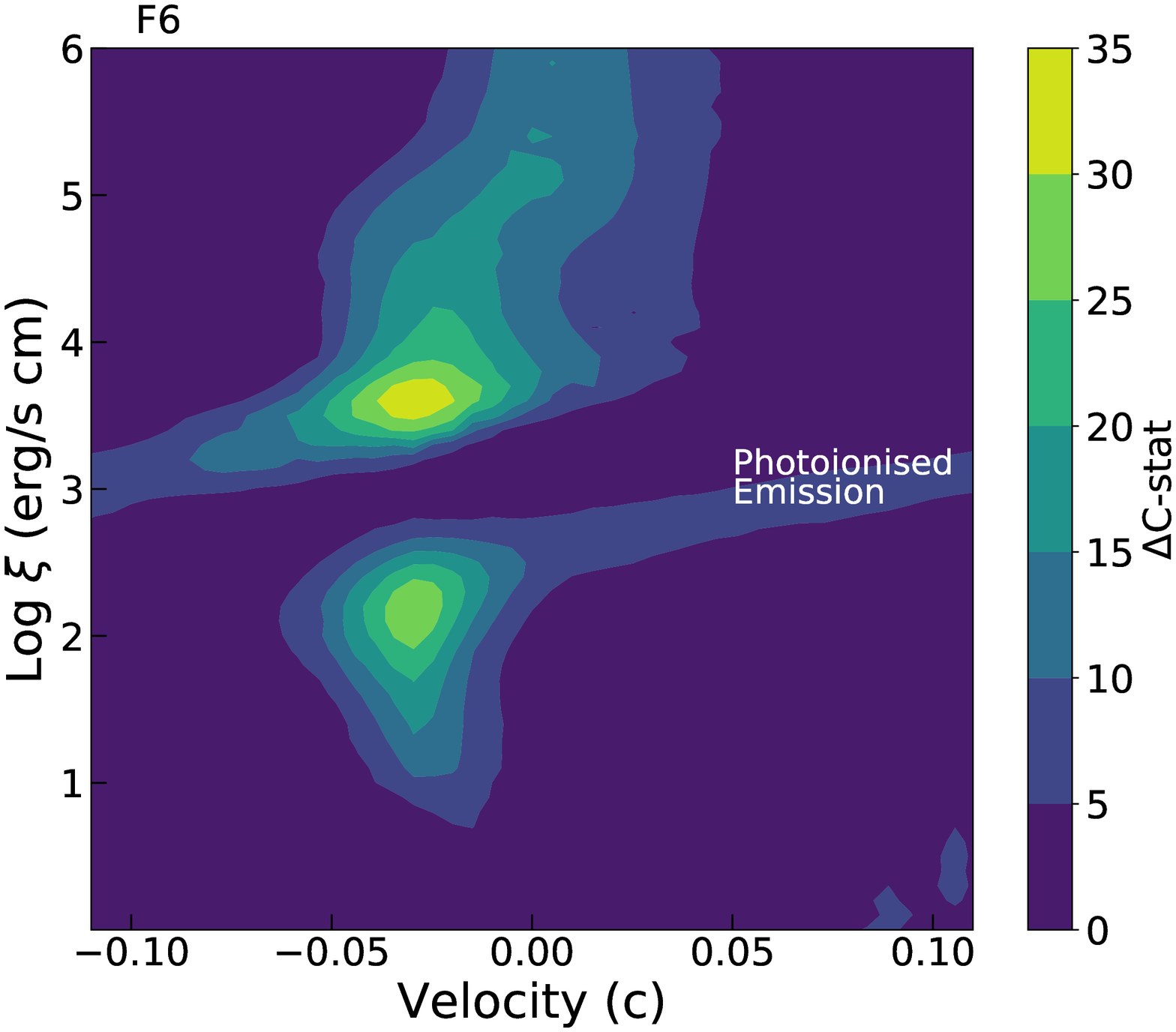}
\includegraphics[width=0.30\textwidth,trim={0 0.0cm 0 0}]{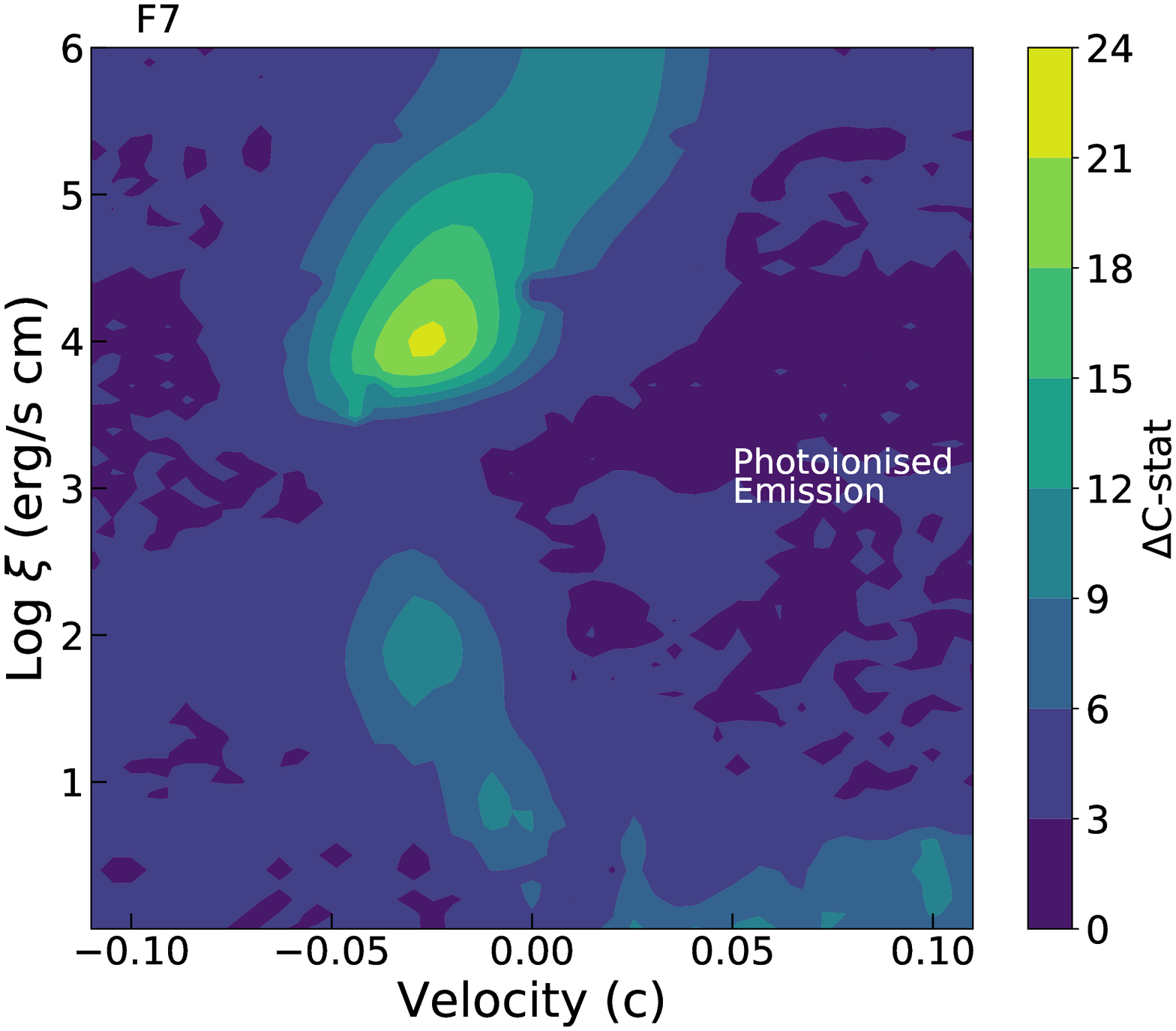}
\caption{
Photoionization emission model search for the time-average and flux-resolved spectrum of 1H 0707. The spectrum is scanned in a grid of ionization parameters $\log\xi$ and the line-of-sight velocities $v_\mathrm{LOS}$, with turbulent velocities of 5000\,km/s. The color shows the statistical improvement $\dcstat$ after adding the emission model to the continuum model. The significance distribution displays two possible solutions of $\logxi<3$ and $>3$ and only the best solution of F1 spectrum falls in the lower ionization zone.
}
\label{fig:PIONflux}
\end{figure*}
\begin{figure*}
\centering
\includegraphics[width=0.49\textwidth,trim={0 0 0 0}]{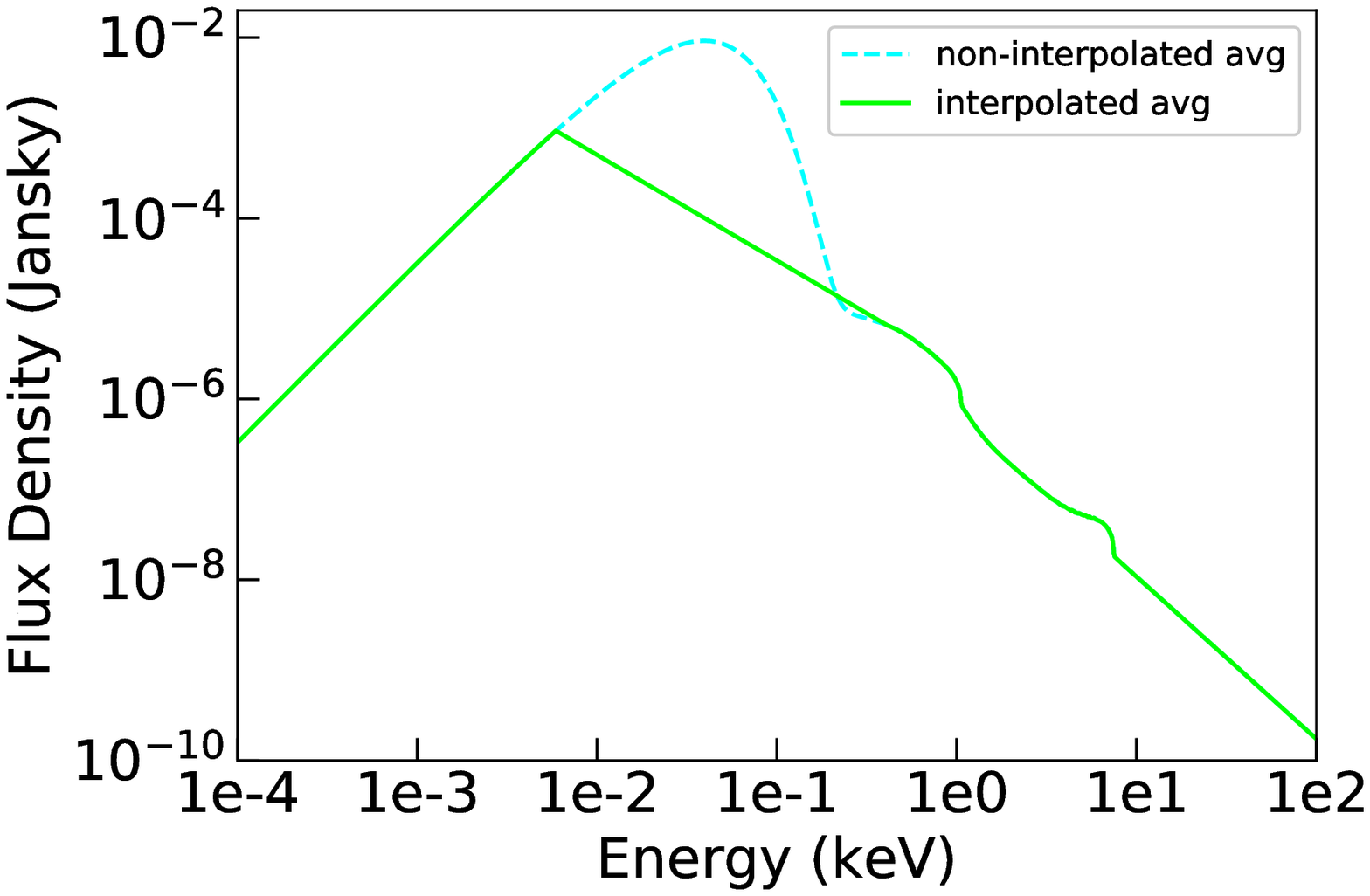}
\includegraphics[width=0.49\textwidth,trim={0 0 0 0}]{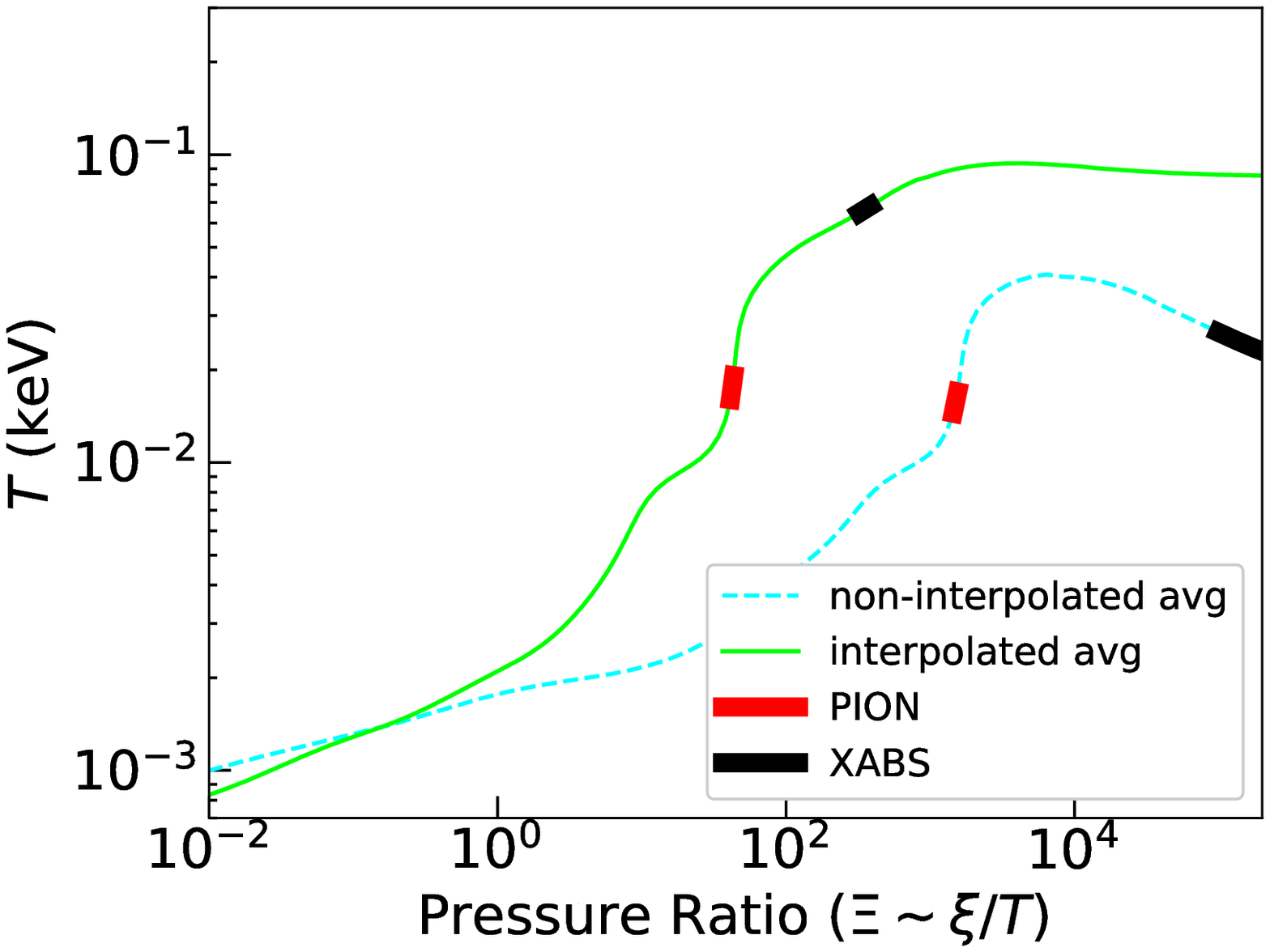}
\caption{
The time-average non-interpolated SED ({\it left}) predicted by models and the corresponding stability curve ({\it right}) compared with the interpolated SED and its stability curve. The ionization parameters of the photoionized absorber ({\it black}) and emitter ({\it red}) for each SED are also shown onto the curves.
}
\label{fig:nonchop}
\end{figure*}

\begin{figure}
\centering
\includegraphics[width=0.49\textwidth,trim={0 0.0cm 0 0}]{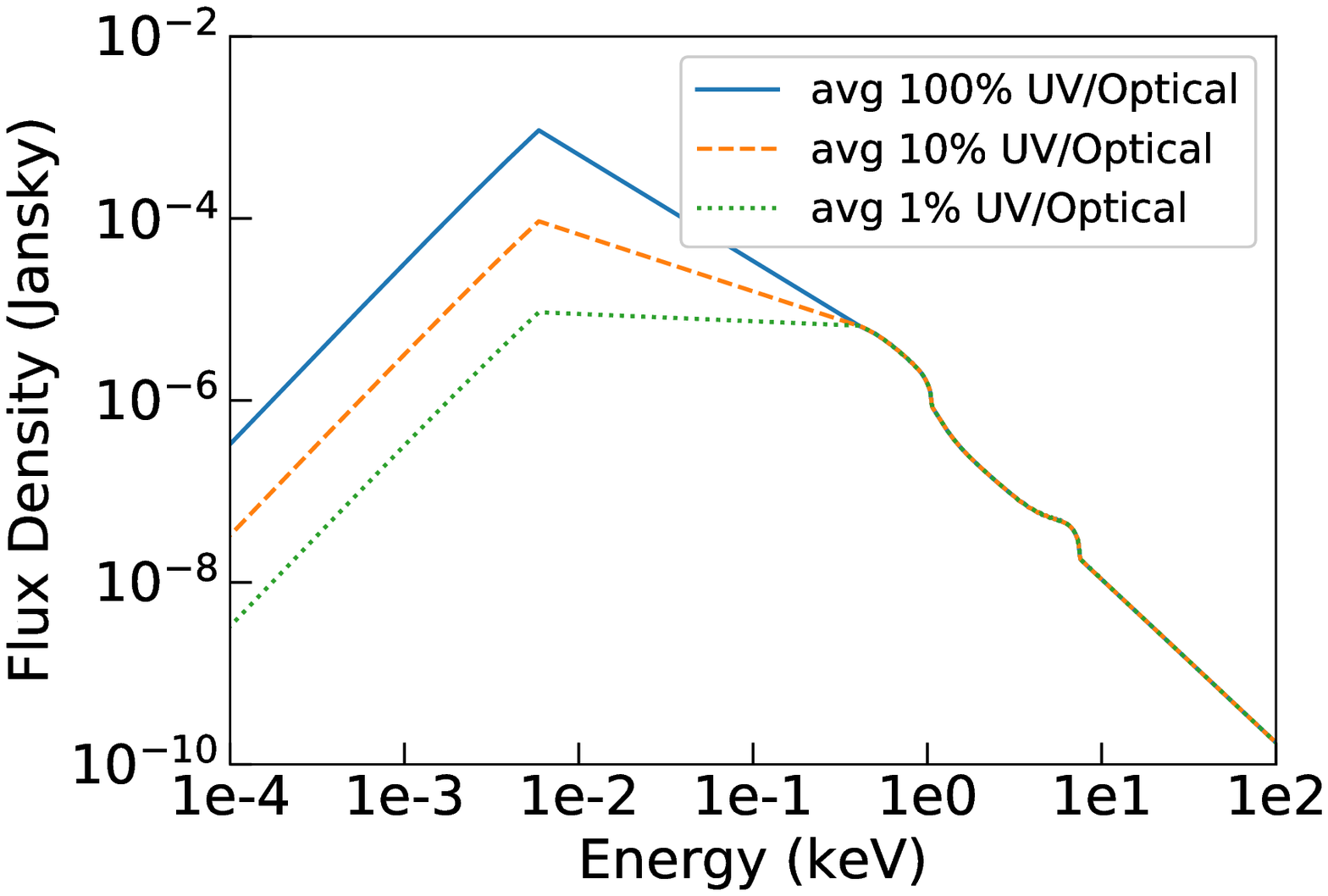}
\includegraphics[width=0.49\textwidth,trim={0 20 0 0}]{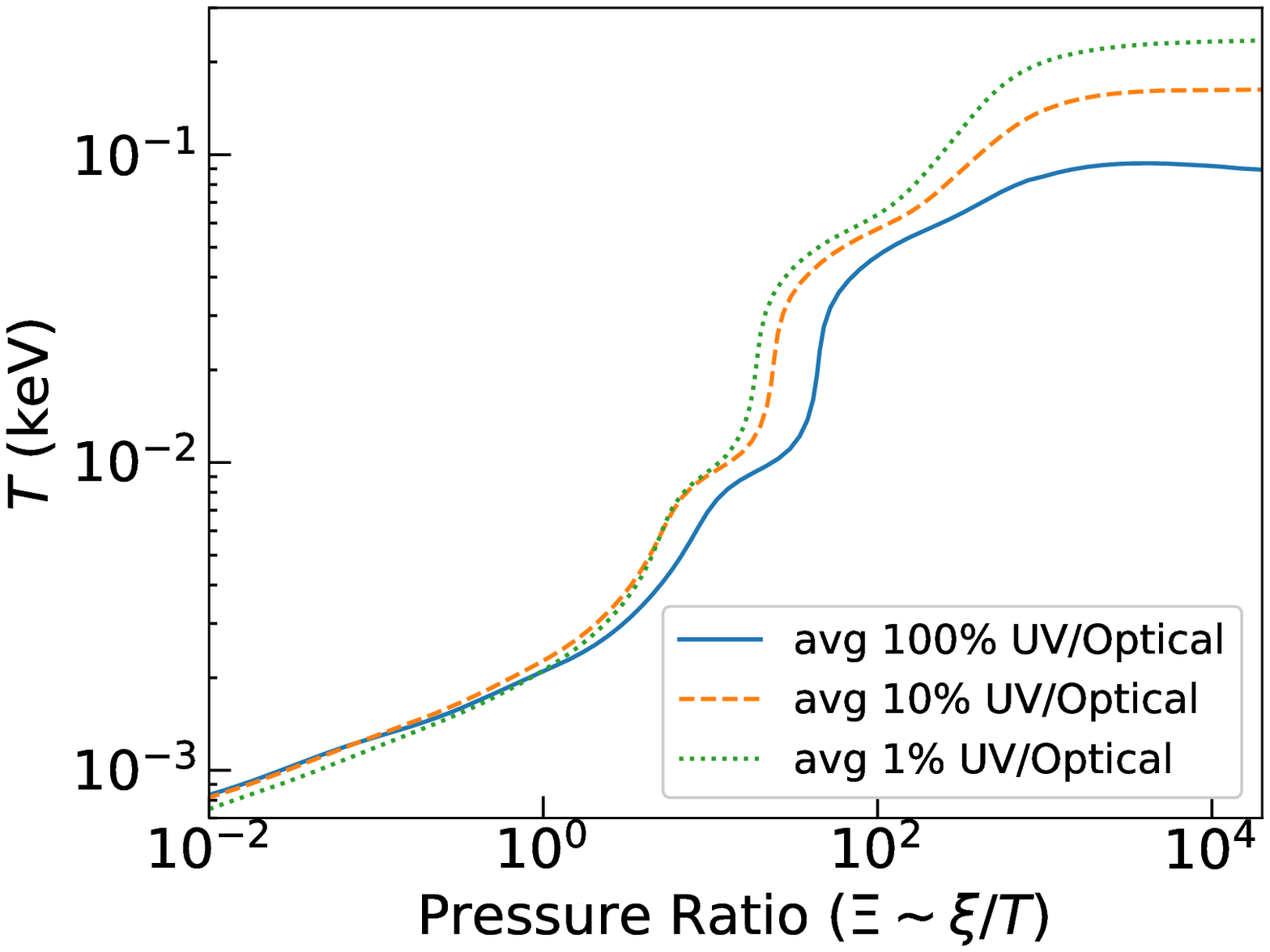}
\caption{
The time-average SED ({\it top}) and stability curves ({\it bottom}) of 1H 0707, compared to those in the cases of 10 and 1 percent UV/optical fluxes observed by the absorbing gas. The stability curves indicate that the winds are thermally stable even if possibly affected by high UV/optical screening.
}
\label{fig:ionUV}
\end{figure}

\begin{figure}
\centering
\includegraphics[width=0.49\textwidth,trim={0 10 0 0}]{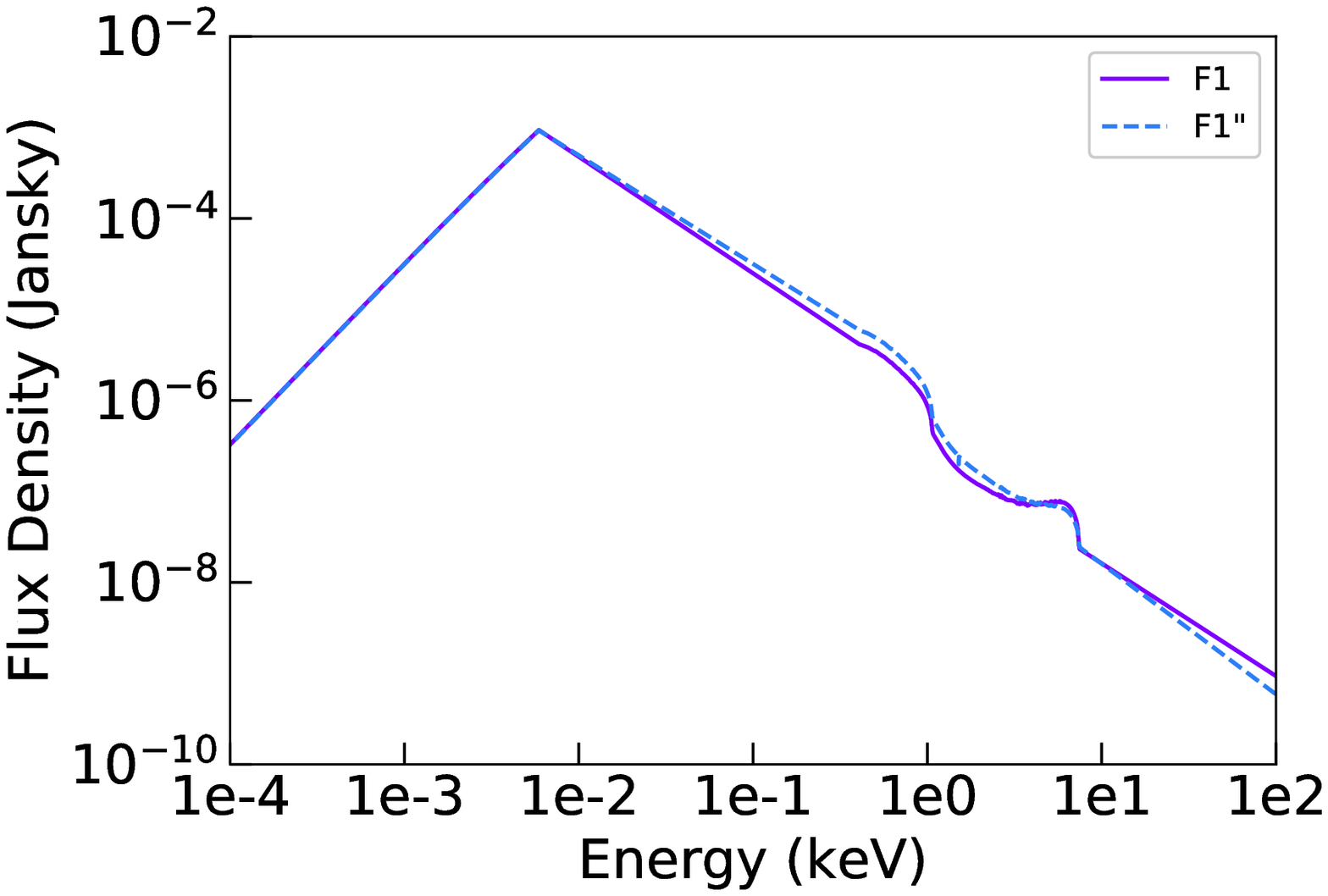}
\includegraphics[width=0.49\textwidth,trim={0 20 0 0}]{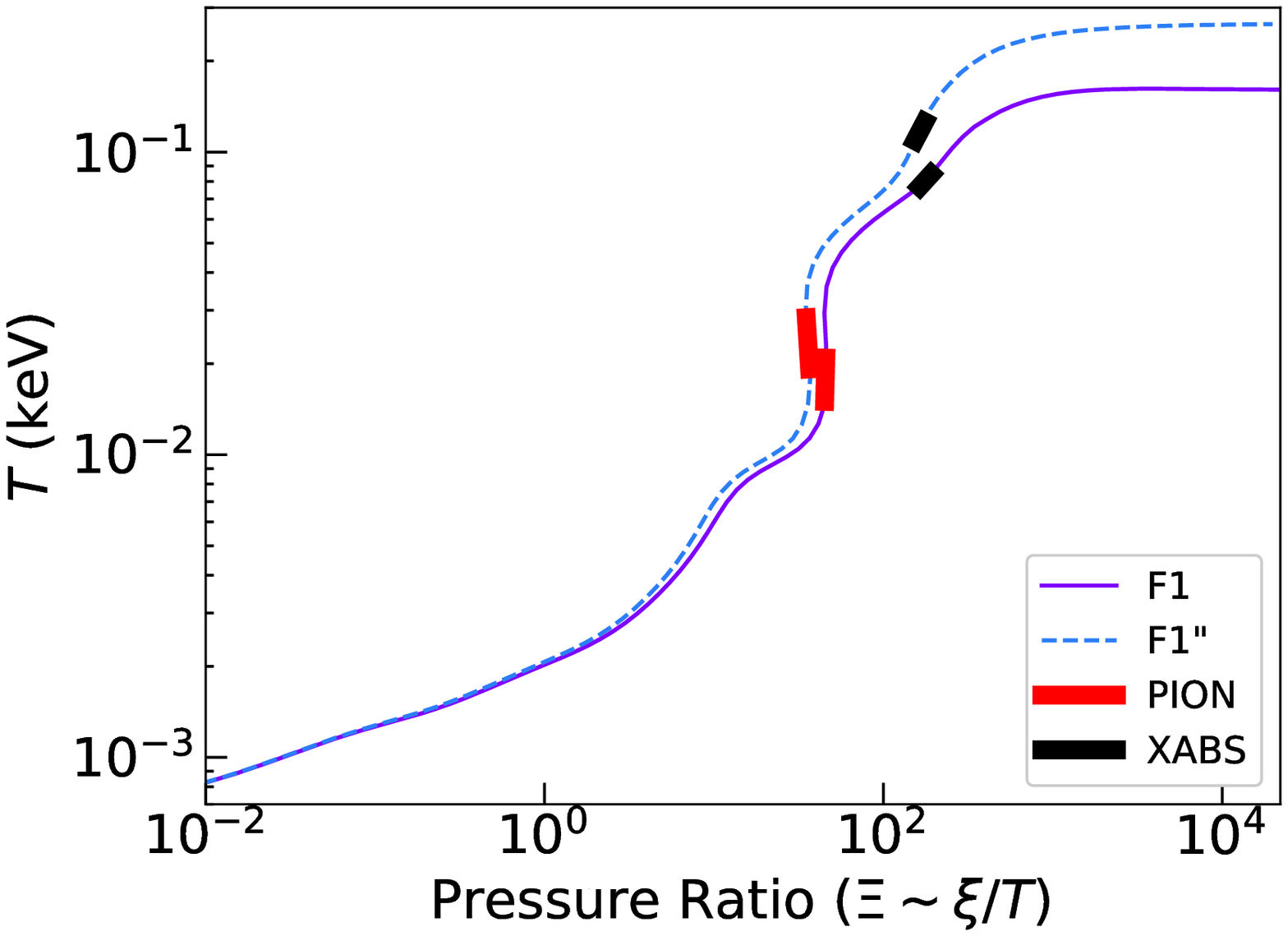}
\caption{
The comparison of the SED ({\it top}) and stability curves ({\it bottom}) between the full F1 and F1" spectrum. The {\it red} and {\it black} bold lines indicate the best solutions of the photoionization emission and absorption components respectively. The emitter and absorber detected in F1" spectrum are slightly hotter than those of F1 probably due to more X-rays. 
}
\label{fig:comparison}
\end{figure}

\bsp	
\label{lastpage}
\end{document}